\documentclass[twocolumn]{aastex62}
\usepackage[latin1]{inputenc}
\usepackage[english]{babel}
\usepackage{amsmath}
\usepackage{amsfonts}
\usepackage{amssymb}
\usepackage{makeidx}
\usepackage{graphicx}
\usepackage{amsmath}
\usepackage{natbib}
\usepackage{subfigure}
\usepackage[flushleft]{threeparttable}
\usepackage{etoolbox}

\usepackage{color}

\newcommand{\au}{~{\rm AU}}
\newcommand{\kms}{~{\rm km}~{\rm s}^{-1}}\newcommand{\Ms}{~{\rm M}_\odot}
\newcommand{\bbh}{{\rm BHB}}
\newcommand{\ev}{{\rm ev}}
\newcommand{\bh}{{\rm BH}}
\newcommand{\smbh}{{\bullet}}
\newcommand{\ARGdf}{\texttt{ARGdf }}
\newcommand{\ARCHAIN}{\texttt{ARCHAIN }}

\newcommand{\gc}{{\rm GC}}
\newcommand{\nc}{{\rm NC}}
\newcommand{\gw}{{\rm GW}}

\graphicspath{{./}{./}}

\shorttitle{Formation of BHBs around SMBHs}
\shortauthors{Arca-Sedda M.}

\begin{document}

\title{Birth, life, and death of black hole binaries around supermassive black holes: dynamical evolution of gravitational wave sources}

\correspondingauthor{Manuel Arca Sedda}
\email{m.arcasedda@gmail.com}

\author{Manuel Arca Sedda}
\affil{Zentrum f\"{u}r Astronomie der Universit\"{a}t  Heidelberg\\
Astronomisches Rechen-Institut\\
M\"onchhofstrasse 12-14\\
Heidelberg, D-69120, DE}

\begin{abstract}
In this paper, we explore the mechanisms that regulate the formation and evolution of stellar black hole binaries (BHBs) around supermassive black holes (SMBHs). We show that dynamical interactions can efficiently drive "in-situ" BHB formation if the SMBH is surrounded by a massive nuclear cluster (NC), while orbitally segregated star clusters can replenish the BHB reservoir in SMBH-dominated nuclei. We discuss how the combined action of stellar hardening and mass segregation sculpts the BHB orbital properties. We use direct N-body simulations including post-Newtonian corrections up to 2.5 order to study the BHB-SMBH interplay, showing that the Kozai-Lidov mechanism plays a crucial role in shortening binaries lifetime. We find that the merging probability weakly depends on the SMBH mass in the $10^6-10^9{\rm ~M}_\odot$ mass range, leading to a merger rate $\Gamma \simeq 3-8$ yr$^{-1}$ Gpc$^{-3}$ at redshift zero. Nearly $40\%$ of the mergers have masses in the "BH mass gap", $50-140{\rm ~M}_\odot$, thus indicating that galactic nuclei are ideal places to form BHs in this mass range. We argue that gravitational wave (GW) sources with components mass $m_1>40{\rm ~M}_\odot$ and $m_2<30{\rm ~M}_\odot$ would represent a strong indicator of a galactic nuclei origin. The majority of these mergers could be multiband GW sources in the local Universe: nearly $40\%$ might be seen by LISA as eccentric sources and, a few years later, as circular sources by LIGO and the Einstein Telescope, making decihertz observatories like DECIGO unique instruments to bridge the observations during the binary inspiral.
\end{abstract}

\keywords{black holes - supermassive black holes - galactic nuclei - gravitational waves}

\section{Introduction}
The vast majority of galactic nuclei, if not all, are expected to host a central supermassive black hole (SMBH), often surrounded by a nuclear cluster (NC). Large masses and densities make NCs excellent factories for the production of stellar-mass black holes, which possibly pair in binaries (BHBs) and occasionally merge releasing gravitational waves (GWs). The mechanisms that favour BHB formation in galactic nuclei are still partly unknown. 
In NCs without a central SMBH, dynamical interactions represent one of the dominant processes for BHBs buildup and merger \citep{miller09}, possibly contributing to the observed population of GW sources \citep{antonini16b,antonini18c}. The picture becomes more complex if the galaxy hosts an SMBH, as this can affect BHBs evolution in two ways. On the one hand, the high velocity dispersions in these environments suppress low-velocity dynamical interactions, particularly three-body scattering, leaving little room in the phase space for BHBs to form. On the other hand, newly formed BHBs can undergo Kozai-Lidov oscillations \citep[KL,][]{lidov62, kozai62} driven by the SMBH, which can induce the binary eccentricity to increase up to values close to unity and facilitate the merger \citep{antonini12,hong15,vanL16,hoang18,ASG17,ASCD17b,leigh18}.
Understanding what mechanisms regulate the formation of stellar-mass BHBs around an SMBH is still a partly open question of modern astrophysics.  
The larger density and escape velocities in galactic nuclei can allow merged BHs retention and recycling \citep{Gerosa17,antonini18c}, possibly leading to GW sources notably different from those originating via other formation channels \citep{ASBEN19}. Moreover, the presence of a SMBH might leave some information in the GW signal, depending on the SMBH-BHB orbital properties \citep{chen17,ASCD17b}.

Recently, a growing number of papers attempted at constraining BHBs merger rates for galactic nuclei environments \citep[see for instance][]{antonini12,vanL16,antonini16,hoang18,leigh18,ASG17,ASCD17,fragione19,hoang19,gourgoulhon19,zhang19}, showing that the range of possible values is particularly wide. The main source of uncertainty in most of the models is the poor knowledge of typical BHB orbital properties.

In this paper, we provide an extensive study aiming at describing all the phases of BHBs life in galactic nuclei: from the formation and evolution to the interaction with the central SMBH. As a first step, we explore two potential BHB formation channels, placing constraints on the maximum number of BHBs that can develop in galactic nuclei with different masses. As a second step, we focus on BHB hardening processes. As a last step, we follow BHB orbits around the SMBH taking advantage of direct N-body simulations. We use an updated version of the \ARCHAIN code, which features post-Newtonian formalism up to 2.5 order \citep{mikkola99,mikkola08} and the {\it algorithmic regularization} scheme to model close encounters. The paper is organized as follows: 
we explore BHBs formation channels in Section \ref{Sec2} and BHBs dynamics in Section \ref{Sec3}; Section \ref{Sec4} focuses on direct N-body simulations modelling the onset of KL effects; Section \ref{Sec5} is devoted to discuss the mergers properties in different GW observational bands; Section \ref{Sec6} summarizes the conclusion of this work.

\section{Binaries birth, life and death in galactic centres }
\label{Sec2}

\begin{figure*}
\centering
\includegraphics[width=\textwidth]{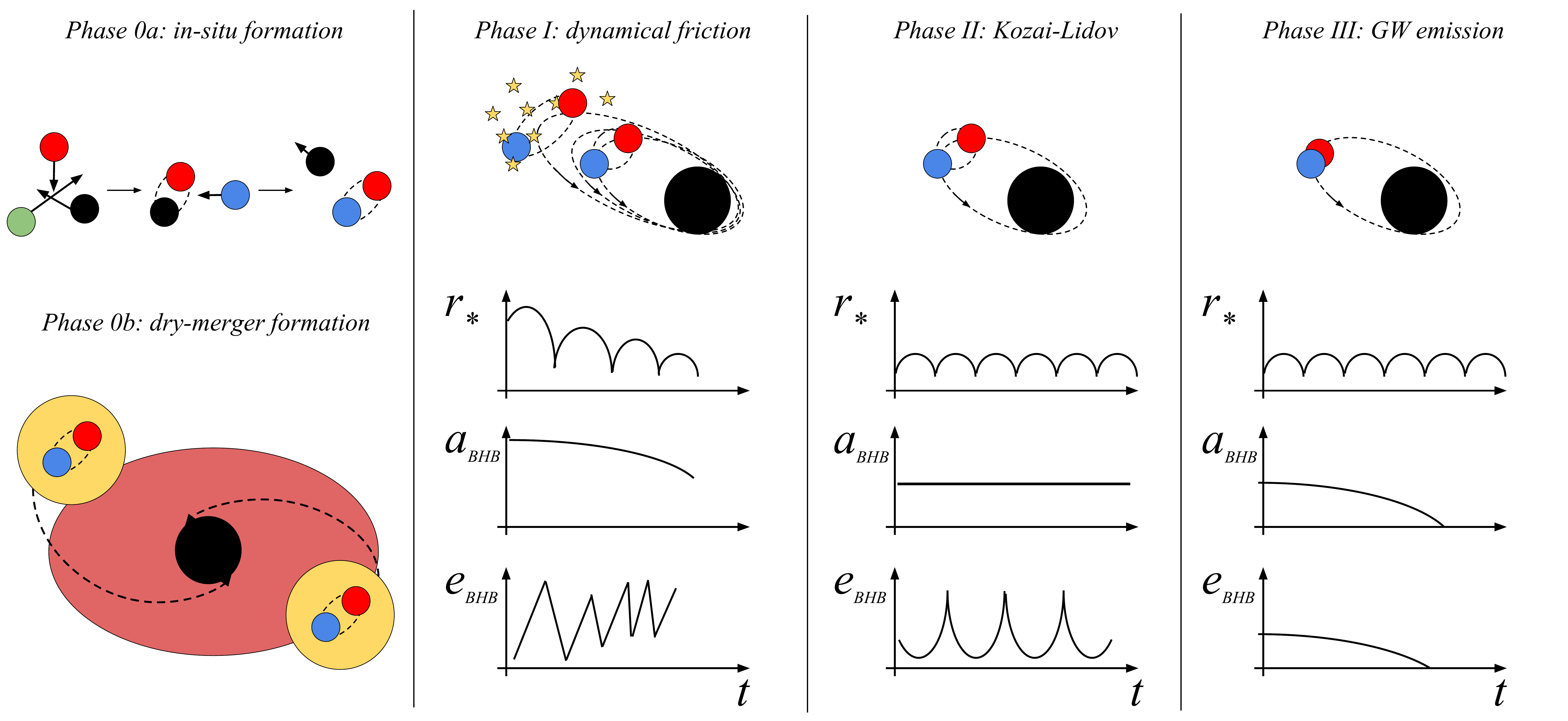}
\caption{Evolutionary phases of a BHBs in a galactic nucleus. The BHB form either via three-body encounters and component swap (Phase 0a) or are delivered in the galaxy centre by spiralling clusters (Phase 0b). Further close encounters drive BHB mass segregation and hardening (Phase I). This effect ceases as soon as the SMBH tidal field becomes dominant, depending on the orbital configuration Kozai-Lidov cycles may initiate and leads the BHB eccentricity to increase (Phase II). The eccentricity increase leads the BHB in the GW-dominated region (Phase III).}
\label{sketch}
\end{figure*}

The birth and life of a BHB orbiting a galactic nucleus can be roughly sketched in four phases, as  
depicted in Figure \ref{sketch}. After formation (Phase 0), the BHB undergoes mass segregation and hardens via stellar encounters (Phase I), until it reaches a point where the effects of the SMBH tidal field becomes significant. The BHB form a hierarchical triple with the SMBH (phase II), possibly undergoing KL oscillations that can drive a periodic increase of the BHB eccentricity. The eccentricity increase causes an enhancement of energy and momentum loss via GW emission that ultimately leads to the BHB merger (phase III).

The timescales associated to these phases are the BHB mass segregation -- i.e. dynamical friction -- time ($t_{\rm DF}$) in Phase I, the KL oscillation timescale ($t_{\rm KL}$) in Phase II and the GW merger time ($t_{\rm GW}$) in Phase III. At the moment of BHB formation (or deposit) in general we expect $t_{\rm DF}<t_{\rm KL}<t_{\rm GW}$, although these inequalities depend strongly on the host galaxy local properties. 

Although an SMBH-BHB triple represents an appealing system to be studied with a secular approximation formalism, Phase I can represent a crucial step to be explored, as the triple is not isolated. The continuous interactions with galaxy stars can either cause the BHB hardening or its disruption. Moreover, since BHBs are the heaviest stellar objects in the nucleus they can undergo mass segregation, thus leading to a progressive reduction of the BHB-SMBH separation.

In the following, we discuss the possibility that BHBs either form {\it in-situ} or are {\it delivered} into the galactic nucleus. In the ``in-situ'' hypothesis, we assume that BHBs form via gravitational encounters. We make use of the classical arguments that describe three-body scattering \citep{lee95} and binary-single interactions \citep{miller05} to calculate how the population of binaries evolve in time. 

In the ``delivery'' hypothesis, instead, we assume that a population of BHBs is deposited in the NC by massive star clusters that spiral toward the galactic centre due to dynamical friction. This mechanism is thought to contribute significantly to NC formation \citep{Trem75,Dolc93}, providing an excellent explanation for the observed galaxy-NC relations \citep{gnedin14,ASCD14b,Antonini15}. 
Like oysters, spiralling clusters drag toward the centre their compact remnants, which are likely segregated into their core. During the phases of cluster dispersal, the remnants are released in the growing NC, and stars moving in the core of the spiralling clusters are most likely deposited in the NC core, where interactions are frequent due to the high densities \citep{perets14,abbate18,ASK18}.
The interaction with NC stars will force the delivered BHBs to further spiral into the NC because of mass segregation, transiting through regions at increasing densities and velocity dispersion. In the next section, we show that this facilitates BHBs hardening and merger, on average, unless they are in a very ``soft'' status when leaving their parent clusters. 

\begin{figure}
\centering 
\includegraphics[width=\columnwidth]{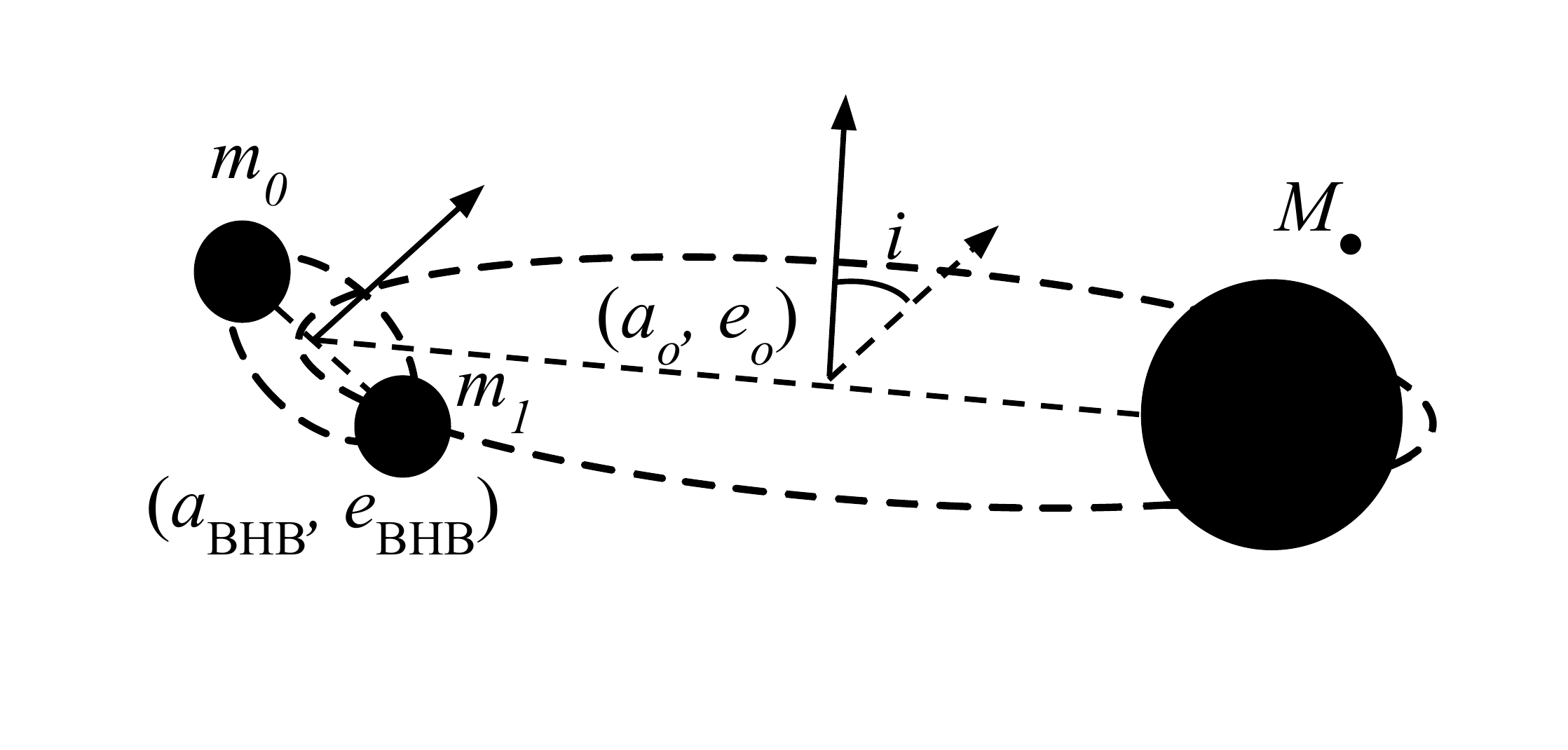}
\caption{Sketch of the BHB-SMBH triple system.}
\label{F111}
\end{figure}

As long as new BHBs are delivered from spiralling clusters, or form in the nucleus via dynamical interactions, their evolution will be inevitably affected by the SMBH tidal field, which can shorten their merger timescale via KL mechanism. Therefore, the BHB-SMBH form a three-body system that can be described in terms of an {\it inner binary} (the BHB), and an {\it outer binary} (the BHB-SMBH system) as sketched in Figure \ref{F111}. We label the inner binary components mass with $m_{0,1}$, the total mass, semi-major axis and eccentricity with $m_\bbh$, $a_\bbh$ and $e_\bbh$, respectively. The outer binary orbital parameters are identified with letter $o$, while the SMBH mass is $M_\bullet$. Accordingly, the inner(outer) binary orbital period is labelled with $P_\bbh$($P_o$). 

\subsection{Black hole binaries formation mechanisms: in-situ vs. dry-merger}
\subsubsection{In-situ scenario}

One of the most efficient channel to form binaries in galactic nuclei is via three-body scatterings, which generally lead to the ejection of one object -- most likely the lightest -- leaving behind a binary \citep{goodman93,lee95}. We define {\it hard} binaries those having a binding energy larger than the mean kinetic energy of the surrounding environment namely $a_\bbh>a_{\rm hard} = Gm_\bbh/2\sigma_g^2$ \citep{heggie75}. This mechanism is thought to be more efficient than two-body gravitational capture, which requires extremely close fly-bys \citep{lee95}, and binary-binary interactions \citep{mikkola84,mcmillan91,miller02,miller09}. 
Alternatively, a fraction of stars in galactic nuclei can form in {\it primordial} binaries, i.e. 
stars that form in the same protostellar cloud and shares a common (stellar+dynamical) evolution. Indeed, although this channel is poorly explored from a theoretical and numerical perspective, with a few notable exceptions \citep[see for instance][]{stephan16,panamarev18,naoz18,stephan19}, observations of the Galactic Centre provided recently clues on the possible binariety of several stars inside the inner 0.1 pc \citep{ott99,pfuhl14} or among the so called S-stars \citep{jia19}.

Regarding dynamically formed binaries, \cite{goodman93} used scattering experiments to derive a formation rate for hard binaries in the form
\begin{equation}
\dot{n}_{\rm bin} = \alpha G^5 m^5 \sigma_g^{-9} n_*^3,
\label{dn}
\end{equation}
being $\sigma_g$ the galaxy velocity dispersion, $m$ the average stellar mass and $n_*$ the stellar number distribution. The expression above relies on the assumption that the binary formation and disruption rates balance each other \citep{goodman93}.
A further assumption is that binaries form and destroy in a uniform sea of single-mass stars following a Maxwellian distribution of velocities. Such approximation might break-down in the closest vicinity of a SMBH, due to the peculiar mass distribution and velocity dispersion profile. However, observations have revealed that the old population of stars inhabiting the Galactic Centre are characterized by a phase space distribution that closely resambles a uniform distribution with Maxwellian velocities \citep{trippe08}, thus suggesting that Equation \ref{dn} can be applied to galactic nuclei, with some caution.
Assuming that the total number of stars remains constant over time, the rate at which binaries form must equal half the rate at which single stars number decreases, i.e. $2\dot{n}_{\rm bin} + \dot{n}_* = 0$.

Upon this condition we can integrate Equation \ref{dn} to determine how the binary number density varies over time:
\begin{align}
n_{\rm bin} &= \frac{n_*}{2}\left(1 - \frac{1}{\sqrt{1+t/t_{\rm 3bb}}}\right),\\
t_{\rm 3bb} &= \alpha G^{-5} m^{-5} \sigma_g^9 n_*^{-2}
\label{ntime}
\end{align}
where $t_{\rm 3bb}$ is the three-body interaction timescale \citep{lee95}. 
The simple assumptions above imply that all the single stars end up in a binary, therefore Equation \ref{ntime} provides an upper limit to the total number of binaries formed in three-body encounters, rather than a precise estimate of this quantity.
Even under optimistic assumptions, in a typical NC characterized by a stellar population with mean stellar mass $m\simeq 1\Ms$, velocity dispersion $\sigma_g \simeq 30\kms$ and stellar density $n\simeq 10^6$ pc$^{-3}$, three-body encounters are expected to take place in quite a long timescale, being $t_{\rm 3bb}\sim 5 \times 10^9$ yr. 

Since both stellar density and velocity dispersion varies at varying the distance from the galaxy centre, the $t_{\rm 3bb}$ timescale must be considered as a local quantity and calculated in different shells centered in the SMBH position. Thus, the number of binaries in each galaxy shell is given by $N_{\rm bin} = n_{\rm bin}V_{\rm shell}$, being $V_{\rm shell}$ the shell volume.

The number of BHs in binaries will constitute only a small fraction $f_{\bh}$ of the whole binary population. 
According to a standard \cite{kroupa01} mass function, and assuming $M_{\rm MS} = 18\Ms$ as the minimum mass for a star to turn into a BH \citep{belckzinski10}, this corresponds to $f_\bh \simeq 0.001$. However, it must be noted that the BH population is most likely strongly segregated within the NC, since BHs are the heaviest objects in a stellar ensamble. For instance, as supported by observational \citep{hailey18} and theoretical \citep{faucher11,ASK18,generozov18} arguments, the MW's inner pc is expected to harbour up to $N_\bh\sim 2.5\times 10^4$ BHs. The observed NC mass enclosed within 1 pc from the SMBH is $M_\nc \sim 2\times 10^6\Ms$ \citep{gillessen09,genzel10,schodel14,feldmeier17}, thus assuming an average BH mass $m_\bh\sim 10\Ms$, the number of BHs over the total number of stars in the MW central pc is roughly $f_\bh \simeq (N_\bh m_\bh)/M_\nc = 0.1$.

Therefore, a NC containing $N_* \sim 10^6$ stars after a time $t_{\rm 3bb}$ will contain, roughly, a number of BHBs $N_\bbh = (n_{\rm bin}(t_{\rm 3bb})/n_*)(N_* / 2)f_{\bh}= 0.075 - 7.5$. The lower(upper) limit corresponds to an unsegregated(segregated) BH population. This simple estimate outlines immediately how hard is for a BHB to form only via three-body encounters, unless BHs are strongly mass-segregated.

\begin{figure}
\includegraphics[width=\columnwidth]{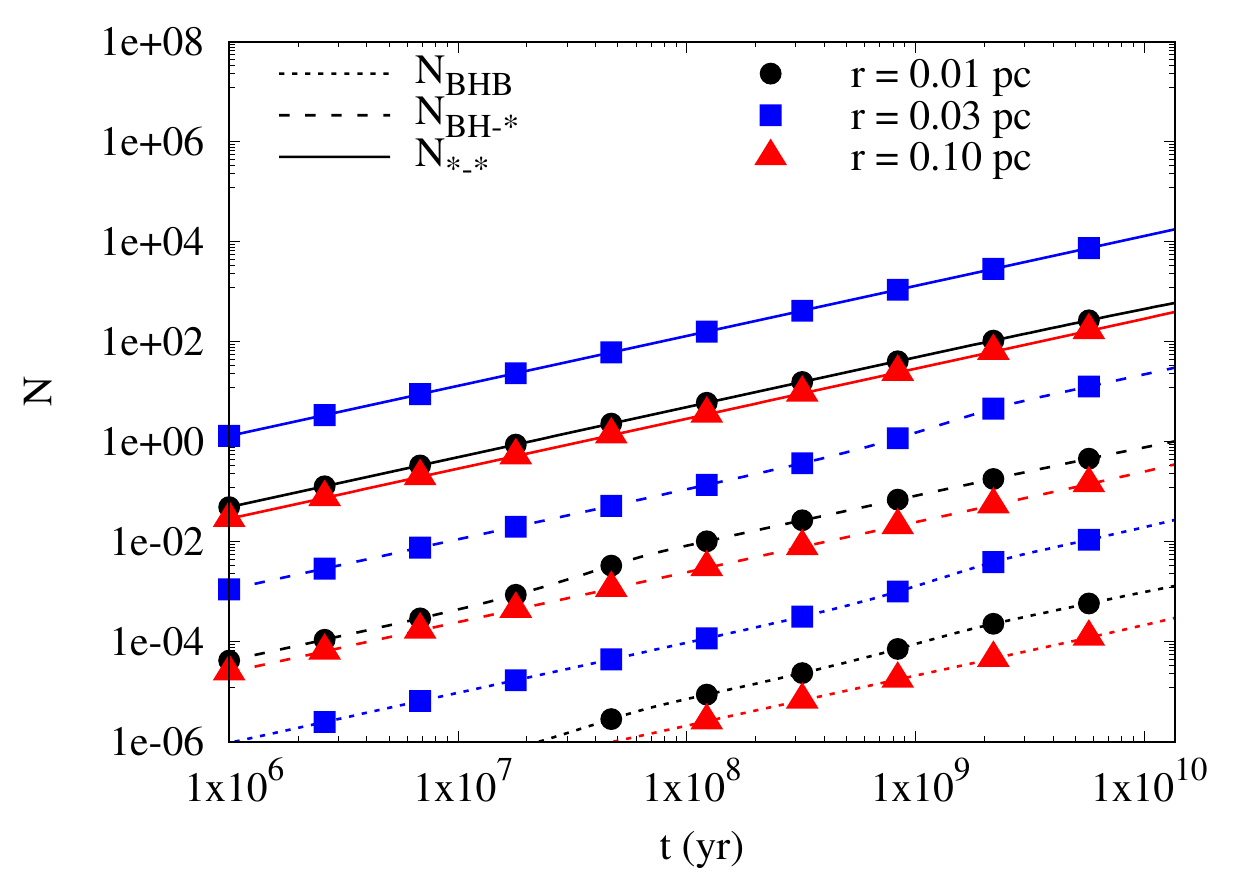}\\
\includegraphics[width=\columnwidth]{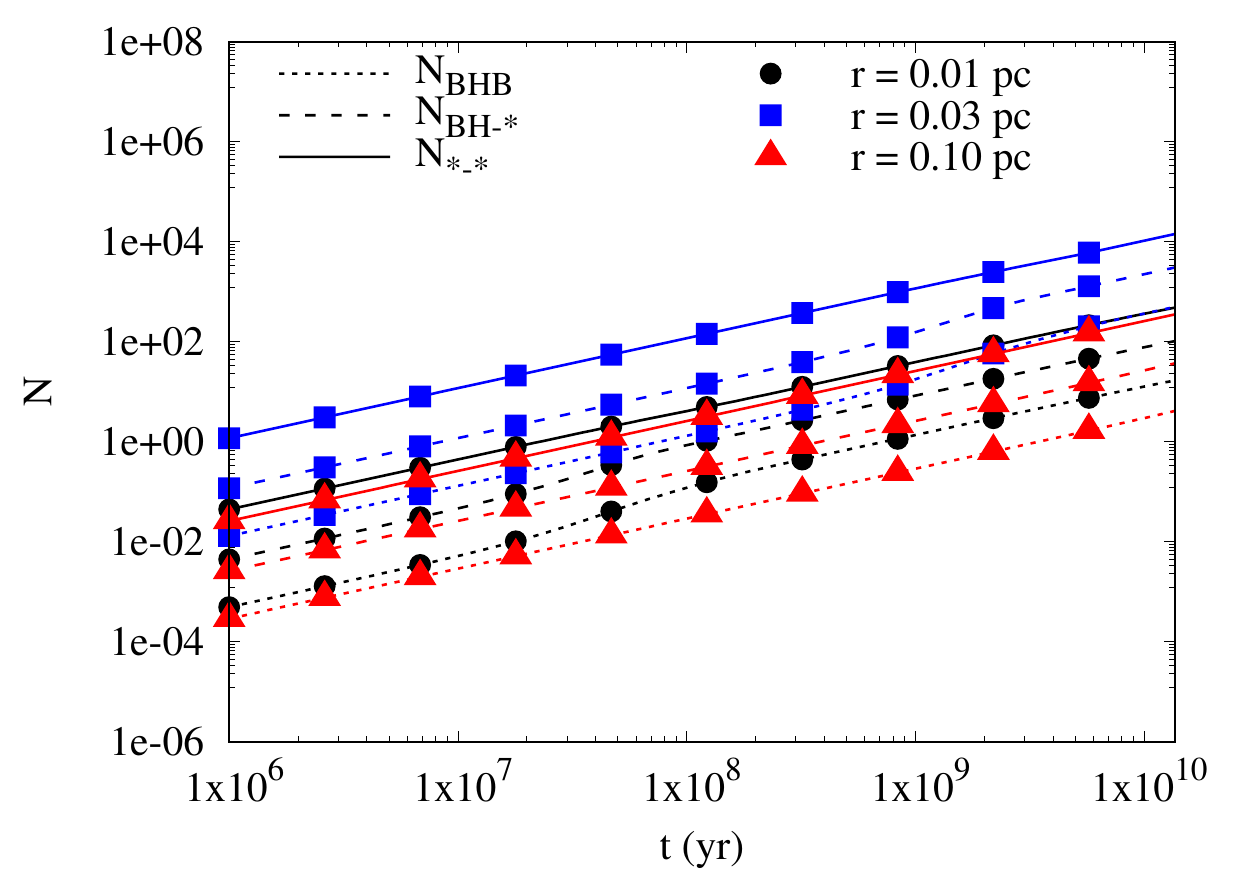}
\caption{Number of BH-BH (dotted), BH-star (dashed) and star-star (straight) binaries as a function of time and in different radial distance bins. The population of BHs is assumed to be either completely unsegregated during the binary formation process (top panel), or fully segregated within 1 pc from the SMBH (bottom panel).}
\label{nbinar}
\end{figure}

A way to further enrich the population of binaries with at least one BH is via component swap, which becomes important as soon as binaries start to form. As discussed by \cite{miller09}, a BH approaching a stellar binary with mass $m_{\rm bin}$, semi-major axis $a$ and eccentricity $e$ will replace one of the components over a typical time-scale 
\begin{equation}
t_{\rm exc} = \left(\sigma_g \Sigma n\right)^{-1}.
\end{equation}
The quantity $\Sigma$ represents the binary interaction cross section at pericentre, namely
\begin{equation}
\Sigma = \pi a^2 (1-e)^2\left(1+\frac{G(m_{\rm bin} +m_\bh)}{2\sigma_g a}\right),
\end{equation}
with $a$ and $e$ being the binary semi-major axis and eccentricity.
Note that at a larger binary mass corresponds a shorter exchange time, thus implying that binary systems containing already a BH have a larger probability to acquire another one. In the following, we use $t_{\rm *}$ to refer to the timescale for a star-star binary to acquire a BH, and $t_{\rm BH}$ to refer to binaries already containing a BH and becoming a BHB via component swap. Note that for a given binary $t_{\rm BH}<t_{\rm *}$, thus BH capture is favoured compared to stellar capture.
Over a time-scale $t_{\rm exc}$ some binaries are expected to acquire either one, or even two, BHs.
This transition from a BH-free to a BH-rich configuration will be regulated by some transfer function $\mathcal{F}$, and the number density of binaries undergoing an exchange will be related to the total number as $n_{\rm exc} = n_{\rm bin}\mathcal{F}$. For the sake of simplicity, in the following we assume that $\mathcal{F}$ is a simple function of the time
\begin{equation}
\mathcal{F}_{\rm t_{\rm exc}}(t) = 1-\exp(-t/t_{\rm exc}).
\end{equation}
Dividing the whole population in BH-BH, BH-star and star-star systems in such a way that $n_{\rm bin} = n_\bbh + n_{\bh-*} + n_{*-*}$ allows us to write the equations that describe how these three types of binaries evolve in time
\begin{align}
\frac{n_\bbh}{n_{\rm bin}}    =& f_\bh^2 \left\{1 + (1-f_\bh)\left[\mathcal{F}_{\rm t_{\rm BH}}(t) + \mathcal{F}_{\rm t_{\rm *}}(t) \right] \right\}; \label{n1}\\                   
\frac{n_{\bh-*}}{n_{\rm bin}} =&	f_\bh(1-f_\bh)\left\{1-f_\bh\mathcal{F}_{\rm t_{\rm BH}}(t) +\mathcal{F}_{\rm t_{\rm *}}(t)\right\}; \label{n2}\\ 
\frac{n_{*-*}}{n_{\rm bin}}   =&	(1-f_\bh)\left[1-f_\bh\mathcal{F}_{\rm t_{\rm BH}}(t)-f_\bh^2\mathcal{F}_{\rm t_{\rm *}}(t)\right];
\label{n3}
\end{align}
note that the $\mathcal{F}_{\rm t_{\rm BH}}(t)$ refers to a binary containing a BH and acquiring a second BH ($t_{\rm exc}\equiv t_{\rm BH}$), while $\mathcal{F}_{\rm t_{\rm *}}(t)$ refers to a star-star binary that acquires one BH ($t_{\rm exc}\equiv t_{\rm *}$). 
Moreover, all the quantities above vary with the distance to the galaxy centre, thus they represent local estimates.
Using equations \ref{n1}-\ref{n3}, we calculate the number of BH-BH, BH-star and star-star binaries in three galaxy shells centered in $0.01$, $0.03$ and $0.1$ pc in a MW-like NC, assuming $M_g = 2.5\times 10^7\Ms$, $R_g=2$ pc, inner slope of the density profile $\gamma = 1.8$ and SMBH, $M_\smbh = 4.5\times 10^6\Ms$. Note that the values chosen for the NC scale radius and slope correspond to an effective radius of $R_{\rm eff} \simeq 4$ pc \citep{Deh93}, compatible with the corresponding observed quantity \citep{schodel14}.
Figure \ref{nbinar} shows how these numbers vary in different shells and at different times for unsegregated and segregated BHs.

Our model predicts that the maximum number of stellar binaries is achieved at distances $\sim 0.03$ pc, while rapidly dropping below and above this radius. If mass segregation is inefficient in dragging BHs into the Galactic Centre, $f_\bh = 0.001$, we find that a handful binaries containing one BH form after 10 Gyr, while no BHBs develop. Efficient mass segregation, however, can change the picture significantly, driving the formation of a few hundreds of BHBs at 0.03 pc from Sgr A*. Our results are summarized in Table \ref{T2}.

\begin{table}
\caption{Binaries number in different Galactic NC shells at $t=10$ Gyr.}
\begin{center}
\begin{tabular}{cccc}
\hline
\hline
$r_{\rm shell}({\rm pc}) $ & $N_{\rm BH-BH} $ & $N_{\rm BH-*} $ & $N_{\rm *-*} $\\
\hline
\multicolumn{4}{c}{unsegregated ($f_\bh = 10^{-3}$)}\\
\hline
$0.01 $ & $1.3\times 10^{-3} $ & $1   $ & $687 $ \\
$0.03 $ & $2.3\times 10^{-2} $ & $32  $ & $17434 $ \\
$0.10 $ & $2.8\times 10^{-4} $ & $0.4 $ & $398 $ \\
\hline
\multicolumn{4}{c}{segregated ($f_\bh = 10^{-1}$)}\\
\hline
$0.01 $ & $17 $ & $106  $ & $462 $ \\
$0.03 $ & $534$ & $3194 $ & $14487 $ \\
$0.10 $ & $4  $ & $39   $ & $344 $ \\
\hline
\end{tabular}
\end{center}
\label{T2}
\end{table}

Our results suggest that the Galactic NC might harbour $N_{\rm BH-*} \sim 10-3000$ BH-star binaries within 0.1 pc. 
Assuming a population of $\sim 2.5\times 10^4$ BHs \citep{hailey18}, our results suggest that up to $12\%$ of BHs in the Galactic Centre might be in a binary \citep[see also][]{faucher11,generozov18}, similar to the fraction of BHs in binaries estimated recently for globular clusters \citep{chatterjee17a,AAG18a,AAG18b}. Some BH-star binaries can undergo a phase of X-ray emission, evolving either into low-mass or high-mass X-ray binaries. In these regards, it is worth recalling the recent observations provided by Chandra satellite, which unveiled the presence of 12 low-mass X-ray binaries inhabiting the inner pc of the Galactic NC \citep{hailey18}. Half of these sources, if not all, might contain a BH, and their presence can be crucial to better understand how binaries form in galactic nuclei \citep{generozov18}. Unfortunately, the number of X-ray binaries in typical clusters seems to be independent on the cluster properties \citep{kremer18a}, making hard to link the total number of BHs, or BH in binaries, to the number of low-mass X-ray binaries. 

Given the observational limits, we infer from our calculations that the fraction of X-ray binaries containing a BH is $\sim 5\times 10^{-4}$ times the total number of BHs or $\sim 4\times 10^{-3}$ times the number of BHs in binaries.
A powerful way to test these predictions is via $N$-body simulations. Unfortunately, simulations that account for different stellar components are extremely expensive from the computational point of view and became affordable only recently, although they still rely on several simplified approximations \citep[see for instance ][]{ASG17,abbate18,ASK18,baumgardt18,panamarev18}. 

Varying the SMBH mass and the NC properties, we use the approach described in this section to calculate the number of BHBs that form over 10 Gyr in different galactic nuclei models. Nuclear clusters mass, scale radius and density slope are selected in the ranges  $M_\nc = 10^6-10^8\Ms$, $R_\nc = 0.8-2$ pc, and $\gamma_\nc = 0.5-2$, respectively, as suggested by observations \citep[see for instance][]{georgiev14}. The corresponding NC scale density is defined as $\rho_{\nc} = (3-\gamma_\nc)M_\nc/(4\pi R_\nc^3)$ \citep{Deh93}.

\begin{figure}
\centering
\includegraphics[width=\columnwidth]{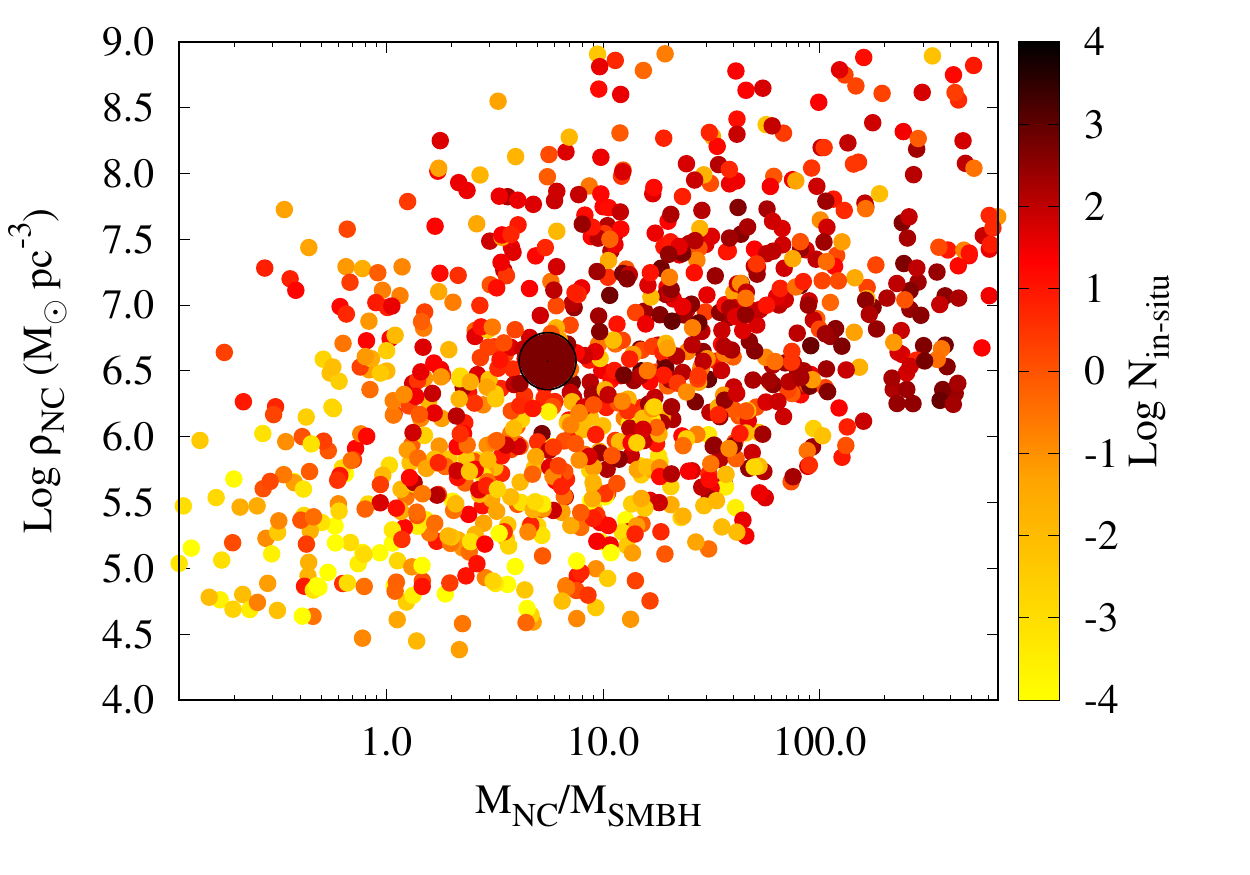}
\caption{Number of BHB (color coded) formed via three-body scattering and components swap as a function of the NC-to-SMBH mass ratio (x-axis) and the NC scale density (y-axis). The big dot identifies the Galactic NC model.}
\label{ncsmbh}
\end{figure}

We find that the NC-to-SMBH mass ratio, $M_\nc /M_\smbh$, and the NC density $\rho_\nc$ affects significantly the number of BHBs, as shown in Figure \ref{ncsmbh}. The dynamical formation of BHBs seem strongly suppressed in low-density NC with masses smaller than the central SMBH. These environments are typical of massive elliptical galaxies, where NCs are expected to be faint and sparse \citep{graham09,Neum12}. Galaxies in which the NC dominates over the SMBH mass, instead, are more suitable to host BHB formation, being in general $N_\bbh \gtrsim 10$ for galaxies with $M_\nc>10M_\smbh$ and $\rho_\nc>10^6\Ms$ pc$^{-3}$. This kind of environment is more common in intermediate mass galaxies like the MW. Note that at fixed $\rho_\nc$, systems having larger $M_\nc/M_\smbh$ are those in which the SMBH is smaller, thus its suppressive effect on the BHB formation is reduced and leads to a larger, on average, number of binaries. At the same time, systems at a fixed $M_\nc/M_\smbh$ value can represent NCs with different structures (inner slope and length-scale of the density profile), thus with different central velocity dispersion and density. The large scatter apparent in the figure is likely a combination of all these factors.

\subsubsection{Dry-merger scenario}
An alternative mechanism, still poorly investigated in the literature, is that the BHB reservoir is incremented or replenished via delivery from orbitally segregated star clusters.  
If the formation and growth of the inner part of a galaxy and its NC partly arises from several star clusters collisions and mergers, the so-called ``dry-merger'' scenario \citep{Trem75,Dolc93}, it is possible that former star clusters' members pollute the galaxy environment. For instance, the observed Galactic Centre high-energy excess \citep{hooper11,perez15,hailey16,hailey18,bartels16,calore16}, can be interpreted as the results of emission coming from compact sources deposited by $\sim O(10)$ spiralling clusters into the growing NC \citep{bednarek13,brandt15,ASK18,abbate18,fragione17b}. 

As recently shown by \cite{belczynski18}, the number of sources delivered by spiralling clusters can be calculated using semi-analytic arguments \citep{ASCD14b}.
The number of decaying clusters can be calculated as
\begin{equation}
N_{\rm GC,dec} = f_{\rm dec} f_{\rm GC} M_g M_{\rm GC}^{-1},
\label{ndec}
\end{equation}
where $f_{\rm dec}$ is the ratio between the number of spiralled clusters and the total number of clusters in the host galaxy, $f_{\rm GC}$ is the fraction of galaxy mass converted into star clusters, $M_g$ is the total galaxy mass and $M_{\rm GC}$ is the mean star cluster mass.

For a typical star cluster, the total number of BHs in binaries can be calculated as the product of the BH retention fraction, the fraction of BHBs ($f_{\rm bin}$), and the total number of BHs. The BH retention fraction depends on the host cluster properties \citep{morscher15,AAG18a,AAG18b}. Using the data provided by \cite{morscher15}, we find that the retention fraction can be connected to the cluster mass via a power-law, $\alpha_r (M_{\rm GC}/10^5\Ms)^{\beta_r}$, being $\alpha_r \simeq 0.16$ and $\beta_r \simeq 0.35$. The  number of BHBs per cluster is then given by
\begin{equation}
N_{\rm BH,bin} = 10^5f_{\rm bin}\alpha_r\left(\frac{f_\bh}{m_\bh}\right)\left(\frac{M_{\rm GC}}{10^5\Ms}\right)^{\beta_r+1}
\label{nret}
\end{equation}
Combining equations \ref{ndec} and \ref{nret}, and rearranging them conveniently, we can roughly estimate the number of BHBs delivered into the galactic centre by spiralling clusters as
\begin{equation}
N_{\rm dry} = f_{\rm dec} f_{\rm GC} f_\bh f_{\rm bin}\alpha_r \frac{M_g}{m_\bh}\left(\frac{M_{\rm GC}}{10^5\Ms}\right)^{\beta_r}.
\end{equation}
Equation above is affected by many sources of uncertainty: the fraction of orbitally segregated clusters $f_{\rm dec}$ may vary with the galaxy total mass \citep{ASCD14b}, the fraction of galaxy mass that is converted in star clusters is thought to range between $f_{\rm GC} = (0.2-2)\times 10^{-2}$ \citep{gnedin14,ASCD14b,webb15}, the fraction of BHs bound in BH-BH binaries depends on the cluster properties, and it is expected to vary between $f_{\rm bin} = 0.01$ and $0.2$, depending on the host clusters properties \citep{AAG18a}. 
We discuss the typical timescales over which this mechanism operates in Appendix \ref{AppA}.
In Figure \ref{nbhbdry}, we show how  $N_{\rm dry}$ varies at varying the galaxy and the average cluster mass $M_{\rm GC}$, assuming $f_{\rm dec}=0.1[M_g/(6\times 10^{10})]^{-1/2}$, $f_{\rm GC}=0.01$, $f_{\rm bin}=0.1$, and $m_\bh=10\Ms$. 
We find that the number of delivered BHBs in low-mass galaxies is relatively small, generally $N_{\rm dry}<10$, due to the small number of clusters that are expected to form  in there and undergo efficient orbital segregation. On another hand, MW-like and heavier galaxies ($M_g \gtrsim 10^{11}\Ms$ and $M_{\rm GC}\sim 3\times 10^5\Ms$) can accumulate a few hundreds of BHBs through this mechanism. This trend is opposite compared to results discussed for the in-situ formation channel.

\begin{figure}
\centering
\includegraphics[width=\columnwidth]{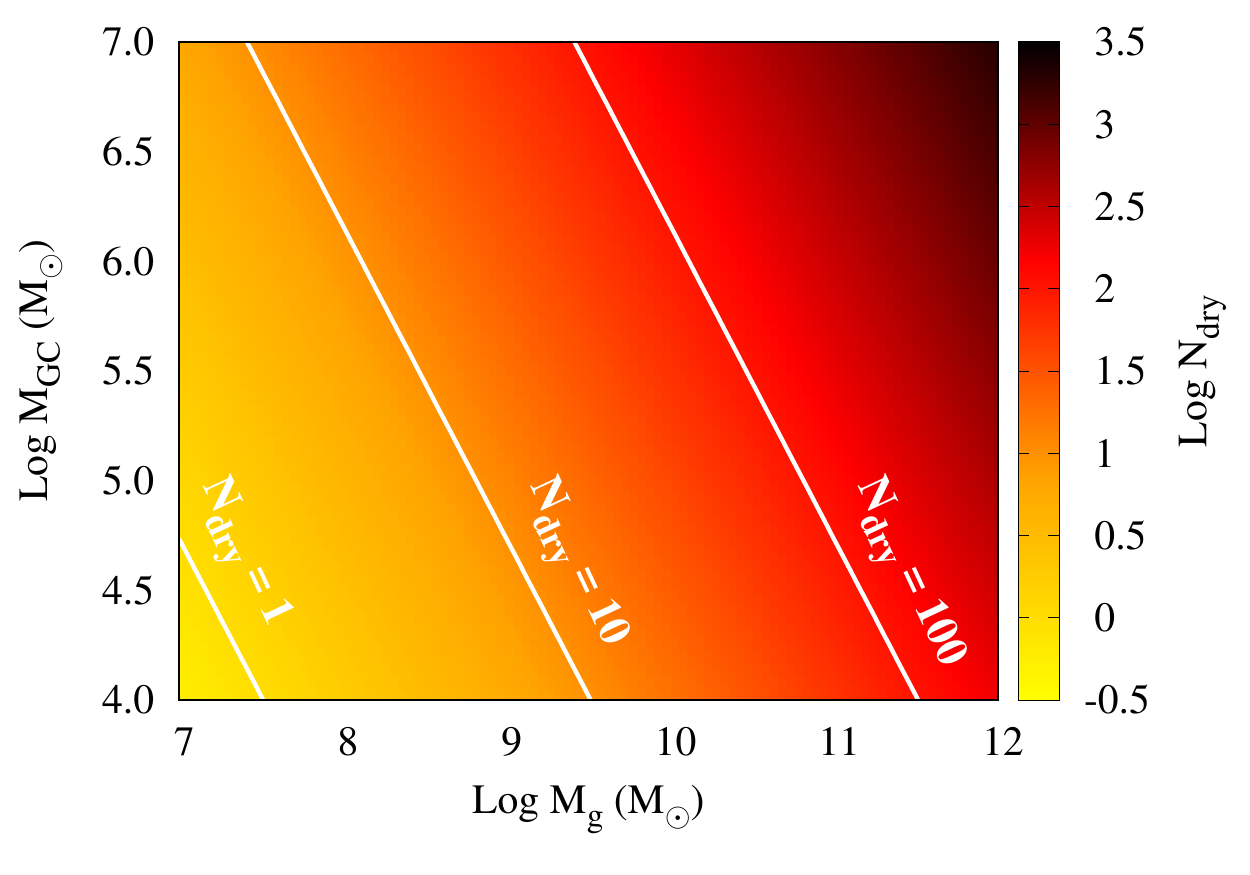}
\caption{Number of delivered BHBs (color-coded) as a function of the galaxy mass (x-axis) and the average cluster mass (y-axis). From left to right, white lines represent the locus of models having $N_{\rm dry} =1, 10, 100$, respectively.}
\label{nbhbdry}
\end{figure}

Therefore, our results suggest that in-situ star formation and dry-merger processes operate in concert to contribute to the global BHB population. The dominant mechanism in low-mass galaxies is most likely in-situ formation, since smaller SMBH masses favour dynamical interactions while the lower number of clusters reduces the probability for BHBs to be dragged into the galaxy centre. 
In MW-like hosts, this mechanism can lead to the formation of $\sim 500$ BHBs, and more than $3000$ BH-star systems, which can possibly explain the observed population of low-mass X-ray binaries inhabiting the Galactic inner pc. On the other hand, in massive galaxies, where tidal forces suppress dynamical binary formation, the dominant contribution to the global BHB population is provided by orbitally segregated clusters that can deposit in the inner regions of the host up to thousand BHBs.

\section{Black hole binaries dynamics}
\label{Sec3}

Regardless of the formation scenario, the evolution of a BHB diving into a galactic centre is mostly influenced by three processes: i) dynamical friction, due to the fact that it is heavier than surrounding stars, ii) hardening (or softening) via stellar scattering, iii) periodic acceleration exerted by the SMBH that can induce KL oscillations \citep{kozai62,lidov62}. We refer to the first two effects as ``NC-driven'', whereas the third can be thought as an ``SMBH-driven'' effect.
This picture gets more complex if the galactic nucleus features an accretion disc. Indeed, stars crossing the gaseous disc are subjected to a drag acceleration that can trigger the formation of a stellar disc \citep{rauch95,kennedy16}, possibly boosting binary formation \citep{vilkoviskij02,baruteau11,panamarev18b} and collisions and mergers among stars \citep{subr05} or BHs \citep{baruteau11,bartos16,yang19}.

In the following, we explore both ``NC-driven'' and ``SMBH-driven'' effects.

\subsection{Nuclear cluster-driven effects: dynamical friction and hardening}

The dynamical friction timescale for a BHB with mass $m_{\rm BHB}$ moving along an orbit characterized by a semi-major axis $a_o$ and eccentricity $e_o$ can be written as \citep{ASCD14b,ASCD15He,AS16}
\begin{equation}
t_{\rm DF} = \tau_n g(e_o,\gamma) \left(\frac{m_{\bbh}}{M_g}\right)^{\alpha}\left(\frac{a_o}{R_g}\right)^{\beta},
\end{equation}
where $\tau_n$ is a normalization factor, $g(e_o,\gamma)$ is a weak function of the BHB eccentricity around the NC and the NC density slope, wherease $\alpha = -0.67$ and $\beta = 1.76$.
The BHB radial distance from the NC centre will decrease at a rate 
\begin{equation}
\frac{{\rm d}a_o}{{\rm d}t} \simeq -\frac{a_o}{t_{\rm DF}} \propto a_o^{1-\beta}m_\bbh^{-\alpha},
\end{equation}
which allows us to write the BHB orbital time evolution around the SMBH
\begin{equation}
a_o(t) = a_o(0) \left(1-\beta\frac{t}{t_{\rm DF}}\right)^{1/\beta}.
\label{adf}
\end{equation}
Note that the dependence on the BHB mass is contained inside the $t_{\rm DF}$ term.
The evolution of the BHB orbit eccentricity is less trivial to predict. However, it is well known that dynamical friction tends to circularize the orbit at a rate that increases at decreasing the distance to the galactic centre \citep{hashimoto03,just11,antonini12,ASCD14a,Petts15}. In order to model the eccentricity reduction driven by dynamical friction, we assume a simple exponential form
\begin{equation}
e_o(t) = e_o \exp\left(-t/t_{\rm DF}\right).
\end{equation}

At the same time, stellar encounters can cause the BHB hardening. 
As pioneered by \cite{heggie75}, a BHB whose binding energy is larger than the mean kinetic energy of the surrounding environments tends to harden in consequence of gravitational scattering, the so-called ``Heggie's law''. 
Using scattering experiments, \cite{quinlan96} showed that the typical hardening rate for a massive binaries subjected to repeated gravitational encounters is given by \citep[see also][]{antonini16}
\begin{equation}
\frac{{\rm d}}{{\rm d}t}\left(\frac{1}{a_{\bbh}}\right) = H\frac{G\rho_g}{\sigma_g},
\label{adec}
\end{equation}
being $H\simeq 7.6$ the adimensional hardening , $\rho_g$ and $\sigma_g$ the NC local density and velocity dispersion. Close to the centre of a \cite{Deh93} sphere, density and velocity dispersion profiles can be written as
\begin{align}
\rho_g(a_o) &\simeq  \rho_g\left(\frac{a_o}{R_g}\right)^{-\gamma} , \\
\sigma_g(a_o) &\simeq  \sigma_g \left(\frac{a_o}{R_g}\right)^{\delta/2},
\end{align}
being $\rho_g = (3-\gamma)M_g/(4\pi R_g^3)$, $\sigma_g = \sqrt{M_g/R_g}$ and either $\delta = 2-\gamma$ (if $\gamma\geq 1$) or $\delta = \gamma$ (if $\gamma<1$).

Plugging Eq. \ref{adf} into Eq. \ref{adec} allows us to calculate the hardening rate as the BHB moves toward the galactic centre
\begin{equation}
\frac{{\rm d}}{{\rm d}t}\left(\frac{1}{a_{\bbh}}\right) = H\frac{G\rho_g}{\sigma_g}\left(\frac{a_o}{R_g}\right)^{-\gamma-\delta /2}\left(1-\beta\frac{t}{t_{\rm DF}}\right)^{-(2\gamma+\delta)/(2\beta)}.
\end{equation}
Integrating over time we obtain
\begin{equation}
a_{\bbh}(t) \simeq a_{\bbh}(0)\left[1+\frac{a_{\bbh}(0)}{\tilde{a}_g(0)}f(t,t_{\rm DF})\right]^{-1},
\end{equation}
being 
\begin{equation*}
\tilde{a}_g(0) = \frac{\sigma_g}{HG\rho_g}\left[\frac{a_o(0)}{R_g}\right]^{\gamma+\delta/2}
\end{equation*}
a scaling factor that depends on the BHB initial position and the galaxy structure, and 
\begin{equation*}
f(t,t_{\rm DF}) = \frac{2t_{\rm DF}}{2\gamma + \delta - 2\beta} \left[\left(1-\frac{\beta t}{t_{\rm DF}} \right)^{(2\gamma + \delta - 2\beta)/2\beta} - 1 \right] 
\end{equation*}
is a function of time.

The fact that the BHB moves across regions with a varying density and velocity dispersion can increase the hardening rate if the ratio $\rho_g/\sigma_g$ increases at decreasing the distance from the galactic centre. Otherwise, the BHB can transit from a ``hard'' to a ``soft'' status, enhancing the probability for stellar scatterings to destroy it. Close encounters can induce the BHB disruption over a typical {\it evaporation time}, defined as \citep{bt,stephan16,hoang18} 
\begin{equation}
t_{\ev} = \frac{\sqrt{3}\sigma_g}{32\sqrt{\pi}G\rho_g\ln\Lambda a_\bbh} \frac{m_\bbh}{m_*},
\label{eva}
\end{equation}
where $m_*$ is the average stellar mass in the nucleus, $\rho_g$ is the stellar density and $\ln\Lambda=6.5$ is the Couloumb logarithm. 

If GW emission is the dominant process, angular momentum removal leads a BHB to merge over a time-scale \cite{peters64}
\begin{equation}
t_\gw = \frac{5}{256}\frac{c^5 a_\bbh^4 (1-e_\bbh^2)^{7/2}}{G^3m_1m_2(m_1+m_2)k(e_\bbh)},
\label{GW}
\end{equation}
with $k(e_\bbh) = 1+(73/24)e_\bbh^2+(37/96)e_\bbh^4$.

\begin{figure}
\centering
\includegraphics[width=\columnwidth]{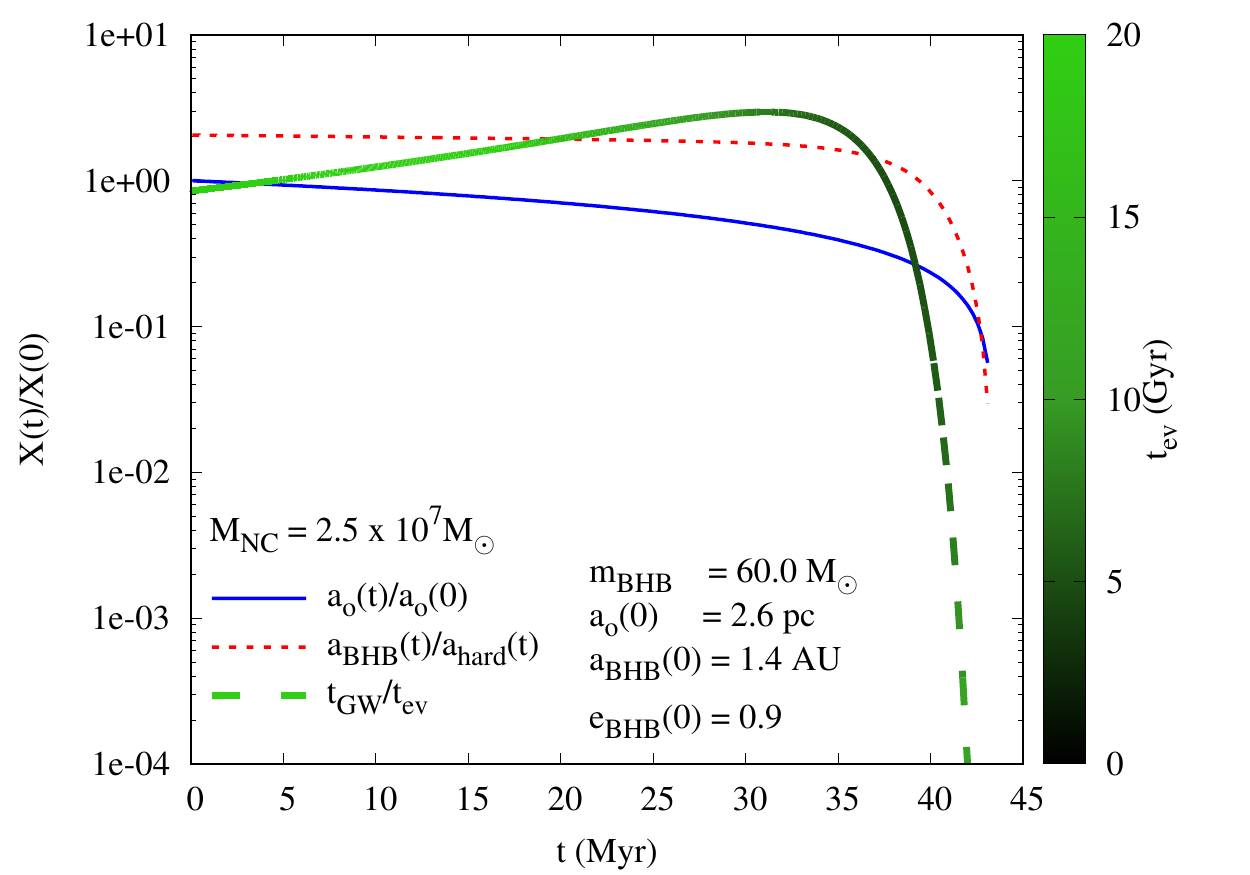}
\caption{Time variation of the BHB distance from the galaxy centre normalized to its initial value (blue straight line) and its semi-major axis normalized to the hard binary separation evaluated locally (red dashed line). The green dotted line mark the ratio between the GW timescale and the evaporation time. The green scale allows identifying the $t_{\ev}$ variation over time. We assume Galactic values for the NC model.}
\label{smaev}
\end{figure}

Using the set of equations above, we follow the evolution of a BHB with mass $m_{\rm BHB} = 69.3\Ms$ an initial semimajor axis $a_\bbh(0)=1.4$ AU and eccentricity $e_\bbh(0)=0.9$, assuming a MW-like NC. The outer orbit is circular and has semimajor axis $a_o(0)=2.6$ pc. The corresponding evaporation and the merging times calculated at $t=0$ are comparable, both exceeding 20 Gyr. Figure \ref{smaev} shows the time evolution of three quantities: a) the outer orbit semi-major axis normalized to its initial value $a_o(t)/a_o(0)$, b) the ratio between the BHB semi-major axis and the corresponding hard binary separation $a_\bbh(t)/a_{\rm hard}$, and c) the ratio between the merger and evaporation time $t_\gw/t_\ev$. All these quantities are calculated along the orbit. Moreover, to calculate the BHB evolution we take into account only dynamical friction and stellar hardening. 

The plot outlines that as long as the BHB spirals toward the centre, its semi-major axis becomes smaller and smaller compared to the limiting value for a binary to be hard. This implies that the BHB hardening rate increases over time. In this specific example, a reduction of the initial $a_o$ value leads to a decrease of the GW time by a factor 100. Therefore, in some cases mass segregation can drive BHBs in a regime where GW emission dominates, making the merger process faster.

Although the figure makes evident how hardening can affect the evolution of a binary, the procedure above does not include the angular momentum and energy loss driven by GW emission, which for compact binaries can become dominant in the last evolutionary phases and accelerate the merging process. To further account for this effect, we evolve binaries semimajor axis and eccentricity by solving the coupled systems of differential equations
\begin{eqnarray}
\frac{{\rm d}a}{{\rm d}t} &=& \left.\frac{{\rm d}a}{{\rm d}t}\right|_* + \left.\frac{{\rm d}a}{{\rm d}t}\right|_\gw, \label{pm63a}
\\
\frac{{\rm d}e}{{\rm d}t} &=& \left.\frac{{\rm d}e}{{\rm d}t}\right|_\gw,
\label{pm63b}
\end{eqnarray}
where the term labelled with $*$ refers to the hardening driven by stellar encounters and described by Equation \ref{adec} and the terms marked with $\gw$ refer to the hardening driven by GW \citep[for further details, see][]{peters63}.

The procedure depicted above is a simplistic approach that allows us to constrain the effect of stellar hardening, mass segregation, and GW emission on the evolution of BHB populations. However, as BHBs get closer to the centre, the SMBH tidal field will start perturbing its evolution, either accelerating the merging process or tearing it apart. In the next section, we will use numerical simulations to model in a self-consistent way the interplay between BHBs and the central SMBH.

\subsection{SMBH-driven effects: Kozai-Lidov oscillations}

In this section we discuss the limit in which the SMBH tidal field becomes dominant in the BHB orbital evolution. 

At zero-th order, the SMBH and the BHB can be treated as an isolated triple system. Clearly, this is an oversimplification, as NC stars can affect BHBs evolution either through ``impulsive'' perturbations, i.e. gravitational scattering, or ``secular effects'' driven by the NC gravitational field. Previous works generally neglect both effects, as they are explected to play a minor role inside the SMBH sphere of influence. Under the assumption of isolation, a triple system is stable if the ratio between the outer and inner semimajor axes exceeds a given value \citep{mardling01}
\begin{align}
\frac{a_o}{a_\bbh} >& \frac{3.3}{1-e_o}\left[ \frac{2}{3}\left(1+\frac{M_\bullet}{M_\bbh}\right) \times \right. \\ \nonumber
& \left. \frac{1+e_o}{(1-e_o)^{1/2}}\right]^{2/5}\left(1-0.3 \frac{i}{\pi}\right).
\label{eqstab}
\end{align}
Here, $i$ indicates the mutual inclination between the directions of the angular momentum vectors of the inner and outer binaries. For instance, a binary with mass $m_\bbh = 50\Ms$ and $a_\bbh = 1$ AU moving on a coplanar, prograde circular orbit around a $10^6\Ms$ SMBH will be stable if $a_o\gtrsim 200$ AU, a threshold that rises up to 0.01 pc for SMBHs with masses $10^9\Ms$. In the majority of simulations presented in the next section, BHBs initially fulfill the criterion above, constituting a stable triple with the SMBH.

According to KL theory, the exchange of angular momentum between the inner binary and the perturber can induce a periodic variation of both the mutual orbital inclination and the inner binary eccentricity, the so-called KL cycles \citep[see ][for a recent review]{naoz16}. In the case of a circular outer binary, the orbital evolution is well described by a three-body Hamiltonian truncated to the lowest order proportional to the ratio between the inner and outer semi-major axis. This is called quadrupole approximation 
\footnote{The quadrupole approximation fails in describing the motion when the outer orbit is eccentric. In this case, the Hamiltonian must be truncated to the next order, so-called the octupole approximation. In the octupole approximation limit, instead, the dynamics is much more complex, leading to the possibility for the triple to flip its configuration and evolve from prograde to retrograde and viceversa \citep{naoz13,li14}. A widely used criterion to discriminate whether the octupole term is comparable to the quadrupole term is \cite{naoz11,naoz13,antognini15,toonen16}
\begin{equation}
\epsilon_{\rm KL} = \frac{|m_1-m_2|}{m_1+m_2}\frac{a_\bbh}{a_o}\left(\frac{e_o}{1-e_o^2}\right),
\end{equation}
being $\epsilon_{\rm KL}\gtrsim 0.01$ the limit in which octupole effects might become important.
However, in the vast majority of our models $\epsilon_{\rm KL}<0.01$, thus we focus mostly on quadrupolar effects in the following}. 
For an initial circular orbit, the quadrupole approximation predicts that, for a circular binary, KL cycles take place if the inclination ranges between $39.2^\circ-140.7^\circ$. In this case, given an initial inclination value $i_{\rm in}$ the inner binary can reach a maximum eccentricity 
\begin{equation}
e_{\rm max} = \sqrt{1-5/3\cos^2(i_{\rm in})}.
\label{emax}
\end{equation}

The typical timescale for KL to take place is \citep{toonen16,antognini15}
\begin{equation}
t_{\rm KL} = \frac{8}{15\mathrm{\pi}}\left(1+\frac{m_\bbh}{M_\bullet}\right)\left(\frac{P_o^2}{P_\bbh}\right)(1-e_o^2)^{3/2},
\label{KL}
\end{equation}
where letter $P$ identifies the orbital period of the inner and outer binary.
Apsidal precession induced by relativistic effects (in case of compact remnants binaries) or tidal effects (in case of stellar binaries) can suppress KL oscillations and limit the SMBH effect on the BHB evolution. The typical timescale
for relativistic precession can be written in form \citep{hollywood97,blaes02,antonini12}
\begin{equation}
t_{\rm GR} = \frac{a_\bbh^{5/2}c^2(1-e_\bbh^2)}{3G^{3/2}(m_0+m_1)^{3/2}}.
\label{GR}
\end{equation}
The damping of KL cycles takes place if $t_{\rm GR}<t_{\rm KL}$.

In order to study in detail how the SMBH affects BHBs orbital evolution, in the next section we present a discuss a large series of direct N-body simulations. 

\section{Black hole binary mergers in galactic nuclei}
\label{Sec4}

\subsection{Evolution of BHB populations in nuclear clusters}

\begin{table*}
\caption{Properties of BHBs in galactic nuclei. Col. 1-4: NC mass, radius, slope, and SMBH mass. Col. 5-9: percentage of surviving, merging, escaping, disrupted, exchanged binaries. Col. 10: percentage of binaries that, at any moment of their evolution, have properties that might trigger KL resonances.}
\centering{}
\begin{center}
\begin{tabular}{cccccccccc}
\hline
$M_\nc$ & $R_\nc$ & $\gamma$ & $M_\smbh$ & $f_{\rm surv}$ & $f_{\rm mer}$ & $f_{\rm disr}$ & $f_{\rm ejec}$ & $f_{\rm exch}$ & $f_{\rm EKL}$\\

$10^7\Ms$ & pc & & $10^7\Ms$ & $\%$& $\%$& $\%$& $\%$& $\%$& $\%$\\
\hline
$1.0$ & $2.0$ & $0.1$ & $10^{-3}$ &$62.2$ &$37.4$ &$0.4$ &$0.002$ &$0.002$ &$55.7$\\
$2.5$ & $2.0$ & $1.8$ & $0.45 $   &$17.3$ &$78.3$ &$4.4$ &$0.006$ &$0.018$ &$69.5$\\
$10$  & $2.0$ & $1.8$ & $10 $     &$13.6$ &$77.2$ &$9.2$ &$0$     &$0$     &$72.4$\\
\hline
\end{tabular}
\end{center}
\label{T3b}
\end{table*}

We apply the treatment described in Section \ref{Sec3} to a population of BHBs inhabiting two types of galactic nucleus: a MW-like environment, assuming the same parameters for the NC and the SMBH used in the previous section, a heavy nucleus (mass $10^8\Ms$) harbouring an SMBH with mass $10^8\Ms$, and a globular cluster (mass $10^6\Ms$) hosting an intermediate-mass BH (IMBH) with mass $10^4\Ms$. For each binary, we distribute the BH progenitors mass according to a \cite{salpeter55} mass function cut between $14\Ms$ and $100 \Ms$ and we associate to the BHB a ``formation'' time drawn randomly between 1 and 10 Gyr. Stars are converted into BHs following the BH mass spectrum depicted by \cite{spera17}. The binary initial eccentricity is drawn from a thermal distribution and the semimajor axis from a logarithmically flat distribution. The lower limit on $a_\bbh$ is set to the maximum between 100 times the sum of the Schwarzschild radii and the value corresponding to a $t_\gw = 10^4$ yr. The upper limit of the distribution, instead, is set by the size of the binary Roche lobe. The eccentricity of the binary orbit inside the NC is drawn according to a thermal distribution as well, while its position is selected following the NC mass profile. 
For each binary, we evolve the trajectory combining Equations \ref{adf}, \ref{pm63a} and \ref{pm63b}. If the time exceeds the three-body encounters time-scale, we create a mock sample of 100 hyperbolic encounters by selecting: the perturber mass $m_p$ (calculated in the same way as for the binary components), the relative velocity $v_\infty$ of the binary and the perturber (whose velocity components are taken from a Maxwellian distribution centered in $\sigma$), and a deflection angle (taken between 0 and $\pi$). For each mock encounter we derive the perturber pericentre $r_p$ and compare the energy transfer $\Delta E\propto (r_p / a_\bbh)^{3/2}$ \citep{heggie75} to a critical value $\Delta E_c = v_\infty^2[m_p(m_1+m_2)]/[2(m_1+m_2+m_p)]$ \citep{sigurdsson93}. To predict the evolution of the binary post-interaction we use the following statistical arguments. In the case $\Delta E < \Delta E_c$ we assume two possibilities: i) $\Delta E < E_b$, with $E_b = Gm_1m_2/2a$ the binary binding energy, the binary hardens or soften according to the Heggie's law \citep{heggie75} and recoils after the interaction with a velocity in the interaction rest-frame $v_{\infty, f} < v_\infty$; ii) $ \Delta E > E_b$, the binary undergoes a component swap if $m_p > m_1 + m_2$ or $m_p > m_{1,2}$\footnote{In this case we associate to the component swap a $50\%$ probability.}. In the case $\Delta E >  \Delta E_c$ we have four further possibilities: i) $v_\infty < v_c$, with  \citep{hut83}
\begin{equation*}
v_c^2 = \frac{G[m_1+m_2+m_p]m_1m_2}{m_p(m_1+m_2) a},
\end{equation*} 
the binary exchange the component if either $ \Delta E > E_b$ or if $ \Delta E < E_b$ and $m_p > m_1 + m_2$ (or $m_p > m_{1,2}$), ii) it undergoes a resonances that, on average, results in the ejection of the lighter component; iii) $v_\infty > v_c$, the binary either exchanges one component if $ \Delta E < E_b$ or iv) it is ionized if $ \Delta E > E_b$. After performing the 100 mock encounters, we associate a probability to four possible categories (ejected, exchanged, disrupted, resonance) given by the ratio between the number of encounters falling in one category and the total number. We thus draw a number between 0 and 1 and compare it with the relative probability of each category. Upon this framework, at each time-step we perform a checklist to verify the status of the binary:
\begin{itemize}
\item {\bf surviving}: the binary survives up to 10 Gyr;
\item {\bf merger}: if $a_\bbh$ is smaller than twice the sum of the Schwarzschild radii of the binary components. 
\item {\bf tidally disrupted}: if $a_\bbh$ exceeds the binary Roche lobe;
\item {\bf disruption via strong encounter}: we halt the integration if the mock procedure returns the tag ``disruption'';
\item {\bf ejected}: we halt the integration if the mock encounters procedure described above returns the tag ``ejection'';
\item {\bf component swap}:  we halt the integration if the mock procedure returns the tag ``exchange'';
\end{itemize}

\begin{figure}
\includegraphics[width=\columnwidth]{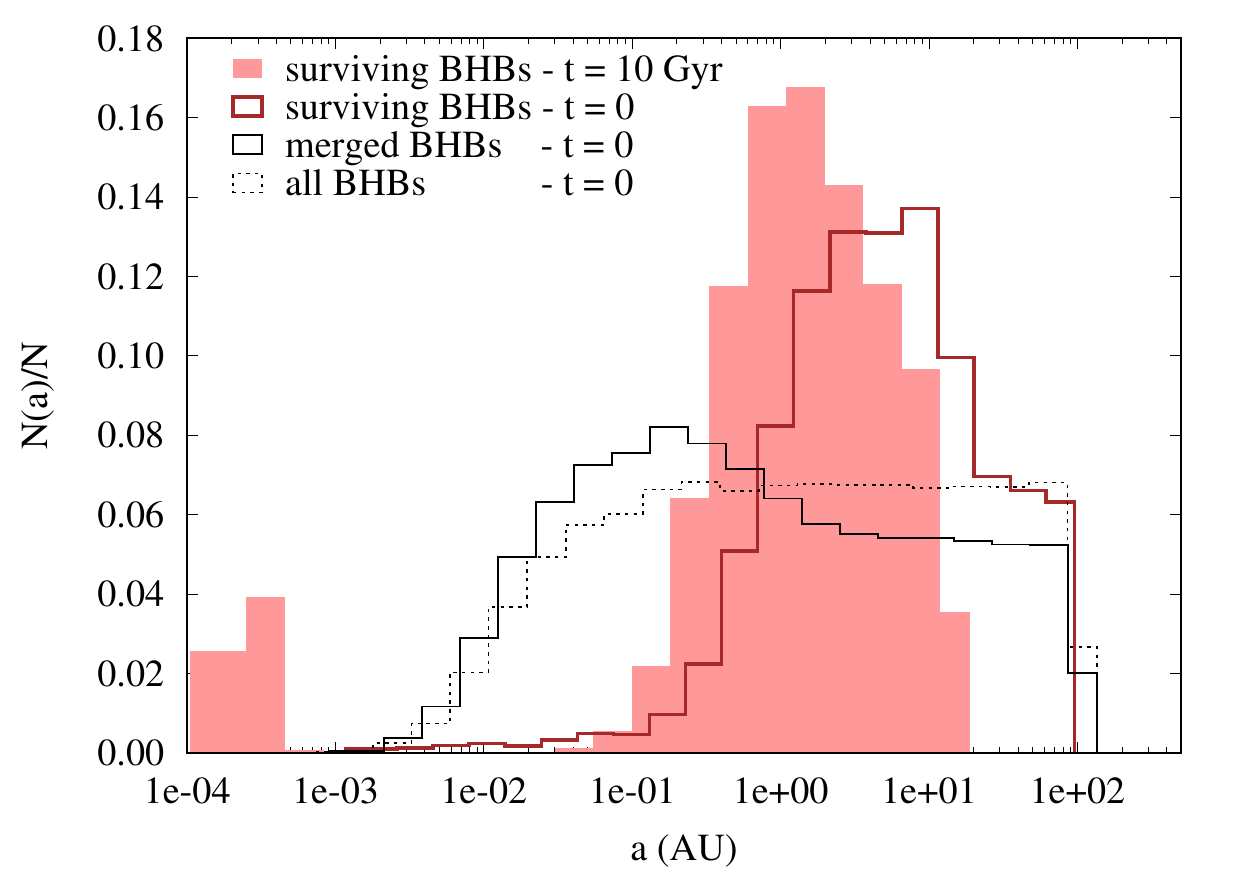}\\
\includegraphics[width=\columnwidth]{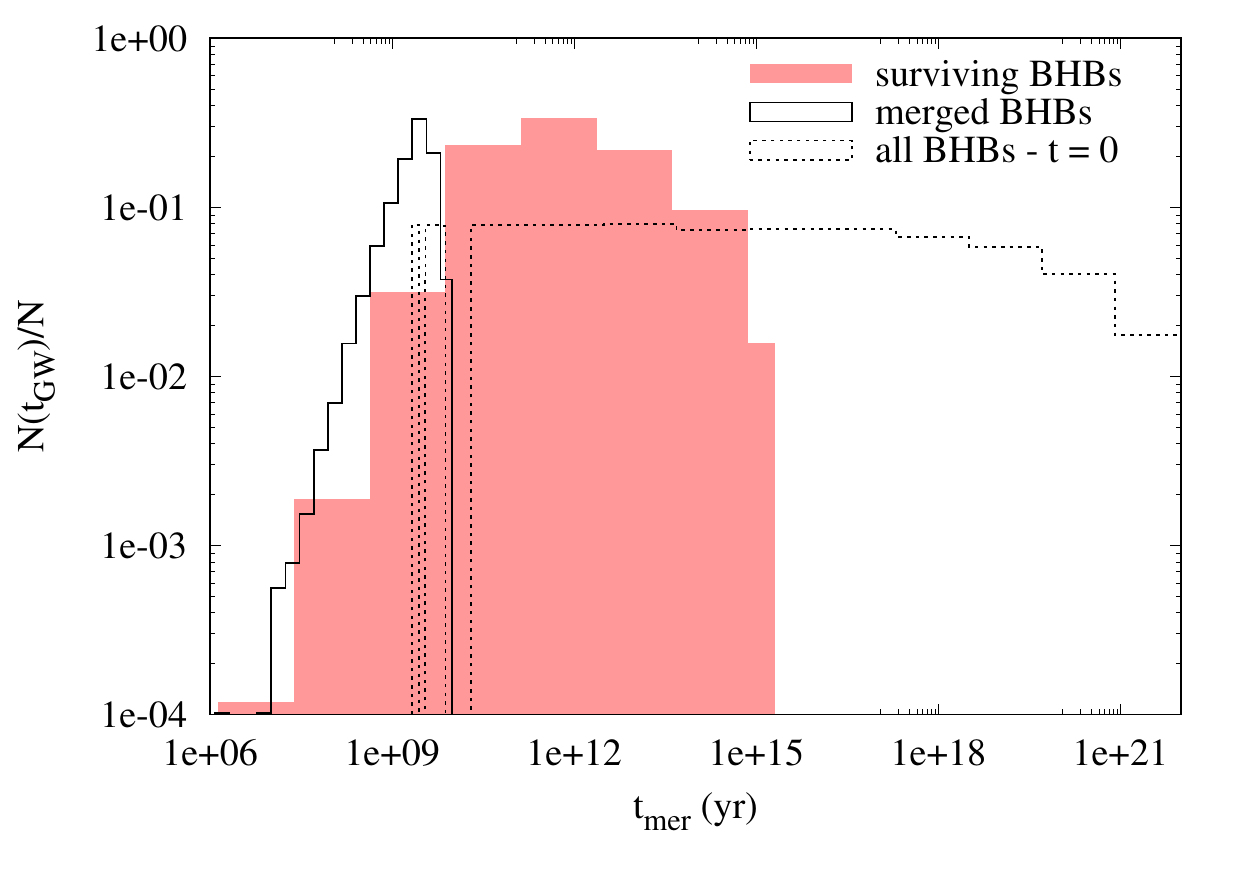}\\ 
\includegraphics[width=\columnwidth]{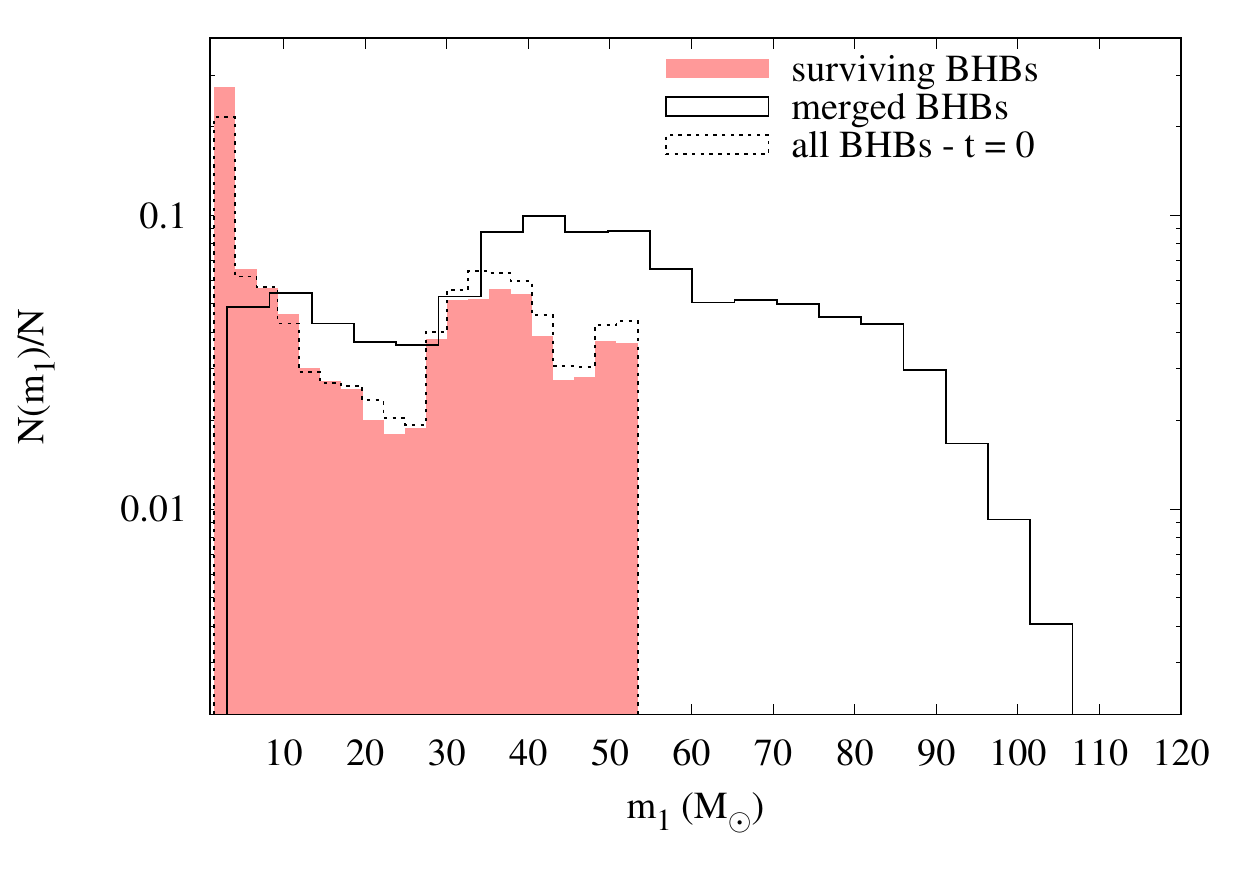}\\
\caption{Top panel: Semimajor axis distribution of all (dotted black steps), merged (thick brown steps), and surviving (red filled steps) BHBs at time $t = 0$, compared to the same distribution for surviving BHBs at time $t = 10$ Gyr (black steps) in a MW-like nucleus. Central panel: merger time distribution for all, merged, and surviving BHBs at time $t=10$ Gyr. Bottom panel: mass distribution for all, merged, and surviving BHBs at $t=10$ Gyr. All results refer to a MW-like galactic nucleus.}
\label{mergers}
\end{figure}

For each type of nucleus, we create a sample of 50000 binaries and calculate the fraction of BHBs falling in one of the categories above as summarized in Table \ref{T3b}. 
We note that statistically our model do not produce ejected BHBs due to the NC large velocity dispersion, neither it produces exchanges due to the long timescale for strong encounters around the SMBH. Both channels are also limited by the fact that the amount of energy transferred during an interaction is on average smaller than the energy of the binary and, in most of the cases, results in a resonance. More than $70\%$ of binaries in our model merge within a Hubble time via the combined action of stellar hardening and GW emission. We note that such large probability depends intrinsecally on the choice of the initial conditions. A more stringent conditions on the processes of disruption and dynamical ejection, for instance, can lead to a final lower number of BHBs that merge inside the cluster. For instance, here we do not account for the potential capture of one of the binary components by the SMBH. The fraction of BHBs that disrupt due to the increasing tidal torque exerted during their migration toward the inner part of the galaxy is limited to 4-9$\%$, while the percentage of BHBs that survive up to 10 Gyr is $\sim 14-17\%$. Note that the fraction of disrupted binaries increases at the expense of the fraction of surviving binaries. In the case of a globular cluster containing an IMBH the fraction of merged binaries decreases significantly together with the fraction of disrupted binaries, leading to a larger probability to observe surviving BHBs around an IMBH. Figure \ref{mergers} shows the distribution of the initial and final value of the semimajor axis in the case of surviving BHB compared to the total initial BHBs population and to BHBs that merge inside the NC. Surviving BHBs have semimajor axes that peak at $a_\bbh = 1$ AU, nearly one order of magnitude compared to their initial distribution. Comparing $a_\bbh$ for surviving BHBs with the total population, we note that the majority of surviving BHBs are characterized by $a_\bbh > 10^{-2} $ AU, while the population of BHBs that merge inside the cluster have $a_\bbh$ values that map the global distribution. Nearly $8\%$ of surviving BHBs have merger times $t_\gw < 2\times 10^8$ yr, $8\%$ have merger times below $10^{10}$ yr, and the remaining have times as large as $t_\gw = 10^{14}$ yr.

Regarding BHBs that merge inside the cluster due to the combined action of GW and stellar hardening, we find that their delay time, namely the time elapsed between the binary formation and the merger, shows a clear peak at aroun $1$ Gyr, with a long tail declining down to $10^6$ yr and a sharp decrease at larger values, as shown in the central panel of Figure \ref{mergers}. If the NC escape velocity is sufficiently high, and depending on the BH spins orientation and amplitude, merged BHs avoid the ejection due to GW recoil and be retained in the nucleus, possibly affecting the overall population of single BHs. The bottom panel of Figure \ref{mergers} shows the mass distribution of merger remnants and single BHs, highlighting the importance of in-cluster mergers in determining an enrichment of BHs with masses in the mass range $50-110\Ms$. These post-merged BHs have a large cross section and, possibly, can capture another companion and undergo a second merger, thus they can have an impact also on the population of GW emitters. 

It must be noted that in this simplistic approach we do not include KL effects, which can further affect BHBs evolution.
To shed light on the secular effects that might be induced by the SMBH, we determine for each binary if its orbital properties are, at any time, potentially stable, i.e. if the following conditions are satisfied \citep{hoang18}:
\begin{align}
\frac{a_\bbh}{a_o} &> \left(\frac{3M_\smbh}{m_1+m_2}\right)^{1/3} \frac{1+e_\bbh}{1-e_o},\\
\frac{a_\bbh}{a_o} &< 0.1 \frac{1+e_o^2}{e_o}.
\end{align}
If the binary satisfies the stability criteria, we compare the timescales for KL ($t_{\rm KL}$, Equation \ref{KL}) and relativistic precession ($t_{\rm GR}$, Equation \ref{GR}), dubbing the binary as ``KL''\footnote{Note that this category is independent on the {\it evolutionary categories} discussed above} if $t_{\rm KL} < t_{\rm GR}$. As outlined in Table \ref{T3b}, up to $55-72\%$ of the modelled BHBs satisfy, at least once, the criteria above, thus this quite large fraction represents an upper limit to the probability for a BHB to be potentially affected by KL oscillations. In the next section we will try to quantify the role of KL using direct $N$-body models.

\subsection{Numerical approach}

In the previous sections, we have shown that BHBs delivery operated by infalling star clusters may constitute a viable channel, altogether with three-body interactions and binary components swap, to populate galactic nuclei harboring an SMBH with BHBs, and that the interactions between the BHB and single stars can significantly affect the BHB properties. 
While BHBs formed in-situ are most likely already hard at formation, those delivered will be either hard or soft, depending on the properties of the parent cluster. Due to this, in what follows we simulate the evolution of both hard and soft binaries as they move around the SMBH. For the sake of clarity, we split simulations in two sets, the first comprised of soft (SET 1) and the second of hard binaries (SET 2) models. As detailed below, we vary the SMBH mass and the BHB initial properties to understand which conditions favour the merger. We provide an estimate of the merger rates for these channel in Section \ref{Sec3}.   

We assume four possible values for the SMBH mass, ${\rm Log} (M_\bullet /~\Ms) = 6,~7,~8,~9$, and distribute the simulations to have approximately the same number of models for each $M_\bullet$ value. 

The gravitational field generated by the NC in which the SMBH and the BHBs are embedded is included in the particles' equations of motion as an external potential modelled as \citep{Deh93}
\begin{equation}
\Phi_{\rm ext}(r) = \frac{GM_\nc}{(2-\gamma_\nc)R_\nc}\left[1-\left(\frac{r}{r+R_\nc}\right)^{1/(2-\gamma_\nc)}\right].
\end{equation} 
This family of potential-density pairs is characterized by three main parameters: the NC mass $M_\nc$, its typical radius $R_\nc$, and the inner density slope $\gamma_\nc$. 
The associated density profile is
\begin{equation}
\rho_{\rm ext}(r) = \frac{(3-\gamma)M_\nc}{4\pi R_\nc^3} \left(\frac{r}{R_\nc} \right)^{-\gamma} \left(1+\frac{r}{R_\nc}\right)^{-4+\gamma}.
\end{equation}

The NC mass is inferred from the galaxy velocity dispersion $\sigma_g$, which can be easily connected to both the NC and the SMBH via scaling relations. Combining the $M_\bullet - \sigma_g$ relation derived by \cite{kormendy13} for SMBH and the $M_\nc - \sigma_g$ relation discussed in \cite{ASCD14b}, we obtain 
\begin{equation}
{\rm Log} M_\nc = 2.509 + 0.521{\rm Log} M_\bullet + \mathcal{F},
\end{equation}
where we define the scaling factor $\mathcal{F} = 2$ in such a way to obtain NC and SMBH masses in agreement with observed values. 
The NC scale radius $R_\nc$ is selected randomly between typical values, namely 0.8 and 2 pc, whereas the density slope $\gamma_\nc$ is randomly assigned between 0.5 and 2.

The outer binary eccentricity $e_o$ is drawn accordingly to a thermal distribution, $P(e_o){\rm d}e_o = 2e_o{\rm d}e_o$ \citep{jeans19}, as suggested for stars orbiting the Galactic SMBH \citep{schodel03,alexander05}. The outer semimajor axis $a_o$ is drawn following the NC mass distribution, $M_\nc(a_o)$, within 0.1 pc from the SMBH. The latter condition mimic the fact that BHBs will be likely segregated inside the NC innermost regions. 

The cosine of the mutual inclination between the inner and outer binaries ($\cos(i)$) is selected randomly between $-1$ and $1$.

\begin{figure}
\includegraphics[width=\columnwidth]{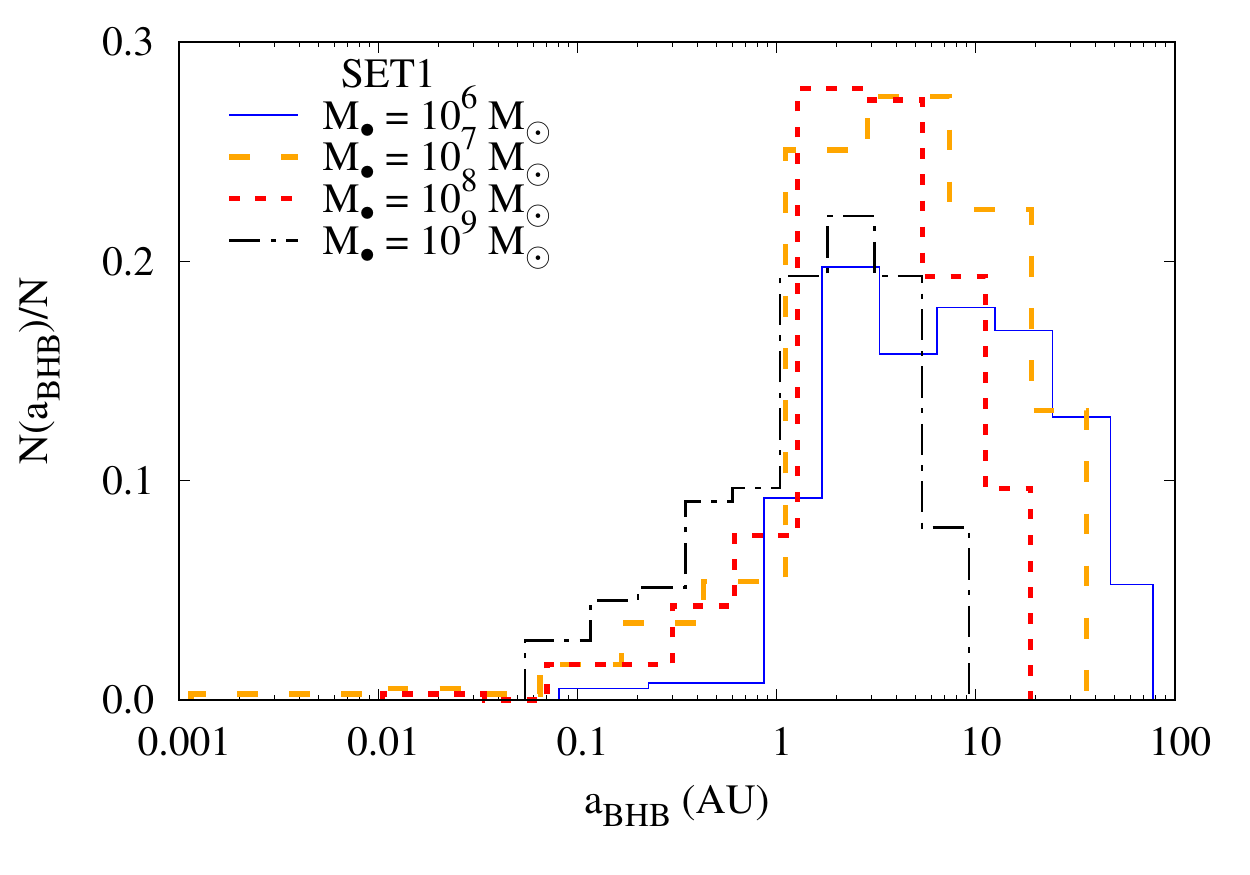}
\includegraphics[width=\columnwidth]{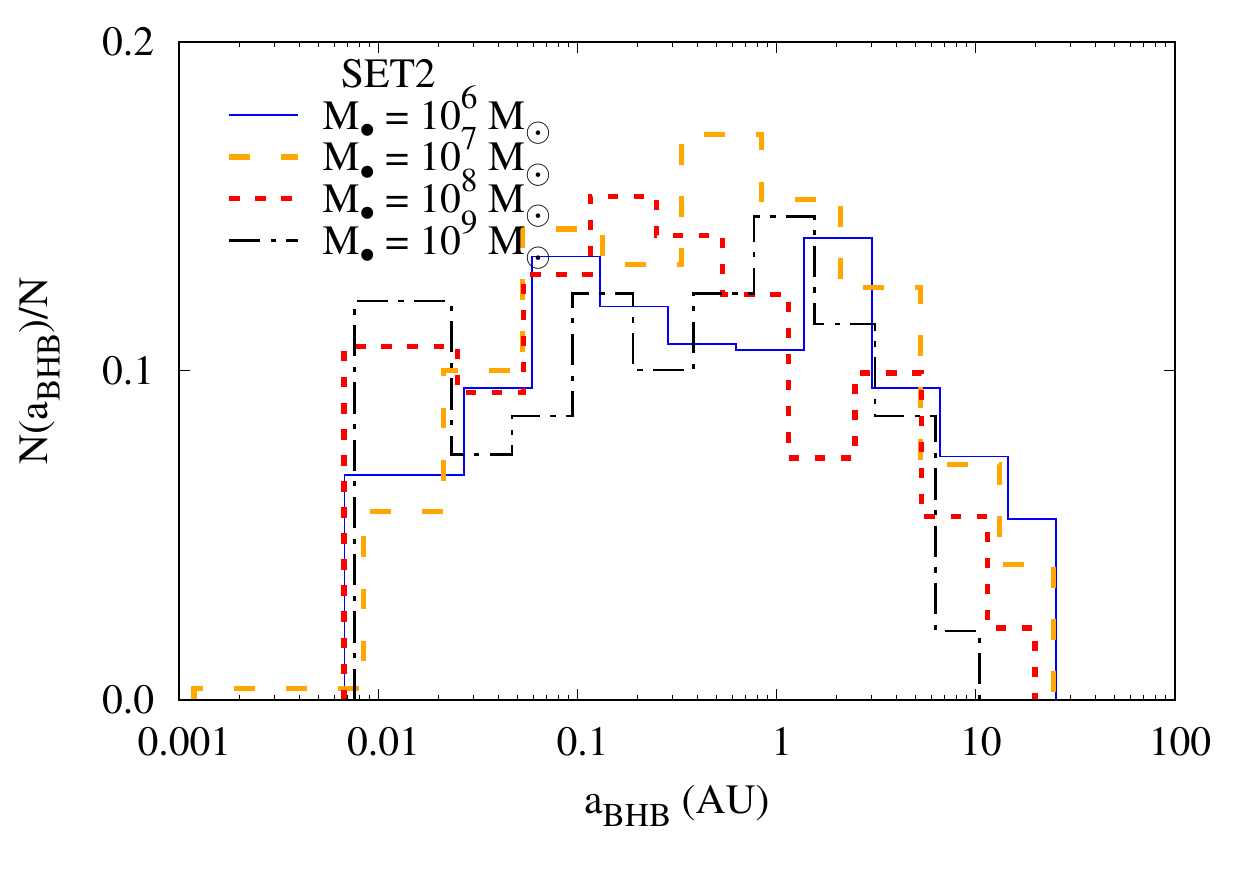}
\caption{Initial distribution for BHBs in SET1 (top panel) and SET2 (bottom panel) at different values of the SMBH mass.}
\label{semi0S1}
\end{figure}

Similarly to the outer binary, the inner binary eccentricity $e_\bbh$ is extracted from a thermal distribution. The semimajor axis $a_\bbh$, instead, is chosen in the range $1 - 100\au$  for SET 1 and $0.01-15 \au$ for SET 2, assuming in both cases a logarithmically flat distribution. 
To ensure that the total NC+SMBH orbital field does not tear apart the BHB at the beginning of the simulation, we check if the BHB orbital apoapsis exceeds a fraction of the tidal radius calculated at the pericentral distance from the SMBH \citep{hill1878}
\begin{equation}
a_\bbh (1+e_\bbh) \leq r_{\rm tid} \equiv \alpha a_o (1-e_o) \left(\frac{m_0+m_1}{3M_\bullet}\right)^{1/3},
\label{hillR}
\end{equation}
where we assume $\alpha = 0.5$ as a conservative value. In the case in which our selection procedure does not fulfil the inequality above, we set $a_\bbh = r_{\rm tid}/(1+e_\bbh)$. 
Due to this criterion, binaries in SET1 have an initial semi-major axis distribution that depends on the SMBH mass, as shown in top panel of Figure \ref{semi0S1}, with the heaviest SMBHs being orbited by tighter BHBs. The distribution is less affected in SET2, as shown in bottom panel of Figure \ref{semi0S1}. Note that the distribution of semimajor axis in SET 1 and 2 is similar to the one obtained with our semianalytic approach described in the previous section for BHBs that survives up to 10 Gyr and that merge inside the cluster (black steps and red steps in the top panel of Figure \ref{mergers}).

The masses of the inner binary components are assumed to follow a power-law, $f(m) \propto m^{-2.2}$, limited between $m_{\rm min} = 10\Ms$ and either $m_{\rm max} = 30\Ms$ in SET 1 or $m_{\rm max} = 100\Ms$ in SET 2. It is worth noting that the mass range considered here take into account both the possibility that the BHs is formed from single stellar evolution or is the remnant of a previous merger, as discussed in the previous section (see central panel of Figure \ref{mergers}).

The difference between SET 1 and SET 2 are meant to be representative of different BHBs formation scenarios. For instance, the hard binary separation for a typical globular cluster ($\sigma \sim 10$ km s$^{-1}$), is relatively large, $a_\bbh<100$ AU \citep{heggie75}. Such values might be typical of delivered BHBs, or binaries that occasionally form in the NC outskirts. In dense NC, instead, the hard-soft threshold value can fall below 10 AU. Close to an SMBH, where the velocity dispersion scales as $\sigma \propto M_\smbh r^{-1/2}$, the hard binary separation can decrease sharply with the distance to the SMBH and its mass, thus leading to significantly different hard-binary regimes even within the same galactic nucleus, because of the $r$ dependence, and across different galactic nuclei, because of the $M_\smbh$ dependence. 
Regarding the mass distribution, BHB merger remnants in galactic nuclei have a large probability to be retained and undergo multiple merger events, leading to BHs with masses up to $\sim 100 \Ms$ \citep{antonini16,antonini18c,ASBEN19}. In the case of soft binaries, either formed in the NC outskirts or delivered by infalling clusters, the binary components will have masses most likely close to the standard BH natal mass distribution \citep{downing10,rodriguez15,morscher15,AAG18a,AAG18b}.
The main properties of SET 1 and 2 are summarized in Table \ref{T1}.

\begin{table*}
\caption{Main properties of direct $N$-body models}
\centering{}
\begin{center}
\begin{tabular}{ccccccccc}
\hline
SET ID & $Log M_\bullet$ & $a_o$ & $e_o$ & $i_o$ & $m_{0,1}$ & $a_\bbh$ & $e_\bbh$ & binary status \\
       & $\Ms$ & pc &  &  & $\Ms$ & AU & &\\
\hline\hline
1      & $6-9$ & $10^{-3} - 100$ & $0-1$ & $0-\pi$ &$10-30$ & $1-100$   & $0-1$ & soft\\
2      & $6-9$ & $10^{-3} - 100$ & $0-1$ & $0-\pi$ &$10-100$& $0.01-15$ & $0-1$ & hard\\
\hline
$f(x){\rm d}x$ & discrete & $M_\nc(a)$ & $f(e)=2e$ & $f(\cos(i))=$const & $f(m)=km^{-2.2}$ & $f({\rm Log}(a))=$const &$f(e)=2e$ & \\
\hline
\end{tabular}
\end{center}
\begin{tablenotes}
\item Col. 1: set ID. Col. 2: SMBH mass. Col. 3-4: semimajor axis and eccentricity limiting values of BHB orbits around the SMBH. Semi-major axis upper limit is given by Equation \ref{hillR}. Col. 5: inclination between the BHB and the SMBH orbital planes. Col. 6: BHB components mass. Col. 7-8: BHB semimajor axis and eccentricity. Col. 9: binary status. 
\item Top lines shows the limiting values assumed for each parameters, whereas the bottom line summarizes the distribution function assumed to select them. 
\end{tablenotes}
\label{T1}
\end{table*}

Under the assumptions detailed above, we run 3000 simulations of the SMBH-BHB triple embedded in the NC external potential, equally divided between SET 1 and 2. 
All the simulations are performed using \ARGdf \citep{ASCD17b}, an upgraded version of the \ARCHAIN $N$-body code \citep{mikkola99,mikkola08}. 
\ARGdf allows the user to include in the particles' equations of motion also the gravitational field generated by NC stars, modelled as an external static potential, and dynamical friction. 

Although our code allows us to follow the BHB evolution with high accuracy, it does not capture the possibility that the BHB dynamics is altered by close interactions with passing by BHs \citep[see][for a discussion on this effect]{trani19}. 
We halt the simulations if one of the following criteria is satisfied: a) the BHB merges, b) one of the binary component is ejected away, c) one component merges with the SMBH, d) the integrated time exceeds the maximum between 500 times the BHB orbital period around the SMBH and 5 times the binary KL time.
On top of these four criteria, we have a computational integration time limit set to 3 hours, extended to up to 2 days for models that do achieve the integration of neither 500 orbits around the SMBH nor 5 times the KL time, after which a simulation is automatically halted. We remove all the models that not fulfill any of the criteria above. Most of these systems are characterised by $t_{\rm KL} / P_\bbh > 10^8$ and require O($10^{10}$) steps to be accomplished. We decided to remove these models because they have a very high computational cost and, at the same time, integrating over such a large number of steps can cause a cumulation of the roundoff error sufficient to bias the results. This lead the actual number of simulations to 1248(1589) in SET1(SET2). 
Potential merger candidates are identified according to different criteria, depending on the SMBH+BHB orbital configuration at the end of the simulation. We define four classes of merger candidates:
\begin{itemize}
\item {\bf fast mergers}: take place before the simulation ends. We record the orbital properties of the BHB at the snapshot before the merger occurs;
\item {\bf hierarchical mergers}: show a clear eccentricity oscillation typical of KL cycles. In this case, $t_\gw$ in Equation \ref{GW} will overestimate the actual merger time, due to the enhancement in GW emission achieved at the BHB maximum eccentricity $e_{\rm max}$. In this case, we calculate the maximum eccentricity along the BHB orbital evolution. This implies that the adopted value is not the actual maximum value allowed if the KL cycle is not fully covered, representing in this case an upper limit to the actual merger time. Following \cite{antonini12}, we define the actual merger time as 
\begin{equation}
t_\gw = \frac{t_\gw(a_\bbh,e_{\rm max})}{\sqrt{1-e_{\rm max}^2}};
\label{AP12}
\end{equation}
\item {\bf non-hierarchical mergers}: show chaotic, but not drastic, variation of the eccentricity. The GW time in this case is calculated through \cite{peters64} formula (Eq. \ref{GW}) using as $e_\bbh$ the average between the initial, final, and maximum value of the BHB eccentricity;
\item {\bf unperturbed mergers}: the BHB orbital properties are unaffected by the SMBH and the NC tidal field. Also in this case, the merging time is calculated through Eq. \ref{GW}.
\end{itemize}
For all the candidates, we compare the merger time calculated following the prescriptions above with the evaporation time in Equation \ref{eva}, and label as ``mergers'' only models for which $t_\gw<t_\ev$.

Depending on the initial configuration, the BHB evolution, and eventually its merger, can be shaped or not by the presence of the SMBH and the surrounding NC. We thus classify the mergers depending on their connections to the galactic nucleus in which they are embedded, distinguishing into:
i) unperturber mergers (unp), for which  $t_{\gw}/t_\gw(0) > 0.9$ and $t_{\gw}(0)<14$ Gyr is the initial merger timescale; ii) perturbed mergers (ext), having $t_{\gw}/t_\gw(0) < 0.9$ and $t_{\gw}(0)<14$ Gyr; iii) driven mergers (driv), having initially $t_{\gw}(0)>14$ Gyr. 

\subsection{Soft black hole binaries}

As anticipated in Section \ref{Sec4}, BHBs in SET1 are characterized by low masses ($10-30 \Ms$), and wide orbits ($a_\bbh \simeq 1-100$ AU). In this case, all the BHBs have initially a merger timescale -- $t_{\gw}(0)$ -- longer than the typical Hubble time, $t_H = 14$ Gyr.

In this set, we find that BHBs merge in $\sim 16.0\%$ of the simulations performed. The $f_{\rm KL} = 11.3\%$ of all mergers is affected by KL variations (note that this class can include  "perturbed", "unperturbed", and "driven mergers"). The combined action of the NC+SMBH accelerate the merger in $f_{\rm ext} = 1.7\%$ of the cases of binaries with $t_\gw(0) < 14$ Gyr, and determine the merger of binaries with a longer initial merger time in $f_{\rm driv} = 8.2\%$ of the cases. The remaining $6.1\%$ are unperturbed binaries, namely the external perturbations induce a reduction of the GW time by a factor smaller than $10\%$ of the initial value. 
The SMBH captures one of the BHB components in the $f_{\rm cap} = 3.7\%$ of the simulations, favouring in some cases the formation of an extreme-mass ratio inspiral (EMRI, a promising class of sources for the laser interferometer space antenna LISA \citep{seoane07}) that merge within a Hubble time ($f_{\rm EMRI} = 0.08\%$). Our main results are summarized in Table \ref{TabRes}. 

\begin{figure}
\centering
\includegraphics[width=\columnwidth]{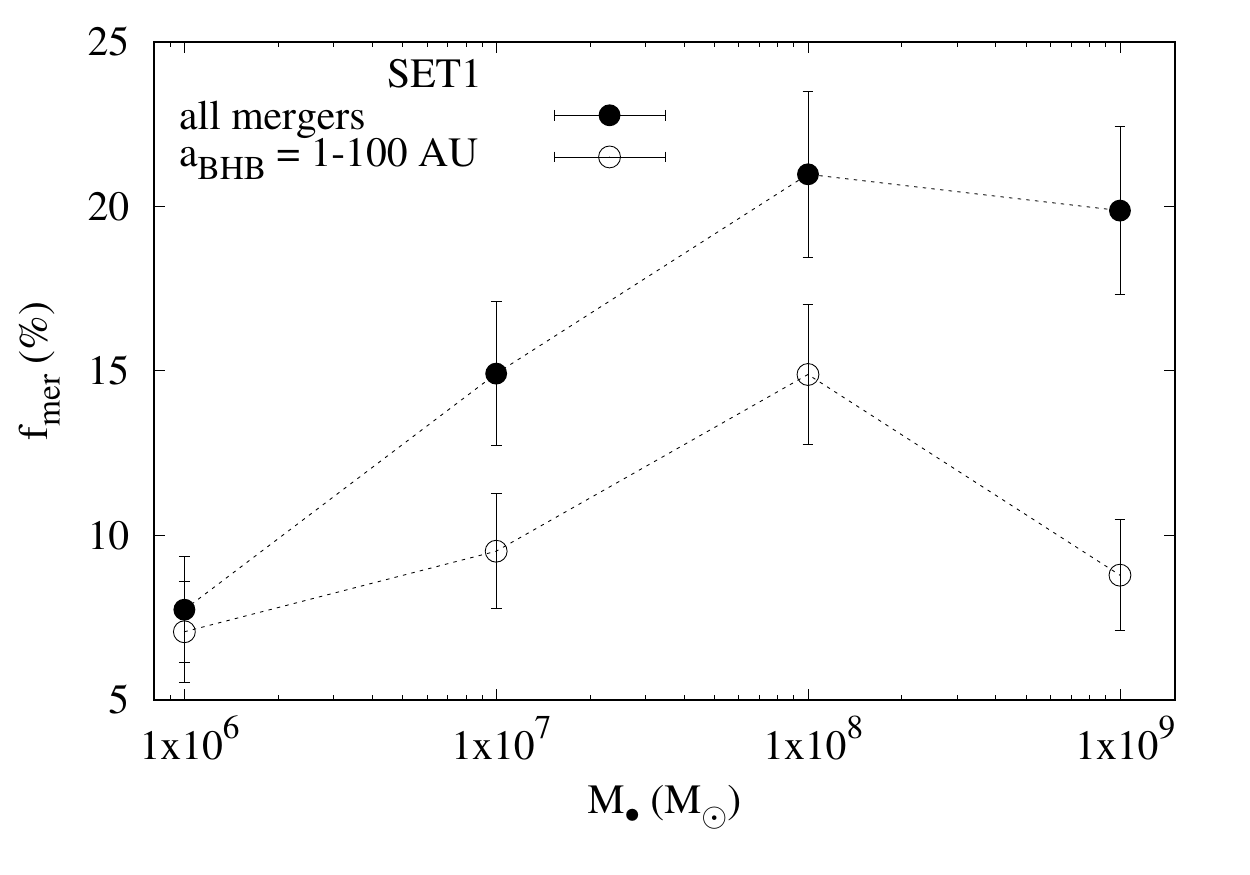}
\caption{Merger probability at varying the SMBH mass in SET1. The error bars are obtained assuming a Poisson distribution of the error.}
\label{F2}
\end{figure}

The BHB merger probability, as shown in Figure \ref{F2}, increases with the SMBH mass, saturating to $f_{\rm mer}\simeq 15\%$ for $M_\bullet>10^8\Ms$. The trend is mostly due to the assumed initial conditions, which in SET1 favours tighter BHBs for heavier SMBHs. To further highlight this effect, we show in Figure \ref{F2} also the merging probability calculated taking into account only mergers having an initial $a_\bbh$ value between 1-100 AU. When this selection criterion is applied, the $f_{\rm mer}-M_\bullet$ dependence weakens significantly. 

\begin{figure}
\centering
\includegraphics[width=\columnwidth]{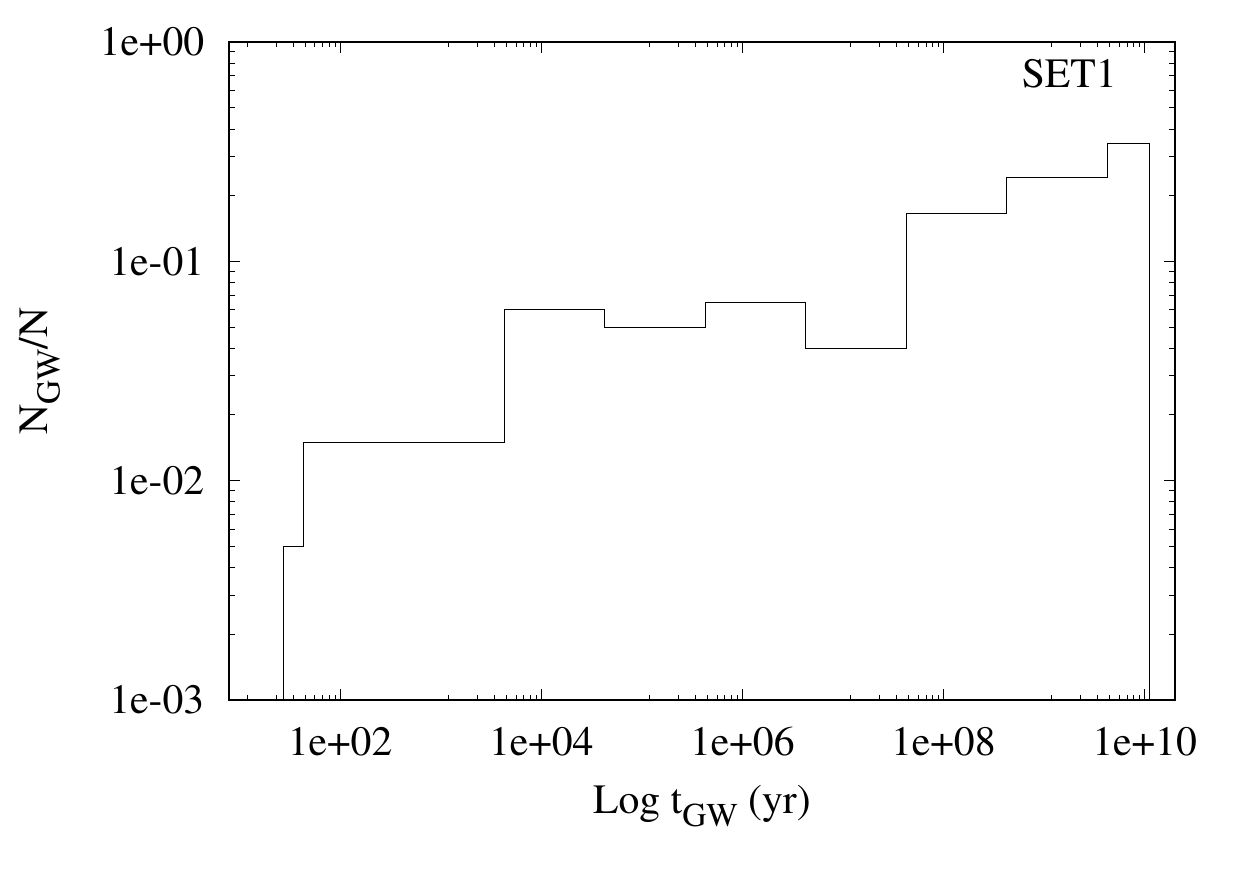}
\caption{Merger time distribution for coalescing BHBs in our SET 1.}
\label{F3}
\end{figure}

Figure \ref{F3} shows the merger time distribution for all mergers in SET1. Nearly $21\%$ of the mergers have $t_\gw<1$ Myr, with a notable percentage ($\sim 8.7\%$) of models undergoing a ``fast merger'', being $t_\gw<10^4$ yr.

\begin{figure}
\centering
\includegraphics[width=\columnwidth]{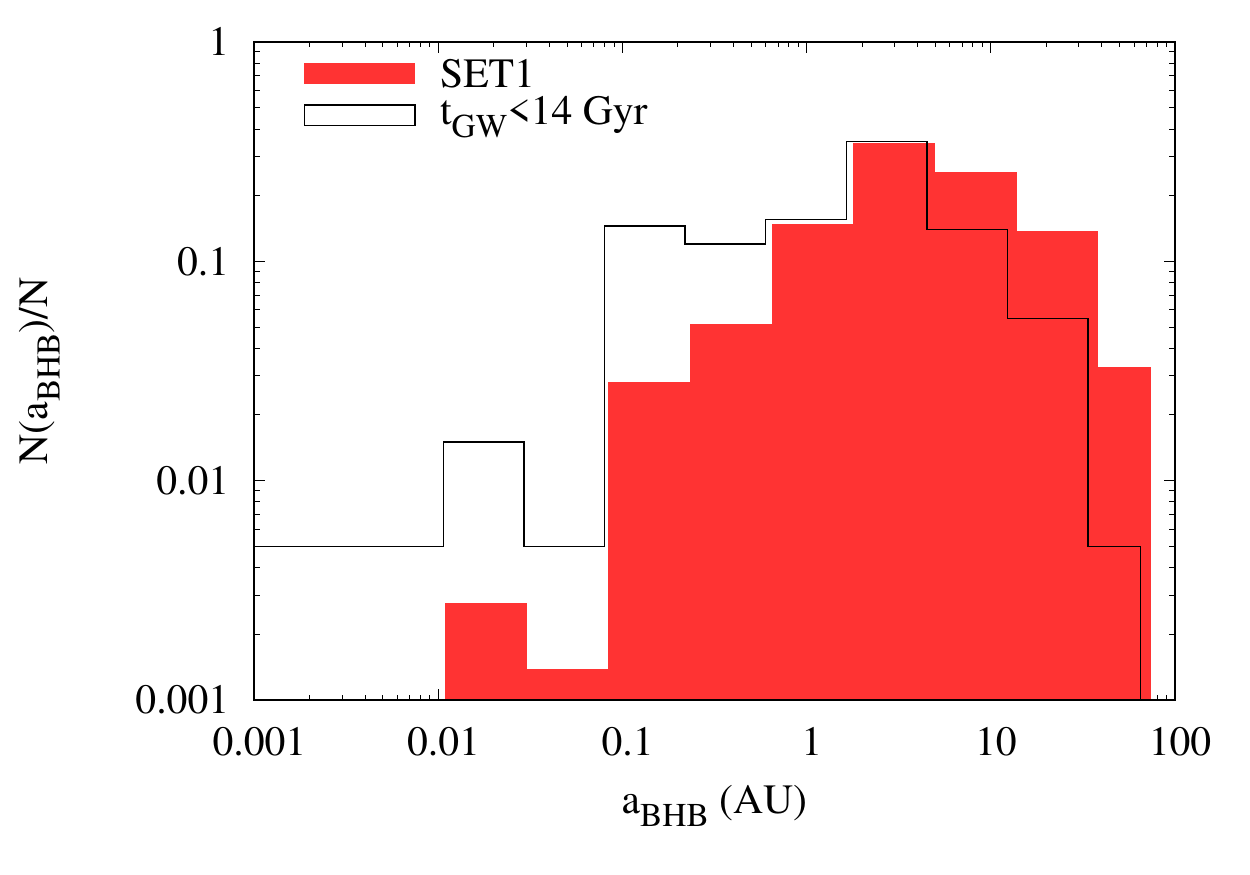}\\
\includegraphics[width=\columnwidth]{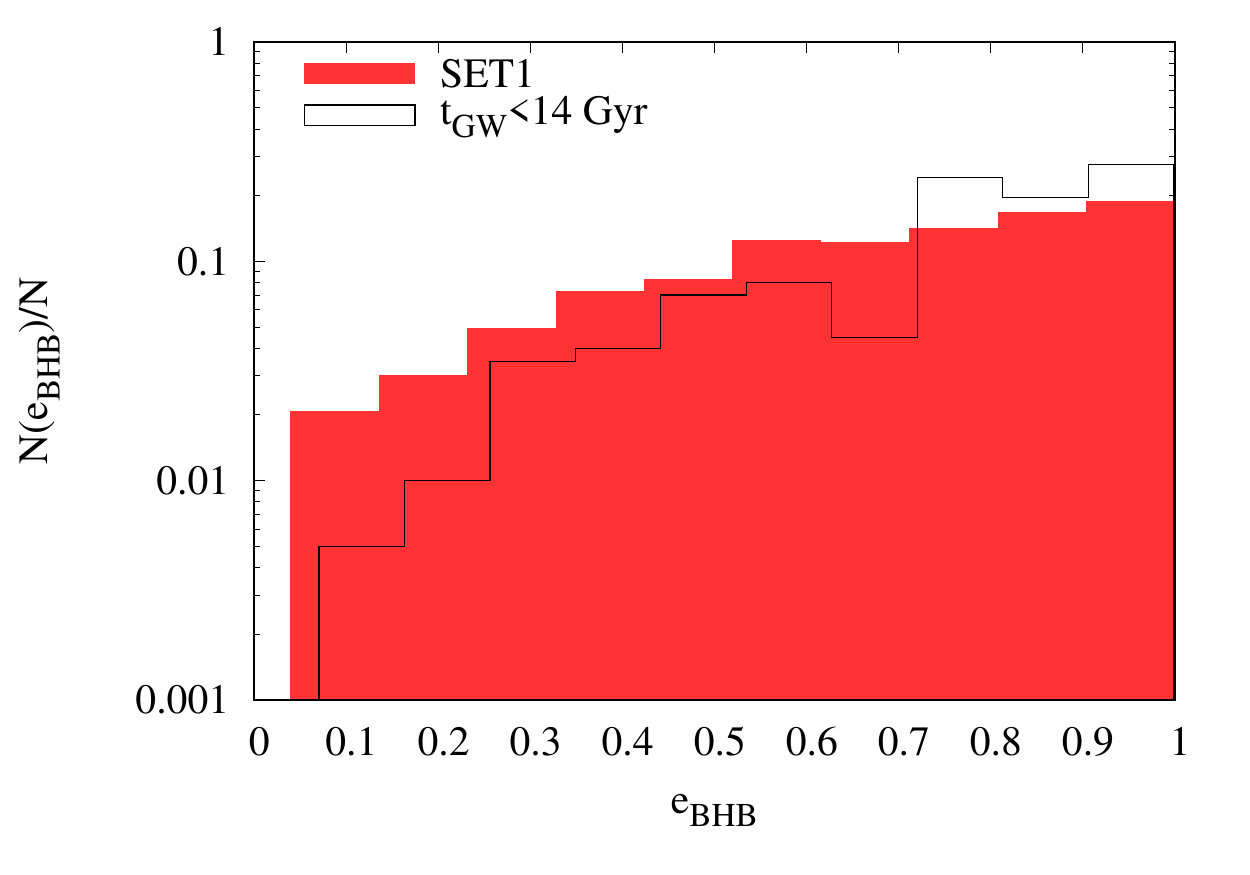}
\caption{Semi-major axis (top panel) and eccentricity (bottom panel) distribution for the whole BHB population (red filled boxes) and the binaries that merge within 14 Gyr (black steps). For the sake of readability, the y-axis are normalized to the total number of objects considered in each distribution.}
\label{F4}
\end{figure}

The distribution of $a_\bbh$ for mergers deviates slightly from the global BHB population, showing a long low-end tail and an abrupt decrease at values $a_\bbh>20$ AU, as shown in the upper panel of Figure \ref{F4}. Similarly, mergers eccentricity distribution is steeper compared to the whole BHB population (see the bottom panel of Figure \ref{F4}).

\begin{figure}
\centering
\includegraphics[width=\columnwidth]{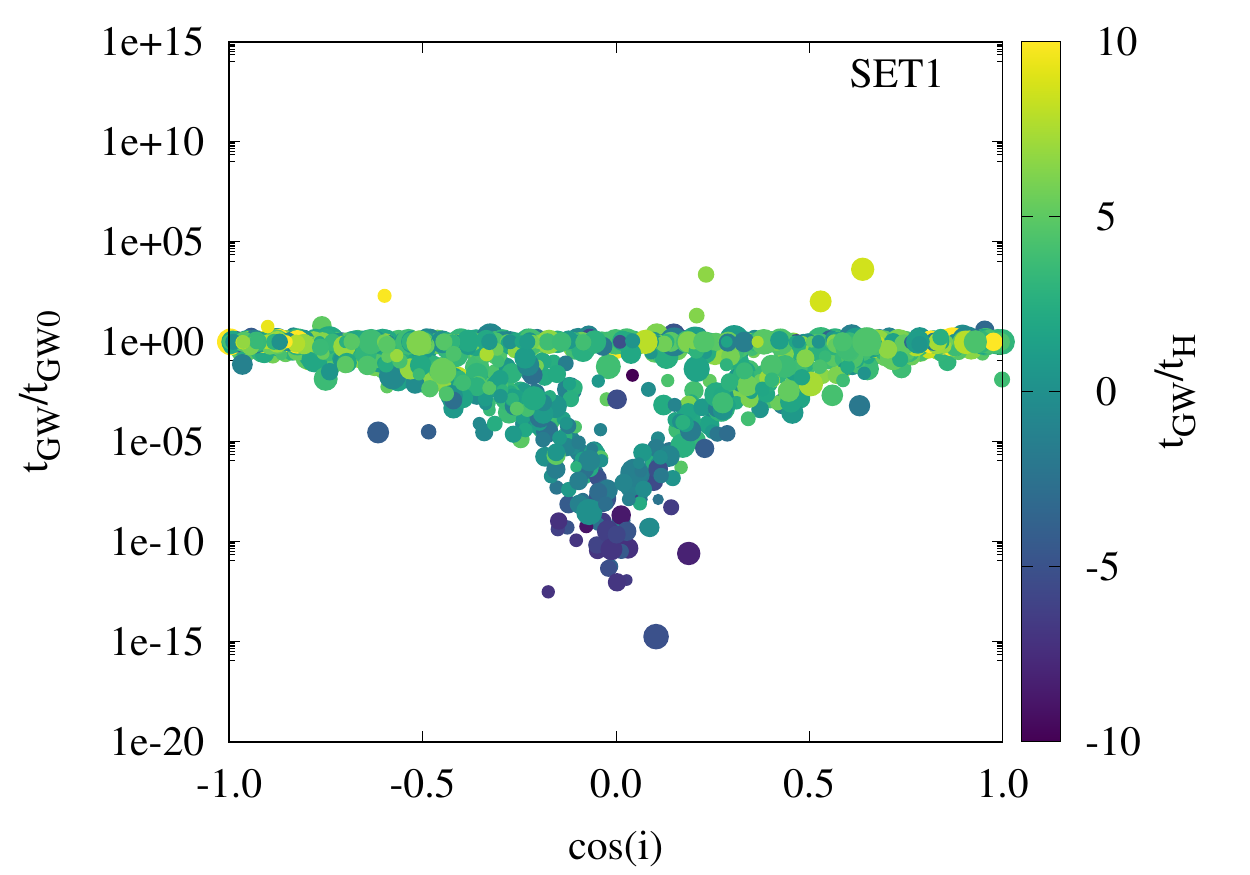}\\
\includegraphics[width=\columnwidth]{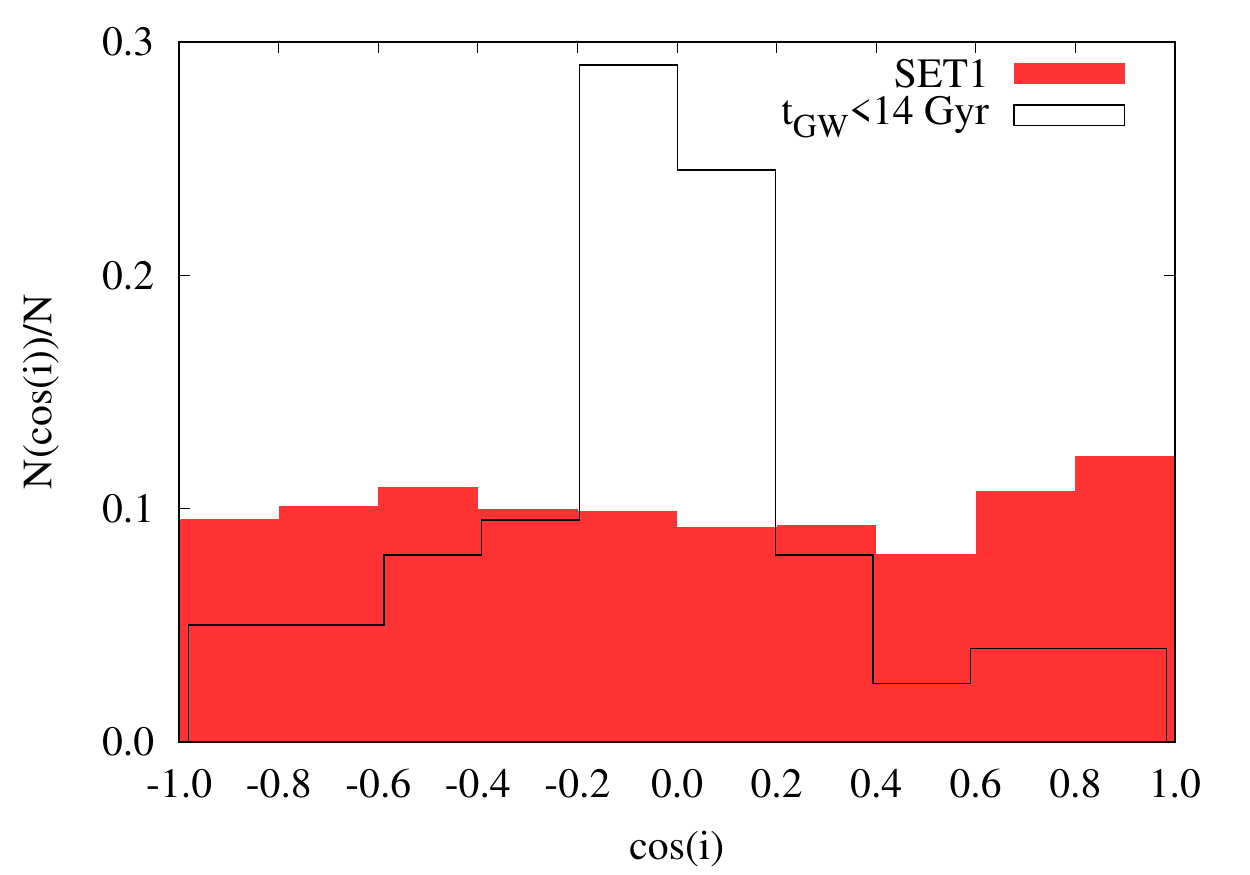}
\caption{Top panel: final value of the merger time-scale as a function of the cosine of the initial inclination. The coloured map identifies the final value of the merger time, normalized to 14 Gyr. Larger dots identify heavier BHBs. Bottom panel: distribution of the initial inclination for all the BHBs in SET 1 (red filled boxes) and for those merging within 14 Gyr (black steps).}
\label{F6}
\end{figure}

The parameter that seems to affect mostly the BHB evolution is the mutual inclination between the BHB orbit and the SMBH-BHB orbital plane, as the most effective reduction of the merger time-scale is achieved when the inner and outer binary orbits are nearly perpendicular, i.e. $i\sim \pi/2$, as shown in Figure \ref{F6}. Note that this is expected from KL mechanism, as the peak of the eccentricity variation is maximized for perpendicular orbits (see Equation \ref{emax}). 

\subsection{Hard black hole binaries}

In SET 2, where BHBs are assumed to be initially hard, we find 652 merger candidates, namely $f_{\rm mer} \simeq 40.8\%$ of the simulated sample. The action of the external perturbations induce a reduction of the GW time larger than 10 in $f_{\rm ext} = 8.5\%$ of binaries with $t_\gw(0)<14$ Gyr, and drives the merger of binaries with longer merger times in $f_{\rm driv} = 3.9\%$ of models. Unperturbed mergers dominate the population ($f_{\rm unp} = 28.4\%$). Among all mergers, $f_{\rm KL} = 30.1\%$ are in a hierarchical configuration. The SMBH captures one of the BHB components in $f_{\rm cap} = 14.3\%$ of models, and triggers the formation of an EMRI in a substanctial fraction of them $f_{\rm EMRI} = 5.7\%$.

\begin{table*}
\centering{}
\caption{Merger probability for different merger types}
\begin{center}
\begin{tabular}{ccccccccc}
\hline
\hline
SET ID & $f_{\rm mer}$ & $f_{\rm unp}$ & $f_{\rm ext}$ & $f_{\rm driv}$ & $f_{\rm KL}$ & $f_{\rm cap}$ & $f_{\rm EMRI}$ & $N_{\rm sim}$\\
       & $\%$ &$\%$ &$\%$ &$\%$ &$\%$&$\%$ &$\%$ \\
\hline
1 & 16.0 & 6.1 & 1.7 & 8.2 & 11.3 & 3.7 & 0.08 & 1248\\
2 & 40.8 & 28.4& 8.5 & 3.9 & 30.1 & 14.3& 5.7  & 1589\\
\hline
\end{tabular}
\end{center}
\begin{tablenotes}
\item Col. 1: model ID. Col. 2: mergers probability. Col. 3-5: probability for mergers unperturbed, induced, or driven by external perturbations, respectively. Note that the sum of columns 3, 4, and 5 returns the value given in Col. 2. Col. 6: probability for hierarchical configurations. Col. 7-8: probability for the formation of an BH-SMBH binary and an EMRI, respectively. Col. 9: actual number of simulations.
\end{tablenotes}
\label{TabRes}
\end{table*}

\begin{figure}
\centering
\includegraphics[width=\columnwidth]{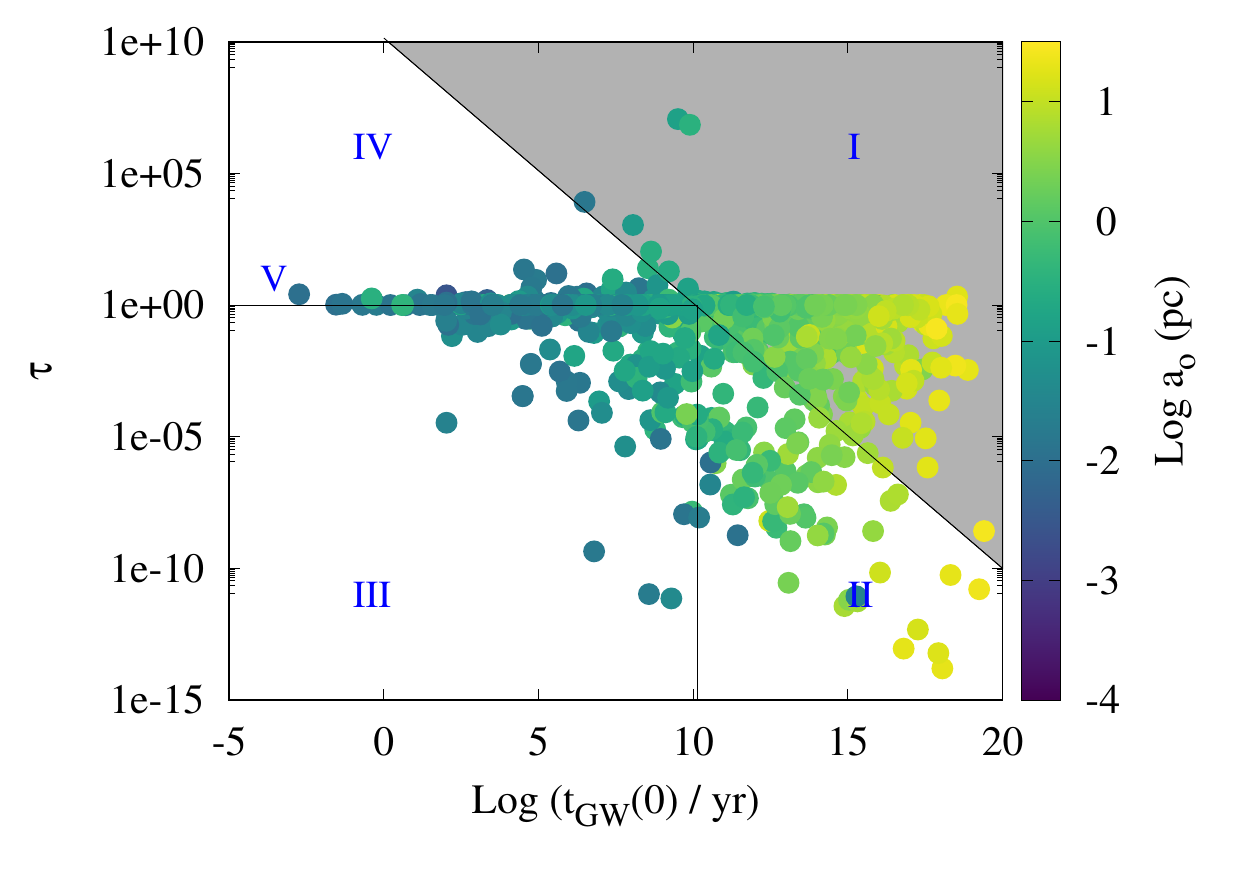}
\caption{Ratio between the initial and final GW time, $\tau$ for BHBs in SET2, as a function of $t_{\rm GW}(0)$. Different colors identify the initial outer semi-major axis. The dotted diagonal separates merging BHBs (in the bottom-left side) from those having $t_\gw>14$ Gyr. The vertical line separates BHBs with $t_{\rm GW}(0)\lessgtr14$ Gyr, while the horizontal separates hardened binaries($\tau>1$) from softened ones.}
\label{F12}
\end{figure}

In order to unveil how the external perturbations affect the BHB dynamics, we plot in Figure \ref{F12} the ratio $\tau = t_\gw/t_{\rm GW}(0)$\footnote{We label with $t_{\rm GW}(0)$ the initial value of the merger timescale.} as a function of $t_{\rm GW}(0)$ for all the BHBs in SET2. 
This plane can be divided in five main sectors, namely I, II, III, IV, and V.

In sector I, BHBs have $t_\gw>14$ Gyr, the external perturbation is not sufficient to drive the BHBs hardening or the eccentricity increase. 

Binaries lying in Sector II have initially $t_{\rm GW}(0)>14$ Gyr, but the perturbations are so efficient to decrease $t_\gw$ below a Hubble time. 

Binaries in Sectors III, IV and V, instead, are initially sufficiently tight to merge within a Hubble time, $t_{\rm GW}(0)<14$ Gyr, but nevertheless the external perturbations affect their evolution. In particular, in Sector III BHBs harden over the simulated time ($\tau<1$), while in Sector IV they soften ($\tau>1$), although conserving a GW time smaller than a Hubble time.

The locus defined by the conditions $t_{\rm GW}(0) < 14$ Gyr and $\tau \simeq 1$ allows us to identify a further Sector V, which gather unperturbed BHBs, whose merger is driven solely by GW emission.
  
\begin{figure}
\centering
\includegraphics[width=\columnwidth]{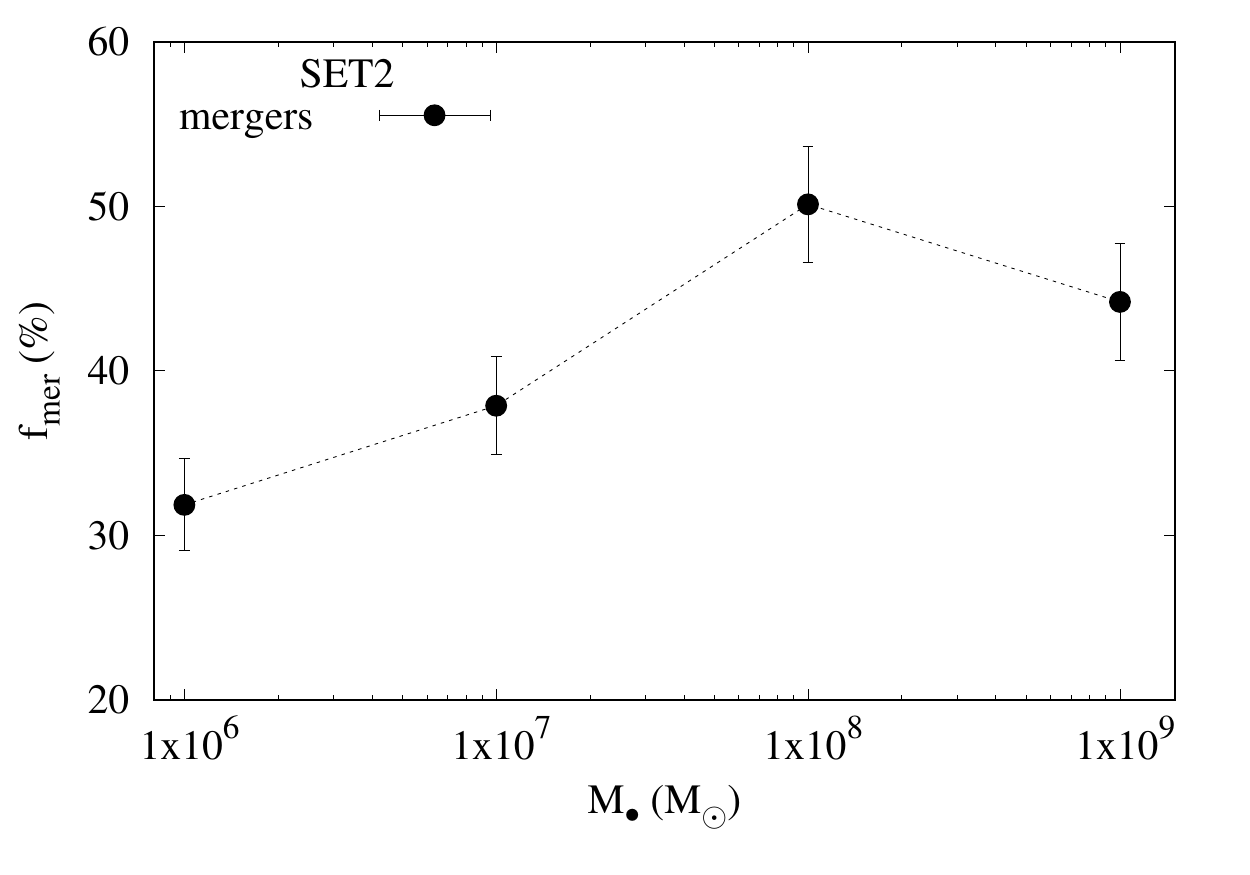}
\includegraphics[width=\columnwidth]{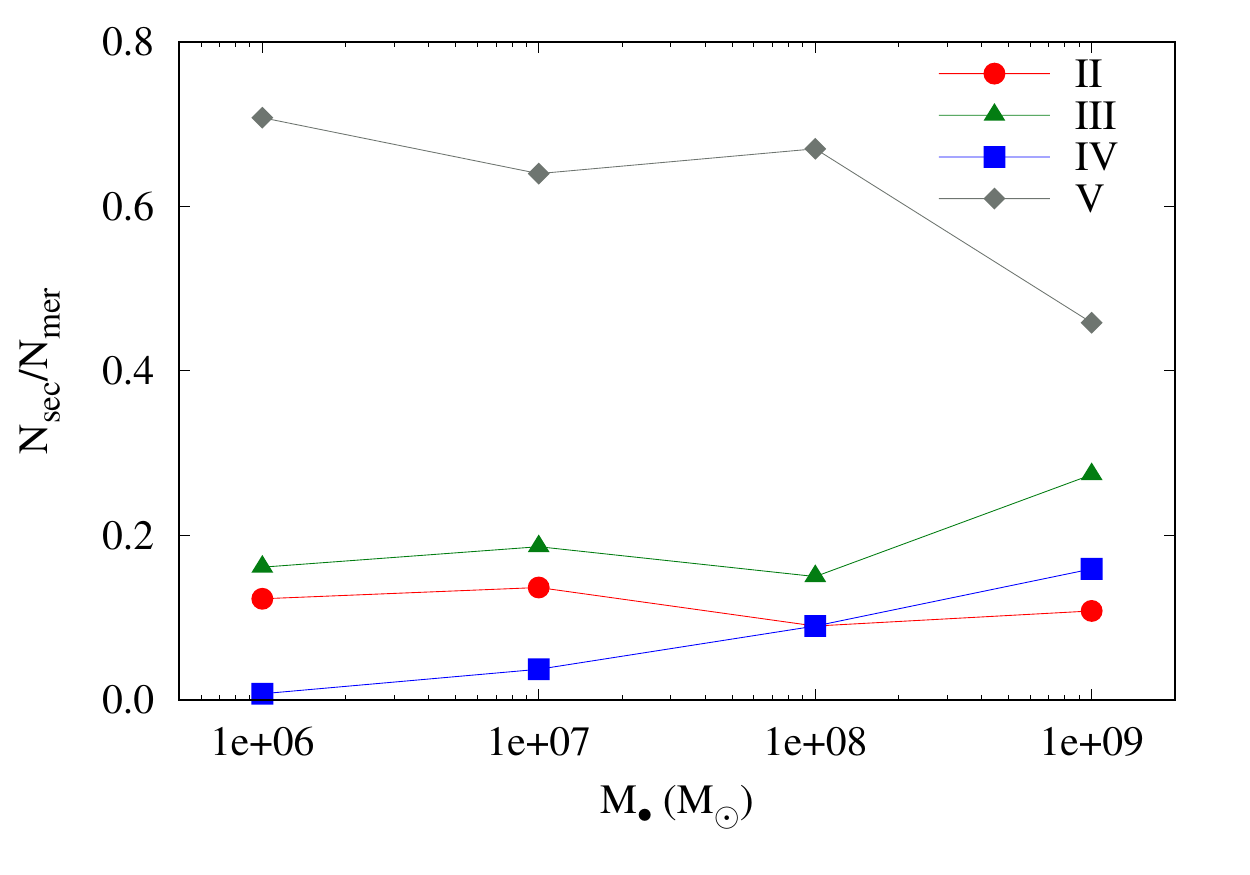}
\caption{Top panel: merger probability for all the merger candidates in SET 2 at varying the SMBH mass. Bottom panel: the same as in the top panel, gathering the mergers in different Sectors as explained in the text.}
\label{F13}
\end{figure}

Considering all the merger candidates, we find that the merger probability $f_{\rm mer}$ attains values around $30-40\%$ and depends weakly on the SMBH mass, as shown in top panel of Figure \ref{F13}.

In the bottom panel, we dissect the mergers population among the different sectors, showing the mergers number in each sector normalized to the total number of mergers, $N_{\rm sec}/N_{\rm mer}$. For BHBs in Sector II the dependence varies weakly with the SMBH mass. The $N_{\rm sec}/N_{\rm mer}$ ratio, instead increases for mergers in Sector III (hardened BHBs) and IV (softened BHBs). This is likely due to the fact that the external perturbation becomes stronger at increasing the SMBH mass. Likely for the same reason, the number of candidates lying on Sector V tends to diminish at increasing the SMBH mass.

\begin{figure}
\centering
\includegraphics[width=\columnwidth]{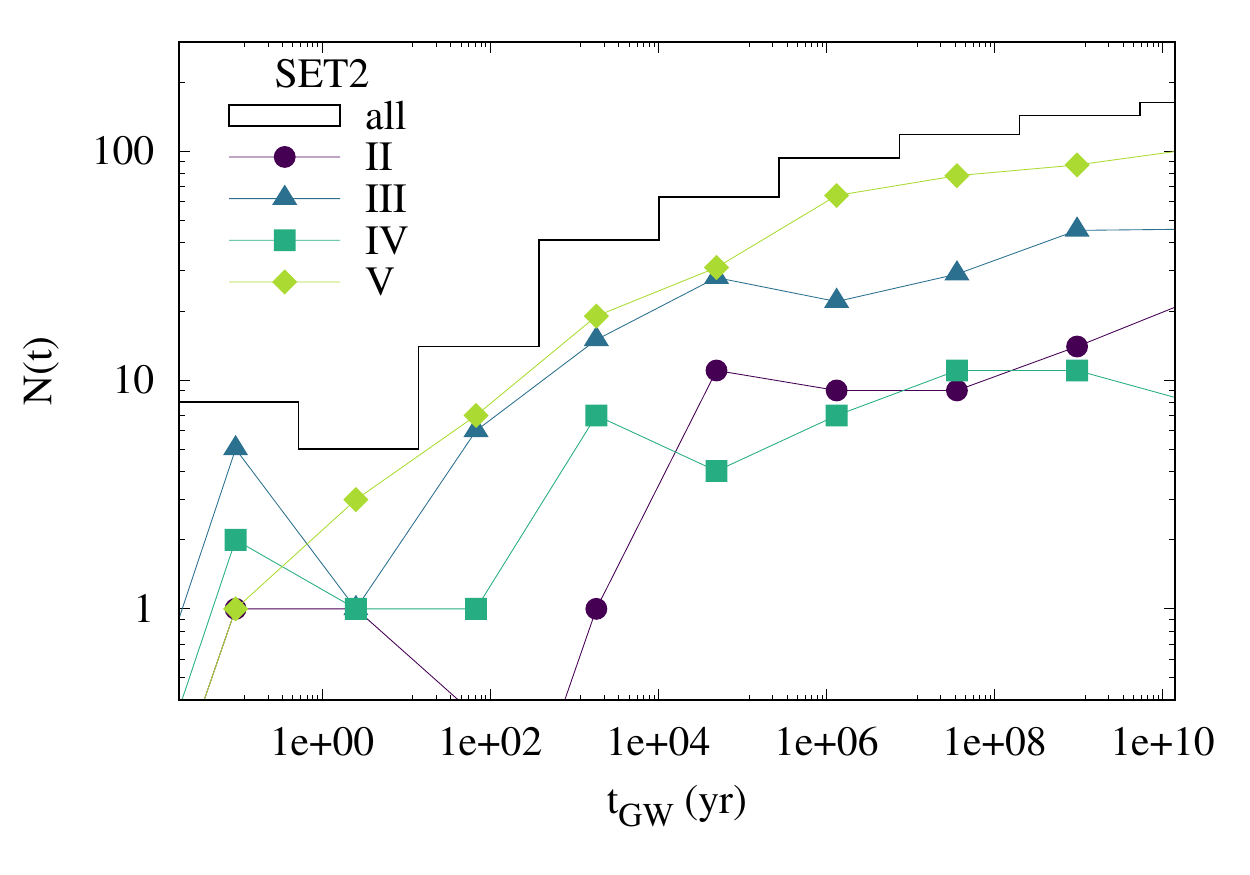}
\caption{Merger time distribution for mergers in different sectors as defined in the text. The black straight steps identify the total merger time distribution. }
\label{F14}
\end{figure}

The population of merger candidates in different sectors have different $t_\gw$ distributions, as shown in Figure \ref{F14}. Binaries in Sector II, for which the external perturbations are sufficiently strong to reduce the GW time below the Hubble time, show two branches, one prominent and broadly distributed between $10^4-10^{10}$ yr and the second in the range $0.1-100$ yr. Mergers in Sector IV show a broad distribution that extends down to $\lesssim 1$ yr and is characterized by a clear rise up to $10^{10}$ yr. Unperturbed BHBs (Sector V) show a monotonic rise that covers the whole $0.01-10^{10}$ yr time range. The GW time distribution for hardened binaries in Sector III shows a smooth increase above $t_\gw = 100$ yr and a small peak around $t_\gw = 0.2$ yr.

\begin{figure}
\centering
\includegraphics[width=\columnwidth]{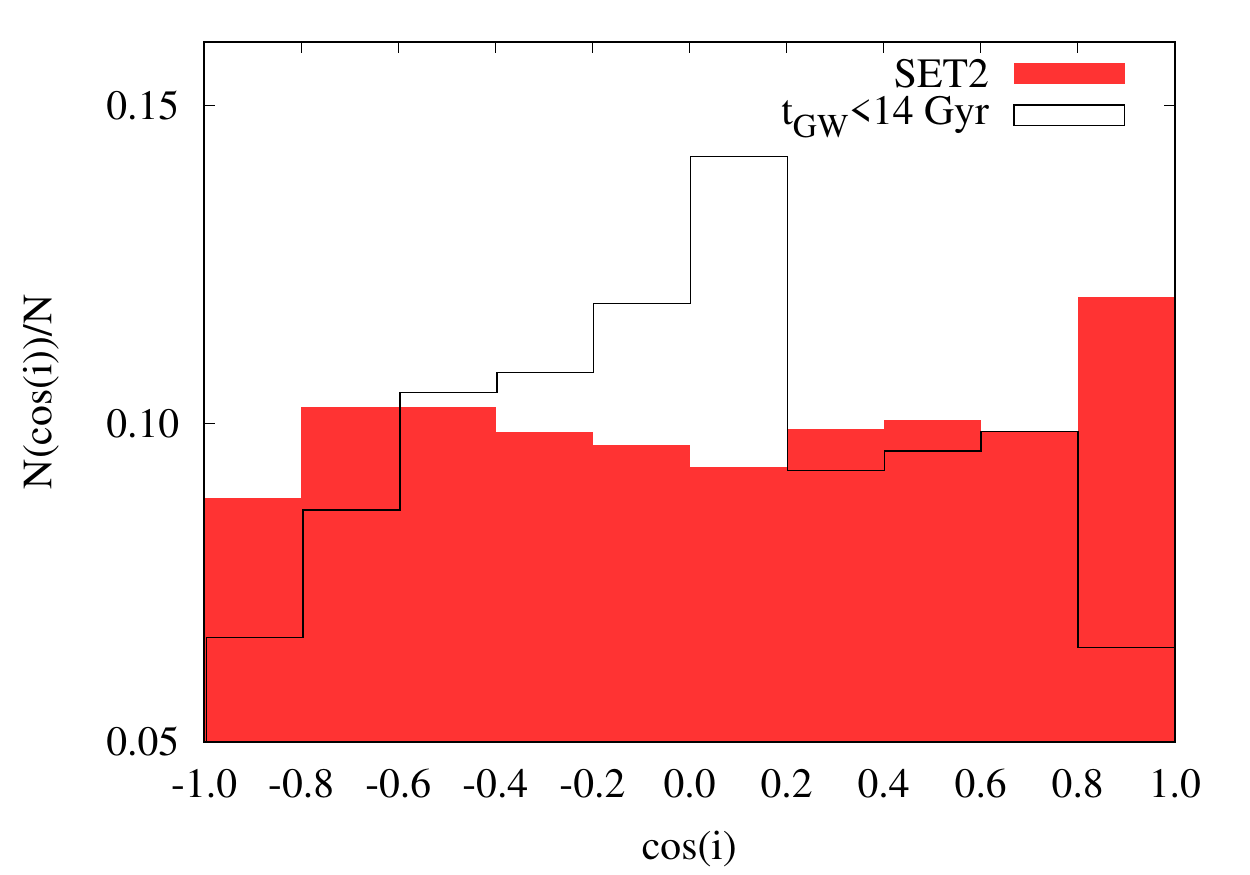}
\caption{Initial (red filled steps) and final (black steps) distributions of the BHB+SMBH mutual inclination in SET 2.}
\label{F16}
\end{figure}

As in SET 1, merging BHBs in SET 2 have a well defined inclination distribution, with a clear peak in correspondence of nearly perpendicular configurations, as shown in Figure \ref{F16}. The peak at low inclinations -- high $\cos(i)$ values -- is likely due to an initial bias of the initial conditions, as suggested by the initial inclination distribution. Nevertheless, is worth noting that a large fraction of nearly co-planar, prograde models ($\cos(i)\sim 0.7$) merge. 

\subsection{Kozai-Lidov oscillations in the nuclear cluster potential well}

\begin{figure}
\centering
\includegraphics[width=\columnwidth]{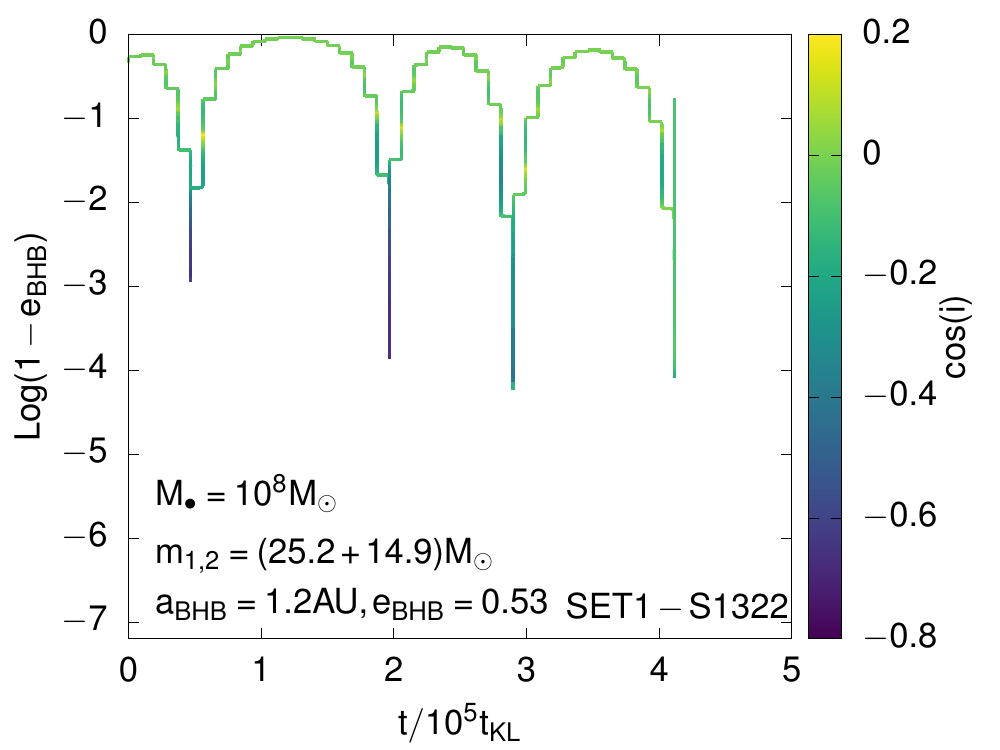}
\caption{Time evolution of the eccentricity for BHB merger model 1321 in SET 1. The color coded map marks the cosine of the inclination. Time is normalized to the KL timescale calculated at $t=0$.}
\label{F8}
\end{figure}

A typical example of merger driven by KL oscillations in SET 1 is given in Figure \ref{F8}, 
which shows the periodic variation of the eccentricity and inclination for a BHB with mass $m_\bbh = (14.9+25.2)\Ms$, semimajor axis $a_\bbh =1.2$ AU, and eccentricity $e_\bbh=0.53$, orbiting an SMBH with mass $M_\smbh=10^8\Ms$. In this specific case, KL mechanisms lead the eccentricity to increase up to $e= 0.999$, inducing the BHB merger in $t_\gw \sim (5\times 10^5~t_{\rm KL}) \simeq 3\times 10^7$ yr, a timescale much shorter than the GW time calculated for the corresponding isolated binary, $t_{\rm GW}(0) = 1.4 \times 10^{13}$ yr. 

As discussed above, our numerical approach includes in particles' equations of motion the contribution coming from the NC gravitational field. This component represents a perturbation term that can alter the evolution of BHBs orbital parameters in a non-trivial way, depending on the distance from the SMBH.

In order to shed a light on the role that the NC gravitational field has on the BHB evolution, we selected two merging candidates and re-simulated them assuming either an isolated BHB-SMBH triple, or adding the external potential $\Phi_\nc(a_o)$.

\begin{figure}
\includegraphics[width=\columnwidth]{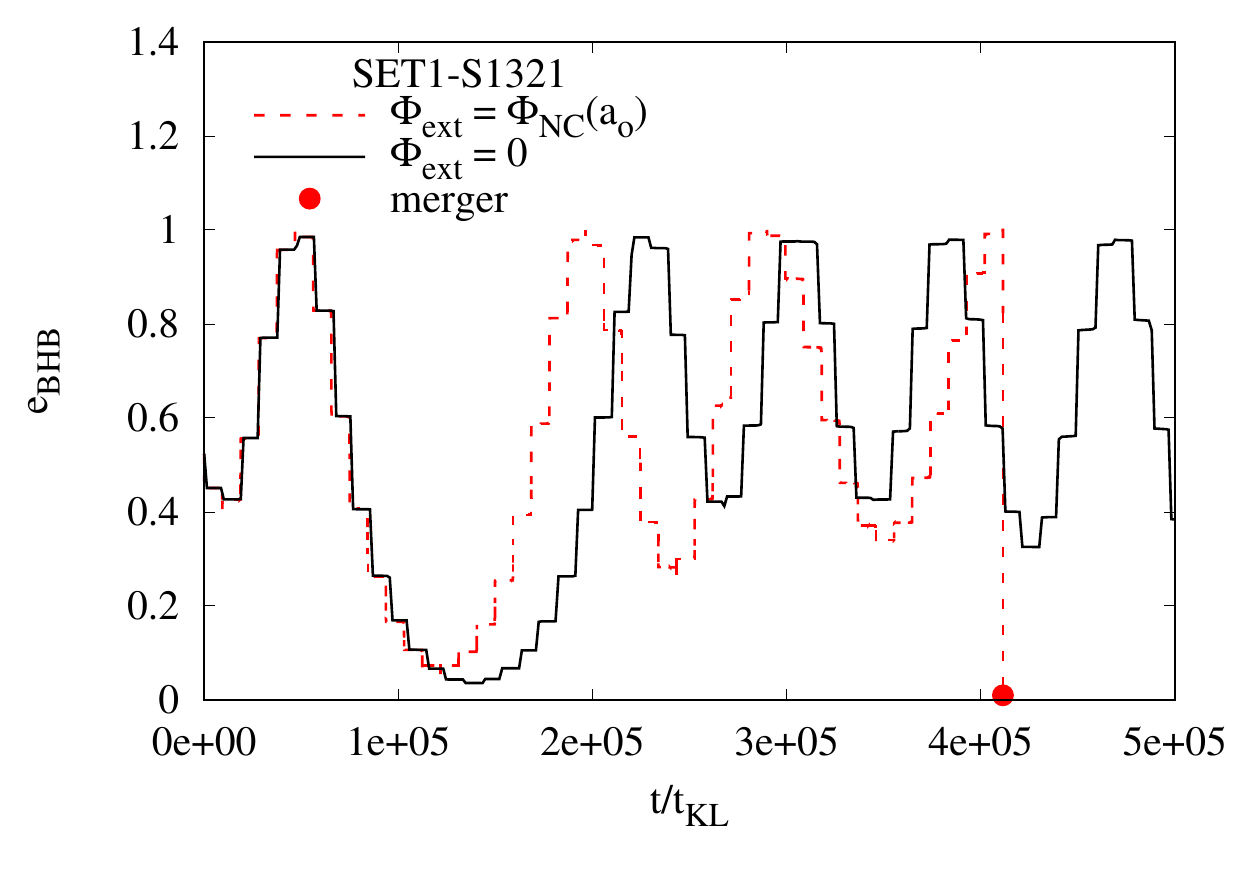}\\
\includegraphics[width=\columnwidth]{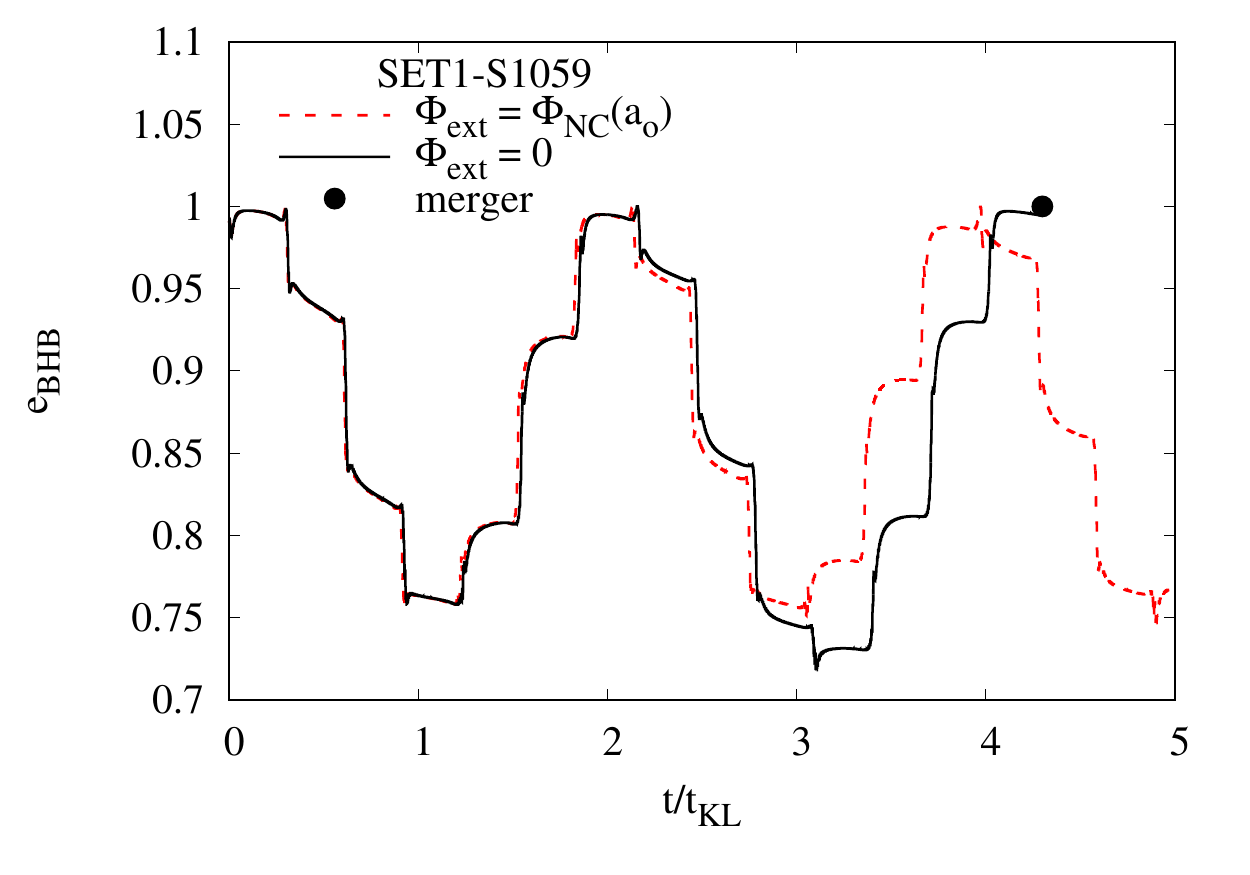}
\caption{Eccentricity variation for models S1321 (top panel) and S1059 (bottom panel) in SET1, assuming $\Phi_{\rm ext} = 0$ or $\Phi_{\rm ext} = \Phi_{\nc}(R_o)$. Coloured points mark the moment at which the BHB merges.}
\label{dd}
\end{figure}

Figure \ref{dd} shows the time evolution of the BHB eccentricity for two particular models in SET 1, namely model S1321 and S1059.

The BHB moves at $a_o\simeq 0.05$($0.1$) pc from an SMBH with mass $M_\smbh = 10^8\Ms$($10^9\Ms$)  in model S1321(S1059). At these distances, the NC contribution to the total mass enclosed within the SMBH orbit is $M_\nc(a_o)/M_\smbh = 2\times 10^{-3}$ for S1321 and $5.6\times 10^{-4}$ for S1059, respectively.

The acceleration impinged by the NC onto the BHB varies over the BHB trajectory, being $10^{-5}-0.01$ times the SMBH acceleration in S1321 and remaining below $6\times 10^{-4}$ in model S1059. Although modest, the NC contribution to the BHB acceleration varies significantly along the orbit, affecting significantly its evolution. In these two particular cases, the $\Phi_{\rm ext}$ term has an opposite effect on the BHB: in S1321, setting $\Phi_{\rm ext}=0$ delays the binary merger by several order of magnitudes, while in S1059 leads to merger in $\sim 4t_{\rm KL}$.

The NC potential has two effects on the BHB overall orbit: first, it reduces the BHB-SMBH apocentral distance, thus implying a larger acceleration impinged on the BHB at apocentre; second, it causes a shift on the orbit compared to the case in which the perturbing field generated by the SMBH is Keplerian. In S1321, the external potential causes an extreme increase of the eccentricity up to $e_\bbh = 0.99999$, which in turn causes a reduction of the semi-major axis because of energy loss due to a GW burst released at pericentre. Subsequently, the BHB undergoes several full KL oscillations until the binary enters the GW dominated regime and quickly merges.
The reverse occurs in S1059: the eccentricity maximizes in the case $\Phi_\nc = 0$, and remains almost constant along a full orbit around the SMBH. This causes the binary to shrink and slowly inspiral down to the merger. When $\Phi\neq0$, instead, the eccentricity increase is less effective, thus avoiding the BHB to fall in the GW regime.

These two examples outline the difficulties in characterizing the actual role of the external potential, which is already effective when the NC mass inside the BHB orbit is a tiny fraction compared to the SMBH mass. The NC field seems to either boost the eccentricity increase, like in S1321, or dump it, like in S1059. However, it is unclear how such effect depends on the full orbital parameters space. We postpone the full exploration of parameter space to a forthcoming work.

\section{Gravitational waves}
\label{Sec5}

The LIGO and Virgo collaboration released the first catalogue of GW sources detected during the O1 and O2 observational campaigns \citep{LIGO19}, consisting of 10 BHBs with total masses up to $\sim 90\Ms$, while more than 10 new potential candidates have been detected during the first months of the O3 observational run. In this section we explore whether the properties of mergers developing in galactic nuclei might be compatible with LIGO sources. In the following, we combine results from both SET1 and SET2, assuming that the whole population of BHB mergers is representative of the typical population harboured by galactic nuclei. 
 
\subsection{Black hole binaries mass}

Our current understanding of stellar evolution suggests that the BHs mass spectrum is severely affected by pair instability and pulsational pair instability supernova \citep{woosley07}. Indeed, these explosive mechanisms leads to a dearth of BHs at low metallicities ($<0.1$ solar values) in the $\sim 50-140\Ms$ mass range \citep{belczynski16a, spera17}. Isolated binary evolution seems to be inefficient at forming BHB mergers with remnants in this ``BH mass-gap'' \citep{spera18,giacobbo18}, although single BHs with such masses can be formed via collision of main sequence stars \citep{spera18}. 

Star clusters, where dynamical interactions are frequent, are unique places to form BHs populating the mass-gap \citep{mapelli16,banerjee16,rodriguez15,banerjee18,rodriguez18,ASKLI18,rastello19,dicarlo19}. In galactic nuclei, the large escape velocities suppress the post-merger BH ejection, thus opening the possibility for BHs to undergo multiple mergers \citep{antonini16,Gerosa17,antonini18c,rodriguez18,ASBEN19}. This makes galactic nuclei appealing systems to chase for BHs in the gap. 

The mass distribution of merging candidates in SET1+2 follows the overall BHB mass distribution, showing a clear peak around $20 \Ms$ and an extended tail up to $120-140\Ms$. This implies the possibility to use observations of merged BHBs to infer information about the global mass spectrum of BHs in galactic nuclei. 

\begin{figure}
\includegraphics[width=\columnwidth]{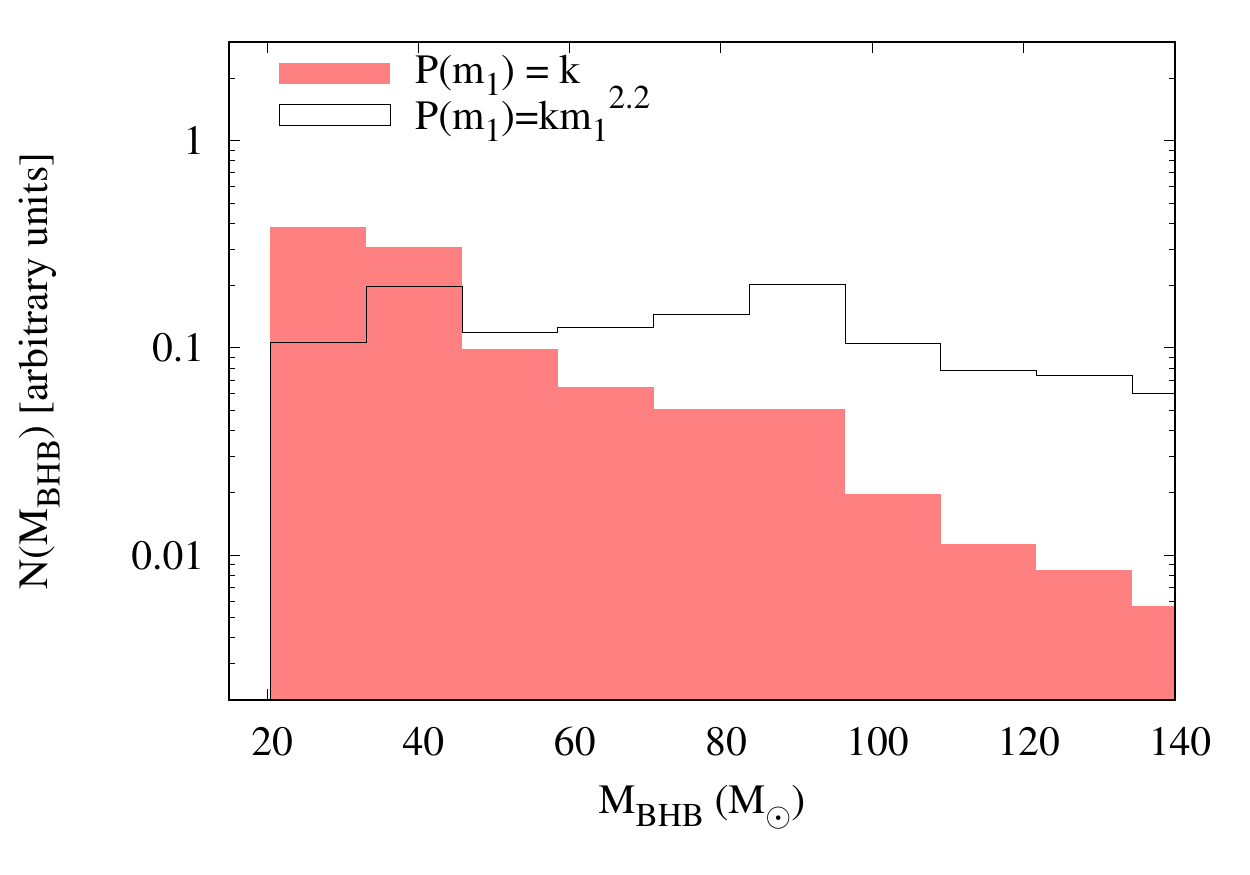}
\includegraphics[width=\columnwidth]{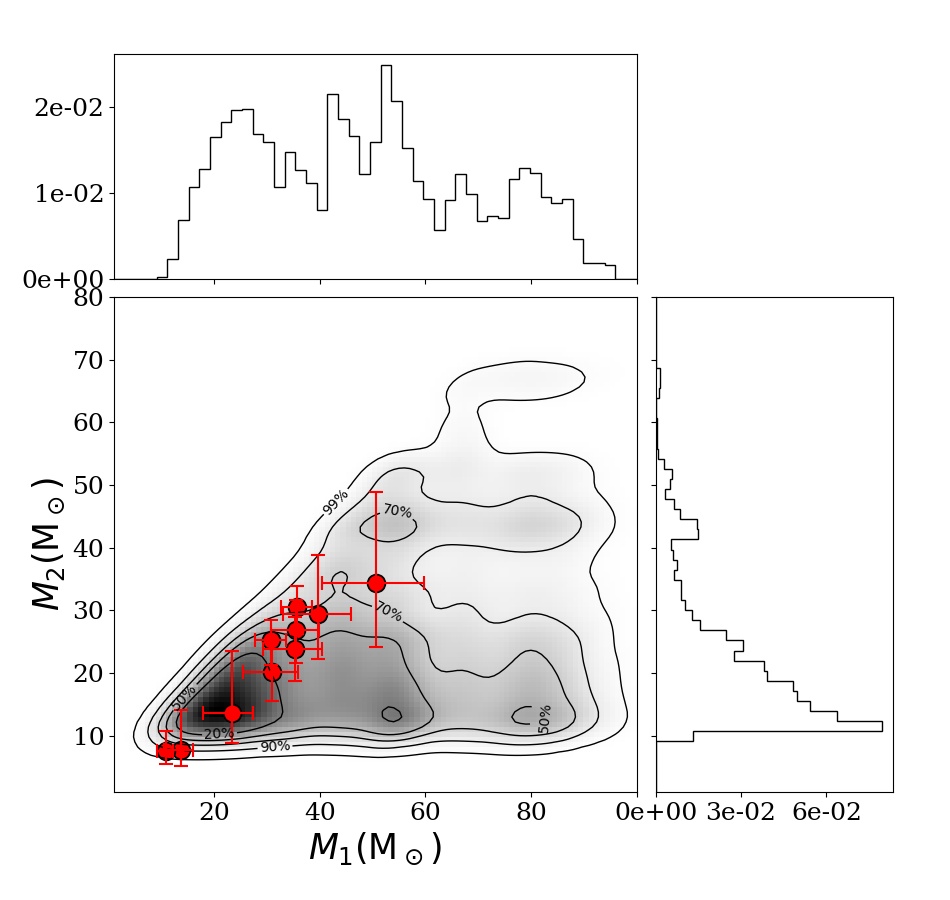}
\caption{Top panel: mass distribution of merger BHBs in SET1+2, assuming no bias on the LIGO sensitive volume (red filled boxes), and assuming that the observed values scales with a power of the primary BH (black steps). Bottom panel: combined mass distribution for merger primary (x-axis) and companion (y-axis).}
\label{fff0}
\end{figure}

To compare our merging BHBs with LIGO observations, we must take into account the fact that the volume to which LIGO is sensitive depends on various parameters, like the mass of the primary component and the spin of both BHs. Recently, \cite{fishbach17a} have shown that the observed volume scales with a power-law of the primary mass, $V \simeq k m_1^{2.2}$. This dependence results in a higher probability for GW detectors to observe heavier BHBs. Such effect might help to reconcile the observed remnant mass and spin distribution with theoretical observations of both isolated and dynamically formed binaries \citep{ASBEN19}. On the other hand, it must be stressed that the sensitive volume depends also on other parameters, like the spins and mass ratio, in a non-trivial way. We take into account the $V-m_1$ dependence in the calculation of the mass distribution by weighting each mass bin with a corrective factor $f_V \equiv km_1^{2.2}$, where $k$ is a normalization constant. Top panel of Figure \ref{fff0} compares the actual BHB mass distribution and the same quantity corrected for the volume-primary mass dependence, which should roughly represent the distribution as seen by the LIGO perspective. Upon this correction, the global mass distribution is roughly flat in the $20-140\Ms$ mass range, thus implying nearly $58\%$ of BHB mergers with masses in the mass-gap ($50-140\Ms$). 
Mergers mass and mass ratio are two important parameters that can be used to constrain their formation channel. Binaries forming in globular clusters tend to be characterized by large mass ratios \citep[see for instance][]{rodriguez15}, those in low-mass clusters have high mass ratios as well \citep{banerjee16} and, on average, lower total masses \citep{dicarlo19}. In galactic nuclei the picture might be slightly different.
The bottom panel of Figure \ref{fff0} shows the combined distribution of primary and companion masses taking into account the correction $f_V$ and how they compare with the 10 known BHB mergers \citep{LIGO19}. We find an interestingly large probability to form mergers with a high primary mass and low mass ratio, namely the region of the plane defined by $m_1>40\Ms$ and $m_2<30\Ms$ that is poorly covered by other dynamical channels \citep{rodriguez16,dicarlo19}. Therefore, observing merging BHs in these mass ranges could indicate a galactic nuclei origin, although it must be noted that, at a fixed primary mass, the GW signal emitted by a merging binary will be fainter for lower mass ratios and might lead to further sources of observational biases that can affect the actual detectability of low-mass ratio binaries.

To understand how our mergers compare to observed BHBs, for each LIGO source we draw 100 mergers from the combined SET1+2 sample and calculate the fraction among them having either the mass or the mass ratio within $10\%$ of the observed value. Upon this selection criterion, we find that galactic nuclei mergers have a probability of $\sim 14-16\%$ to have masses similar to LIGO sources, and $\sim 17\%$ to have similar mass ratios. However, it must be noted that the error on LIGO estimated mass ratios can be as high as $90\%$, thus the comparison for this quantity is much less significant.

\begin{figure}
\includegraphics[width=\columnwidth]{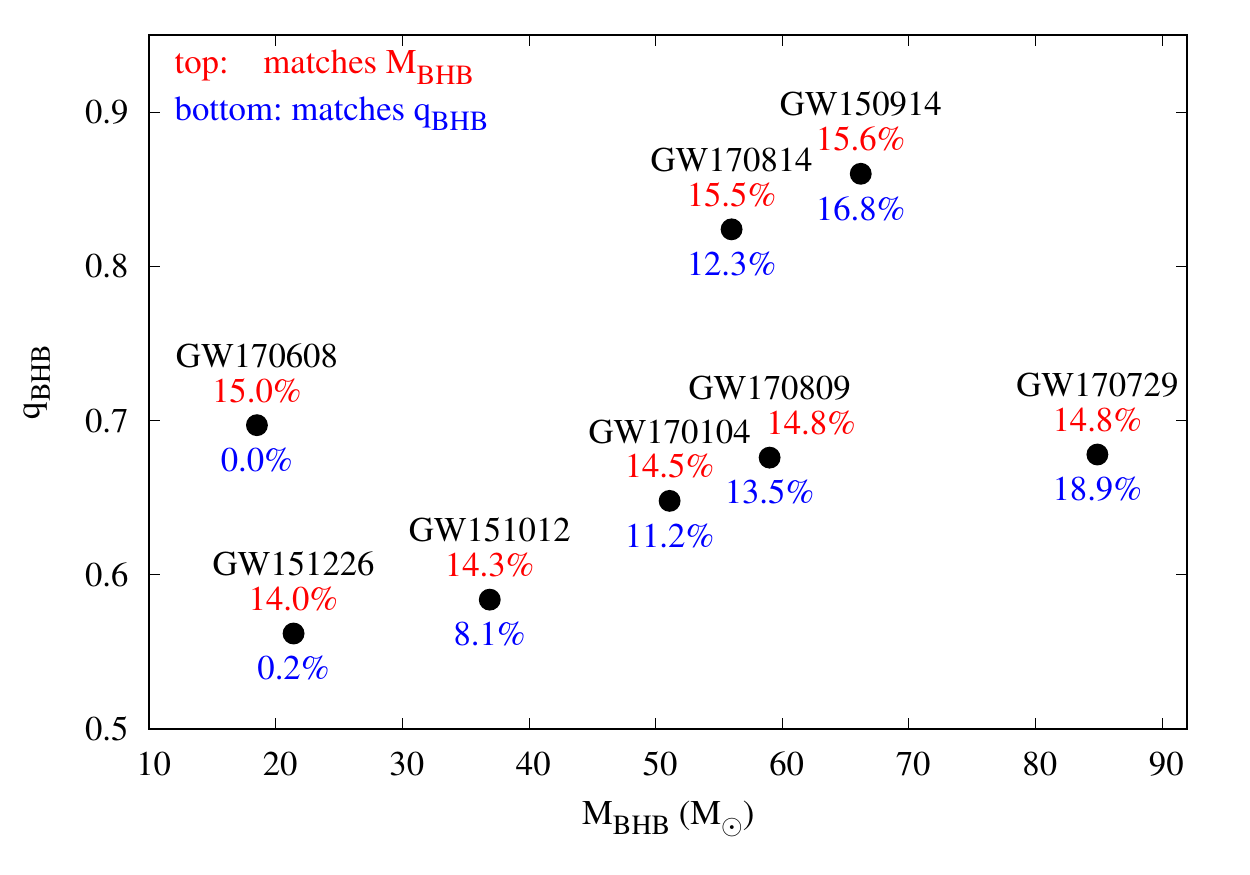}
\caption{Mass ratio as a function of the mass for the known LIGO sources. Blue labels (below the points) identify the probability to obtain a merger in SET1+2 with a mass close to LIGO sources mass. Red labels (on top of points) identify the probability to obtain a merger in SET1+2 with a mass ratio close to LIGO sources.}
\label{fff2}
\end{figure}

\subsection{Merger rates}

To roughly estimate at which rate BHBs merge around an SMBH, we define a merger rate \citep{hoang18}
\begin{equation}
\Gamma = f_{\rm mer} N_{\rm std} n_g f_{\bullet} \delta,
\label{merg}
\end{equation}
where $N_{\rm std}$ is the number of BHBs inhabiting the galactic centre, $n_g$ is the galaxy number density in the local Universe and $f_\bullet$ the fraction of galaxies hosting an SMBH. 
Note that $N_{\rm std}$ represents the steady-state number of BHBs, namely the number of BHBs inhabiting the galactic centre at any time. This is the most uncertain parameter in our treatment, as it depends on the timescale associated to BHBs reservoir replenishment.  However, as discussed in Section \ref{Sec2}, our treatment suggests that in-situ and delivery channels can lead to up to $10^4$ BHBs in galactic nuclei, depending on the NC and SMBH properties. In the following we either assume $N_{\rm std}=200$, to compare with previous works \citep{hoang18}, or $N_{\rm std}=1000$, which provides us with an optimistic estimate. The $\delta$ parameter, defined as
\begin{equation}
\delta = \frac{1}{N_\bbh}\frac{{\rm d}N_\bbh}{{\rm d}t},
\end{equation}
measures the merging frequency. 

\begin{figure}
\includegraphics[width=\columnwidth]{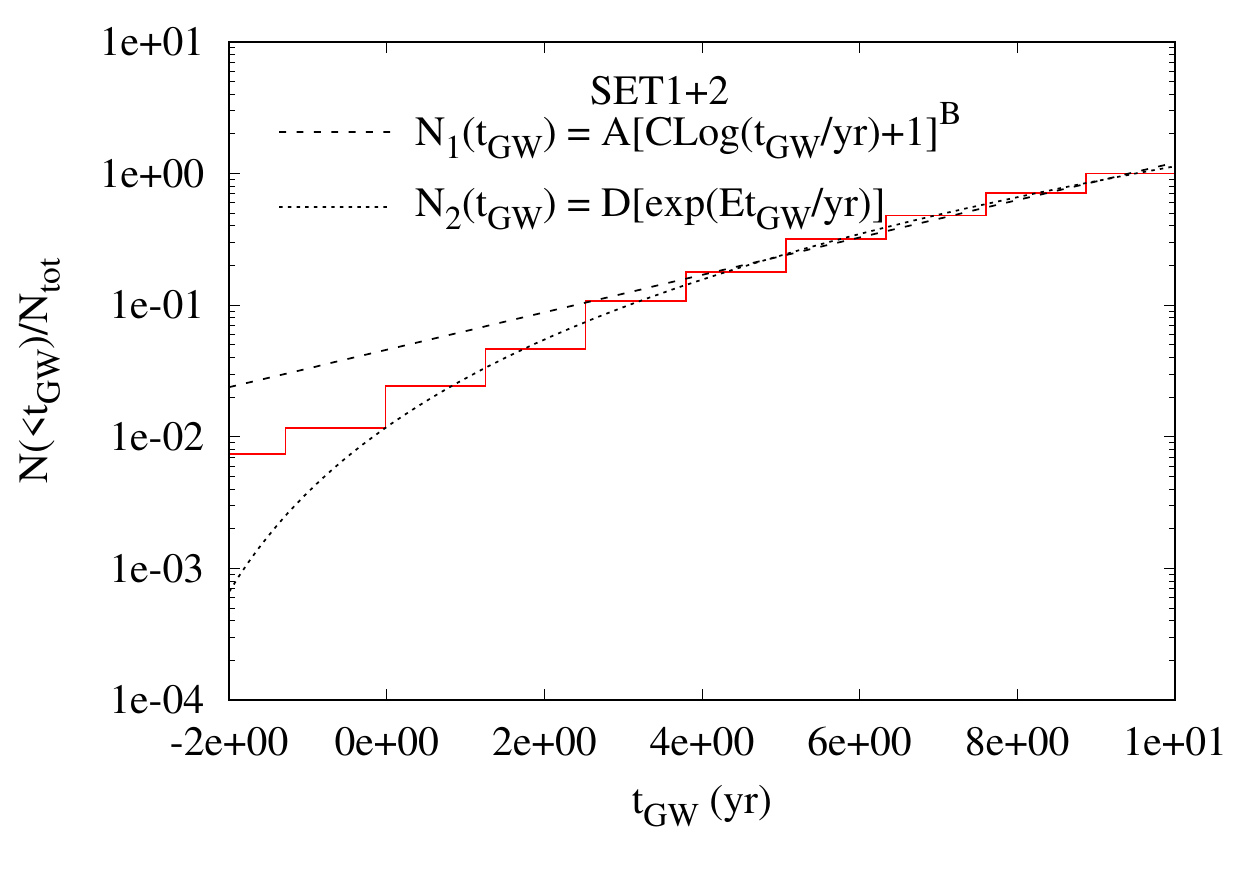}
\caption{Cumulative distribution of merger times for BHBs in SET1+2.}
\label{fitNtgw}
\end{figure}

In order to estimate $\delta$, we resample our mergers ensamble in SET1+2 using the merger times cumulative distribution, similarly to \cite{hoang18} analysis. We create a ``mock'' sample of 50000 mergers that we use to reconstruct the $t_\gw$ cumulative distribution. We find two suitable fitting formula for this quantity (see Figure \ref{fitNtgw} for a comparison between the two expressions), namely
\begin{align}
N_1(t_\gw) & =   A[C{\rm Log} t_\gw + 1]^B, \\
\frac{{\rm d}N_1}{dt_\gw}& =  \frac{ABC[C{\rm Log} t_\gw +1]^{B-1}}{\left[t_\gw \ln(10)\right]}, 
\end{align}
and
\begin{align}
N_2(t_\gw) & =  D\exp\left(E{\rm Log}t_\gw\right), \\
\frac{{\rm d}N_2}{dt_\gw}& =  \frac{DE\exp\left(E{\rm Log}t_\gw\right)}{\left[t_\gw \ln(10)\right]}. 
\end{align}

To calculate the merger rate, we calculate the $\delta$ parameter at the ``half-life time'' $t_{1/2}$, defined as the time over which half of the merging BHBs in our sample actually merge.  
In the following, we assume that half of the galaxies in the local Universe host an SMBH \citep[$f_\bullet = 0.5$,][]{Antonini15}, that the galaxy number density at low redshift is $\sim 0.02$ Mpc$^{-3}$ \citep{conselice16}, and the number of BHBs in the galactic centre is $N_\bbh=200$ \citep[following][]{hoang18}. This allows us to directly compare our results with other works.
In SET1+2, we find ${\rm Log}(t_{1/2}/yr) = 7.1-7.32$, being the lower(upper) value associated to the fitting formula $N_2(t)$($N_1(t)$). This implies $\delta_{1/2} \sim (1.4-2.2)\times 10^{-8}$ yr$^{-1}$. If we repeat the same calculations for only delivered, softer BHBs (SET1), or tighter BHBs (SET2), we find a half-life time slightly different, being longer for SET 1 (${\rm Log}(t_{1/2}/yr) = 8.03$) and shorter for SET 2 (${\rm Log}(t_{1/2}/yr) = 6.82$).
Replacing in Equation \ref{merg} the merger fraction calculated from simulations in both sets, $f_{\rm mer} \simeq 0.21-0.34$, we obtain a merger rate 
\begin{align}
\Gamma =&  (3.3 - 8.6) {\rm ~ yr}^{-1} {\rm ~ Gpc}^{-1} \times \\
        &  \left(\frac{N_\bbh}{200}\right)\left(\frac{n_g}{0.0116{\rm ~Mpc^{-3}}}\right)\left(\frac{f_\smbh}{0.5}\right) ,
\label{GMerg}
\end{align}
with the limiting values set by the limiting values of $\delta$ and $f_m$. The estimate above is inferred assuming that ``soft'' and ``hard'' binaries contribute equally to the overall population of binaries. However, if we restrict the analysis to only one class, we obtain a merger rate in the range $0.8-1.4{\rm ~ yr}^{-1} {\rm ~ Gpc}^{-1}$ for SET1 and $6.3-21{\rm ~ yr}^{-1} {\rm ~ Gpc}^{-1}$ for SET2. Therefore, the relative amount of soft and hard binaries is crucial to assess the actual merger rate. Note that the quantity $1/\delta_{1/2}$ provides an estimate of the ``replenishment time'', i.e. the time over which the mechanism that maintain the BHB reservoir in a nearly steady state operates. In our models, we find a replenishment time-scale $1/\delta_{1/2} \simeq 46-73$ Myr. In Appendix \ref{AppA}, we discuss how this parameter relate to the BHB delivery scenario. We note that these results nicely agree with previous estimates for galaxies containing a central SMBH \citep{ASCD17b,ASG17,fragione19,hoang18} or a massive NC \citep{antonini16}, although predicting a larger upper limit. Also, the inferred merger rate is comparable to values obtained for globular clusters \citep{rodriguez16,askar17}, or open clusters \citep{ziosi14,rastello19,banerjee16}. The most recent estimates based on LIGO sources catalogue place the BHBs merger rate in the range $9.7 - 101  ~{\rm yr}^{-1}~{\rm Gpc}^{-3}$ \citep{LIGO19}. Therefore, our results suggest that galactic nuclei BHBs might constitute a small fraction of the global mergers population, likely dominating, as suggested above, a well defined region of the plane defined by component masses, namely $m_1 > 40\Ms$ - $m_2<30\Ms$.

\subsection{Gravitational wave signal}

In this section we discuss how the merging BHBs orbital parameters evolve in the last stages preceding the merger.
In order to do this, we solve the coupled system of differential equations that regulate the evolution of the BHB semi-major axis and eccentricity, following the formalism pioneered by \cite{peters63} and \cite{peters64}.

The peak frequency of GWs emitted by an eccentric BHB is given by\footnote{To make the notation easier to digest, in the following we remove the pedix BHB from mass, semi-major axis and eccentricity.} \citep{wen03,antonini12,kocsis12}
\begin{equation}
f_p = \frac{1}{\pi}\sqrt{\frac{Gm}{a^3}}\frac{\left(1+e\right)^{1.1954}}{\left(1-e^2\right)^{3/2}},
\end{equation}
which represents the frequency of the GW dominant harmonic. As the BHB inspirals frequency increases at a rate \citep{peters64}
\begin{align}
\dot{f}_p =& \left(-\frac{3}{2}\frac{\dot{a}}{a} - k(e)\dot{e}\right)f_p,\\
k(e) =& \frac{1}{(1-e)^{1/2}(1+e)^{3/2}}-\frac{3}{2}\frac{(1+e)^{1/2}}{(1-e)^{3/2}}.
\end{align}

\begin{figure}
\centering
\includegraphics[width=\columnwidth]{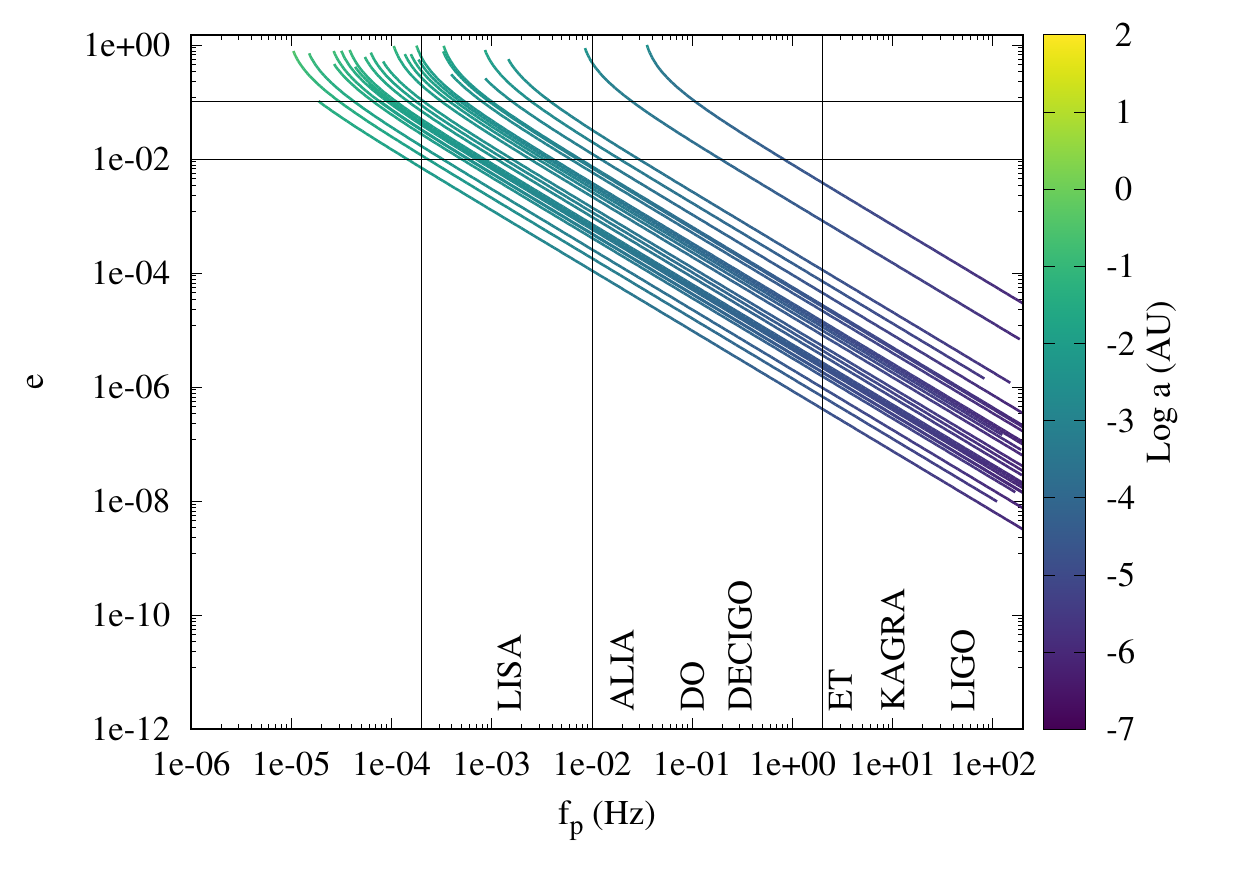}\\
\includegraphics[width=\columnwidth]{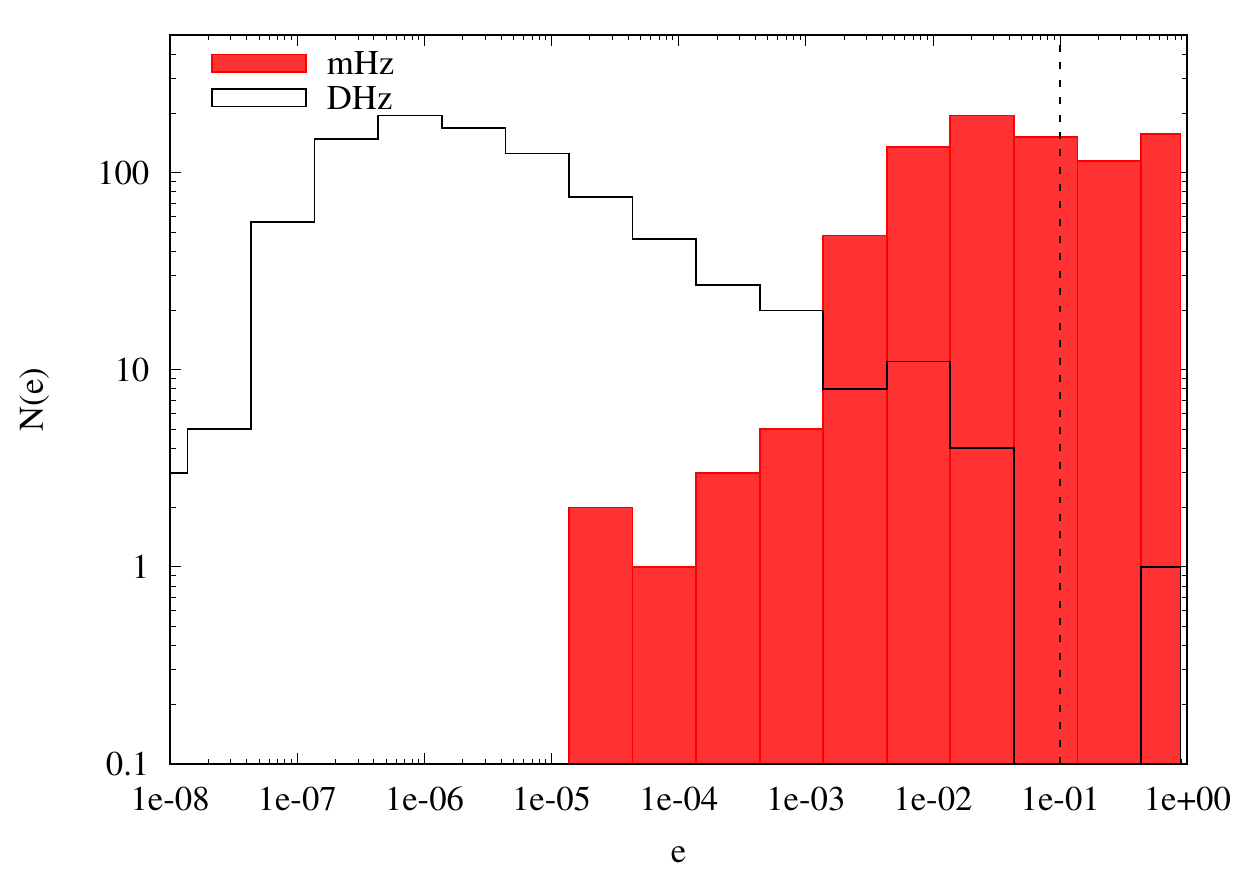}\\
\caption{Frequency (x-axis) and eccentricity (y-axis) variation for all the merging BHBs in SET 1 and 2. The color coded map marks the BHBs semi-major axis evolution. Bottom panel: eccentricity distribution of merging BHBs when they achieve the mHz (filled red steps) and Hz (open black steps) bands in SET 1 and 2.} 
\label{F10}
\end{figure}

The BHB hardening and circularization cause a progressive increase of the GW frequency. It might happen that during this process, the BHB enter an observational frequency band with a still noticeable eccentricity. Top panel of Figure \ref{F10} shows how the merging BHBs frequency varies during the BHB inspiral. 

Comparing the binary evolution with the frequency bands in which GW observatories are, or will be in the future, sensitive, we find that a merger enters the LISA \citep{LISA17} band with an eccentricity $e_\bbh > 0.1$ in $\sim 40\%$ of the cases. The probability to find eccentric mergers drops down to $5\%$ shifting in the 0.1-0.5 Hz regime, the domain of decihertz observatories like ALIA \citep{Bender13}, DO \citep{ArcaSeddaDO19}, or DECIGO \citep{Kawamura11}, and to only $2\%$ in the 0.5-10 Hz window, where LIGO \citep{LIGO14}, KAGRA \citep{Kentaro12}, and the Einstein Telescope \citep[ET,][]{Punturo10} will operate.
Figure \ref{F10} shows the eccentricity distribution calculated when BHBs (in both SET1 and 2) cross the frequency range 0.5 - 5 mHz and 5 - 10 Hz.

The almost complete absence of eccentric sources in the LIGO band is likely due to the fact that when GWs emission kicks in and dominates the binary evolution the typical semimajor axis of the merger candidate is still relatively large, $a_\bbh \gtrsim 0.05-1$ AU, thus the binary is circular by the time enters the Hz frequency window. 

Combining the information on the eccentricity distribution with the merger rates calculated in the previous section, our results suggest that galactic nuclei should contribute to LISA BHB mergers with $\sim 5-6$ sources per yr and Gpc cube.

In order to explore whether these mergers are actually visible to GW detectors, we use the dominant frequency to calculate the GW strain, which can be directly compared with instruments' sensitivity curves.

Clearly, also lower-order harmonics contribute to the GW signal too. 
\cite{oleary09} showed that the $90\%$ of the total GW power emitted  is due to harmonics with frequencies in between 0.2$f_p$ and 3$f_p$.

In each frequency bin, the GW strain for a BHB observed over a time $T$, can be calculated as
\begin{equation}
h_{n}^2(a,e;f) = h_0^2fT\frac{4}{n^2}g(n,e),
\label{strain}
\end{equation}
where $g(n,e)$ is a function of the eccentricity\\ \citep{peters63,gondan18,kocsis12,oleary09}, and $h_0$ is the characteristic strain for a circular orbit \citep{sesana16}
\begin{equation}
h_0(a) = \frac{\sqrt{32}}{5}\frac{G^2}{c^4}\frac{M_{z}\mu_{z}}{Da}.
\end{equation} 
In the equation above, $M_{z} = (1+z)(m_1+m_0)$ is the observed BHB mass, $\mu_{z} = (1+z)(m_1m_0)/(m_1+m_0)$ its reduced mass, while $D$ is the distance from the observer and $z$ the corresponding redshift, which we assume to be $z=0.05$.
Equation \ref{strain} is valid as long as the binary inspiral time is longer than the observation time, namely $f/\dot{f}<T$, and we assume a $T=5$ yr long mission for LISA.
In the case in which this condition is not satisfied, like during the last stages preceding the BHB merger, we scaled down the strain by a factor $\sqrt{Tf/\dot{f}}$ \citep[see][and reference therein]{ASKLI18}. 

\begin{figure}
\centering
\includegraphics[width=\columnwidth]{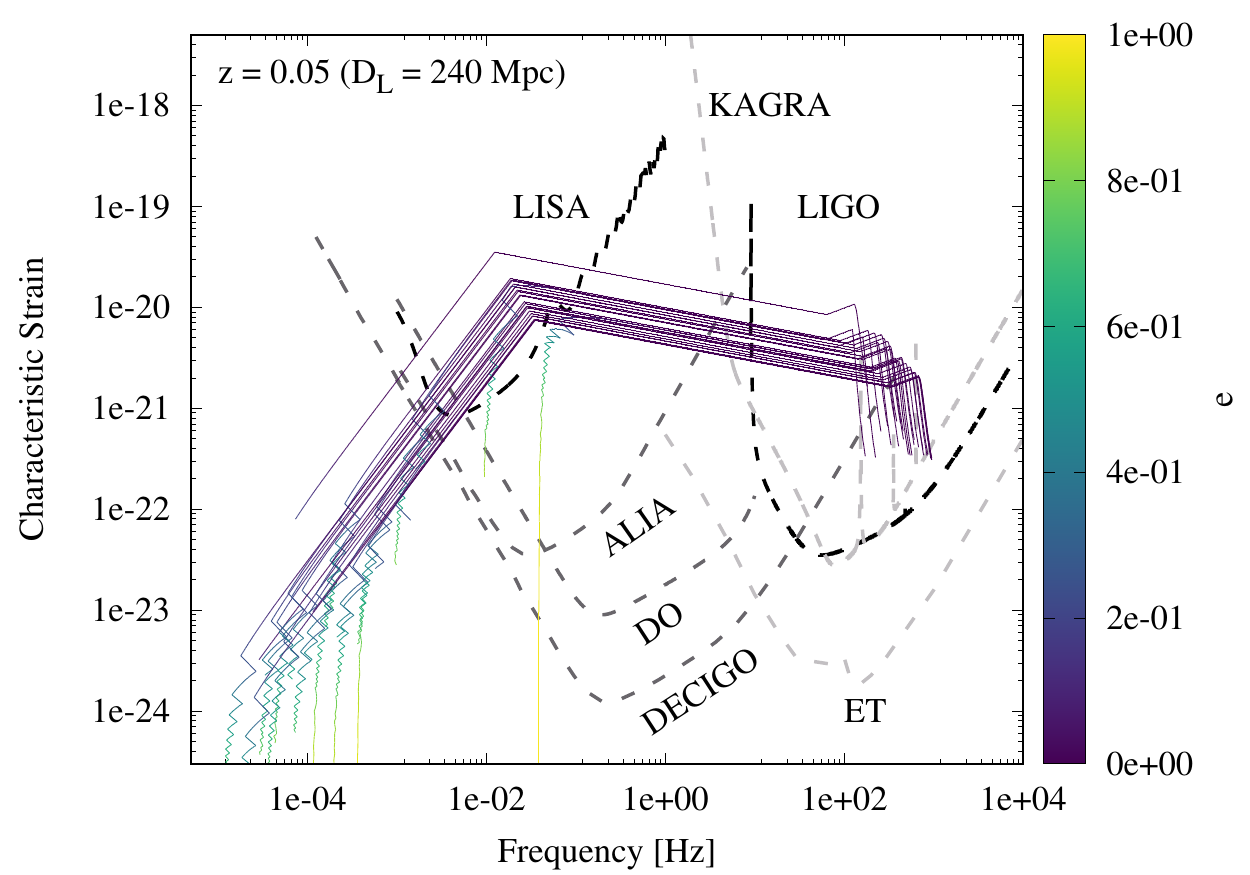}
\caption{Strain-frequency evolution for a subsample of mergers in SET 1 and 2. The color coded maps identify the BHB eccentricity. Our models are overlapped to sensitivity curves of ground- (LIGO, KAGRA, ET) and space-based (LISA, ALIA, DO, DECIGO) GW observatories.}
\label{F11}
\end{figure}

We compare the strain-frequency evolution for some typical BHBs in SET 1 and 2, calculated following the procedure depicted above and only for the dominant frequency, with the sensitivity curve for both low-frequency (LISA, ALIA, DO, DECIGO) and high-frequency (LIGO, KAGRA, Einstein Telescope - ET) detectors, as shown in Figure \ref{F11}. At each moment, we calculate the strain corresponding to the dominant frequency and to lower-order harmonics, not shown in the plot for the sake of readability. Note that in all the models shown, the BHB inspiral crosses at least two observational windows, most of them having a non-zero eccentricity at least in one band. Several mergers transit from $1$ mHz to $100$ Hz during the inspiral phase, possibly being audible in the LISA observational band a few years before they merge and becoming audible to LIGO in the last phases preceding the merger. These ``delayed coincidence'' sources represent the perfect prototype for multiband GW astronomy, as they can be used to validate ground- and space-based detectors, to exquisitely probe general relativity, and to put robust constrains on the cosmological BHB formation and merger rates \citep{sesana16}.  

\section{Conclusions}
\label{Sec6}
In this paper we investigated the formation and evolution of BHBs in galactic nuclei. The main results are summarized in the following.
\begin{itemize}
\item We explore two different formation processes for BHBs in galactic nuclei: {\it in-situ} formation and {\it delivery} from spiralled star clusters. In-situ formation seems to dominate galactic nuclei with high NC-to-SMBH mass ratio, while the delivery formation process is more typical of galaxies hosting massive SMBHs, where dynamical scatterings are prevented by the high-velocity dispersions [Figures \ref{ncsmbh} and \ref{nbhbdry}].
\item Binaries orbiting inside a dense NC will undergo both mass segregation and dynamical scatterings with passing by stars. Due to mass segregation, BHBs move toward regions with an increased scattering rate. We show that in some cases, this leads a BHB to get harder and harder. Overall, this process can significantly shape the global population of galactic nuclei BHBs, potentially leading to sizeable BHB populations that merge only via dynamical hardening. These mergers can enrich significantly the population of BHs with masses above $50\Ms$ in galactic nuclei [Figures \ref{smaev} and \ref{mergers}]. 
\item We perform $N$-body simulations at varying BHBs orbital properties and SMBH and NC masses, taking into account the NC field and post-Newtonian terms. We find that KL mechanism plays a crucial role in determining the properties of merging binaries, causing $\sim 11-30\%$ of all the mergers in our sample. The NC gravitational field has a non-trivial effect on BHBs evolution, as it can either trigger or prevent merger [Figure \ref{dd}]. 
\item In $0.08-5.7\%$ of our models, the SMBH captures one of the BHB components, forming a tight EMRI that merges within a Hubble time.
\item The inferred merger rate for galactic nuclei BHBs is $\Gamma \sim 3.3-8.6$ yr$^{-1}$ Gpc$^{-3}$ at redshift 0 but can increase to up to $\Gamma \sim 20$ yr$^{-1}$ Gpc$^{-3}$ if the population of mergers is dominated by hard binaries. These estimates are compatible with other dynamical channels and falls in the low-end tail of the LIGO merger rate prediction [Equation \ref{GMerg}]. 
\item The combined mass distribution of merger primary and secondary components shows an extended tail in the semi-plane $m_1>40\Ms$ - $m_2<30\Ms$, a region poorly populated by BHBs formed via isolated channel, or via dynamical interactions in young or globular clusters. Observations of GW sources with component masses in these ranges could indicate a galactic nuclei formation channel [Figure \ref{fff0}].
\item BHB mergers forming in galactic nuclei have masses compatible with observed sources in $14-19\%$ of the cases [Figure \ref{fff2}].
\item We calculate the frequency-strain evolution for all merger candidates in our sample, showing that $\sim 90\%$ of them pass through the LISA observational band and merge in the LIGO band. These sources can represent potential candidates for GW multiband observations. In $\sim 40\%$ of the cases, binaries are eccentric in the LISA band, while in a fewer cases binaries are eccentric in the DECIGO band. These binaries spend a short time in LISA band, thus their detectability can be hard, but during the inspiral phase last in the decihertz band for $\sim 1-4$ yr, thus representing potentially bright multiband sources in the 0.01-10 Hz frequency band [Figures \ref{F10} and \ref{F11}].
\end{itemize}

\section{Acknowledgements}
I am grateful to the anonymous referee for the careful reading of the paper and for the suggestions and comments provided, which helped to improve an earlier version of this manuscript. I warmly thank M. Donnari, A. Mastrobuono-Battisti, B. M. Hoang, B. Kocsis, F. Antonini for their helpful comments and suggestions that allowed me to significantly improve an earlier version of this manuscript. I gratefully acknowledge Alexander von Humboldt Foundation, for financial support under the research program ``Formation and evolution of black holes from stellar to galactic scales''. 
Part of this work benefited from support provided by the Sonderforschungsbereich SFB 881 "The Milky Way System" (subproject Z2) of the German Research Foundation (DFG), the COST ACTION CA16104 ``GW-verse'', and the ISSI (Bern), through its Intern. Team prog. ref. no. 393 {\it The Evolution of Rich Stellar Populations \& BH Binaries} (2017-18). Most of the numerical simulations presented here are performed on the Kepler supercomputer, hosted at the Universitat Rechen Zentrum (URZ) of the Heidelberg University, and the Milky Way supercomputer, which is funded by the Deutsche Forschungsgemeinschaft (DFG) through the Collaborative Research Center (SFB 881) "The Milky Way System" (subproject Z2) and hosted and co-funded by the J\"ulich Supercomputing Center (JSC).

\newpage
\footnotesize{
\bibliography{ASetal2015}
}

\newpage
\appendix
\section{The star cluster infall rate}
\label{AppA}

A cluster with mass $M_\gc$, orbiting at a distance $r_\gc$ from the centre is characterized by a dynamical friction time-scale \citep{ASCD14b}
\begin{equation}
\tau_{\rm DF} = \tau_0 g(e_\gc,\gamma) \sqrt{\frac{R_g^3}{M_g}} \left(\frac{M_\gc}{M_g}\right)^{\alpha} \left(\frac{r_\gc}{R_g}\right)^{\beta},
\label{TGC}
\end{equation}
being $\tau_0$ a normalization factor, $g(e_\gc,\gamma)$ a weak function of the cluster eccentricity and the galaxy slope, $\alpha = -0.67$ and $\beta = 1.76$ \citep[see also][]{ASCD15He}.

The time variation of the number of star clusters falling into the galactic centre due to dynamical friction can be written as
\begin{equation}
\dot{N}_{\rm GC} = N_{\rm GC}/\tau_{\rm DF},
\label{eqNrate}
\end{equation}
being $\tau_{\rm DF}$ the average dynamical friction time-scale. To estimate $\dot{N}_{\rm GC}$, we assume that the cumulative spatial distribution of clusters and stars coincide, thus the number of clusters within a given radius is given by \cite{Deh93}
\begin{equation}
N_\gc(r) = N_{\rm GC,t}\left(\frac{r_\gc}{r_\gc+R_g}\right)^{3-\gamma},
\label{NGCc}
\end{equation}
where $N_{\rm GC,t}=0.01M_g/M_\gc$ is the total number of cluster in a galaxy with mass $M_g$ and assuming a cluster average mass $M_\gc$  \citep{ASCD14b,gnedin14,webb15}.

The majority of clusters with infall time smaller than a Hubble time typically formed within the galaxy scale radius $R_g$ or, at most, its half-mass radius $R_h$. The galaxy mass and its length scale are linked by a simple scaling relation, namely \citep{ASCD14b} 
\begin{equation}
\left(\frac{R_g}{{\rm kpc}}\right) = 2.37(2^{1/(3-\gamma)}-1)\left(\frac{M_g}{10^{11}\Ms}\right)^k,
\end{equation}
with $k=0.14$, while the scale radius is connected to the half-mass radius $R_h$ via the relation
\begin{equation}
\nonumber
R_g = R_h \left(2^{1/(3-\gamma)}-1\right).
\label{gh}
\end{equation}
The latter relation implies that the number of clusters and the dynamical friction time calculated at $R_h$ is simply 
\begin{align*}
N_\gc(R_h) & =  2^{2-\gamma} N_\gc(R_g)\\
\tau_{\rm DF}(R_h) & =  \left(2^{1/(3-\gamma)}-1\right)^{-\beta}\tau_{\rm DF}(R_g)\\
\end{align*}

Combining Equations \ref{TGC} and \ref{NGCc} and exploiting these scaling relations, it is possible to show that the clusters infall rate calculated at $R_g$ and $R_h$ is given by
\begin{align}
\dot{N}_{\rm GC}(R_g) &= 0.027{\rm Myr}^{-1} \frac{2^{-3+\gamma}}{\left(2^{1/(3-\gamma)}-1\right)^{3/2}} \left(\frac{M_g}{10^{11}\Ms}\right)^{3/2(k+1)-\alpha} \left(\frac{M_\gc}{10^{11}\Ms}\right)^\alpha, \\
\dot{N}_{\rm GC}(R_h) &=  2^{2-\gamma}\left(2^{1/(3-\gamma)}-1\right)^\beta \dot{N}_{\rm GC}(R_g)
\end{align}

\begin{figure}
\centering
\includegraphics[width=12 cm]{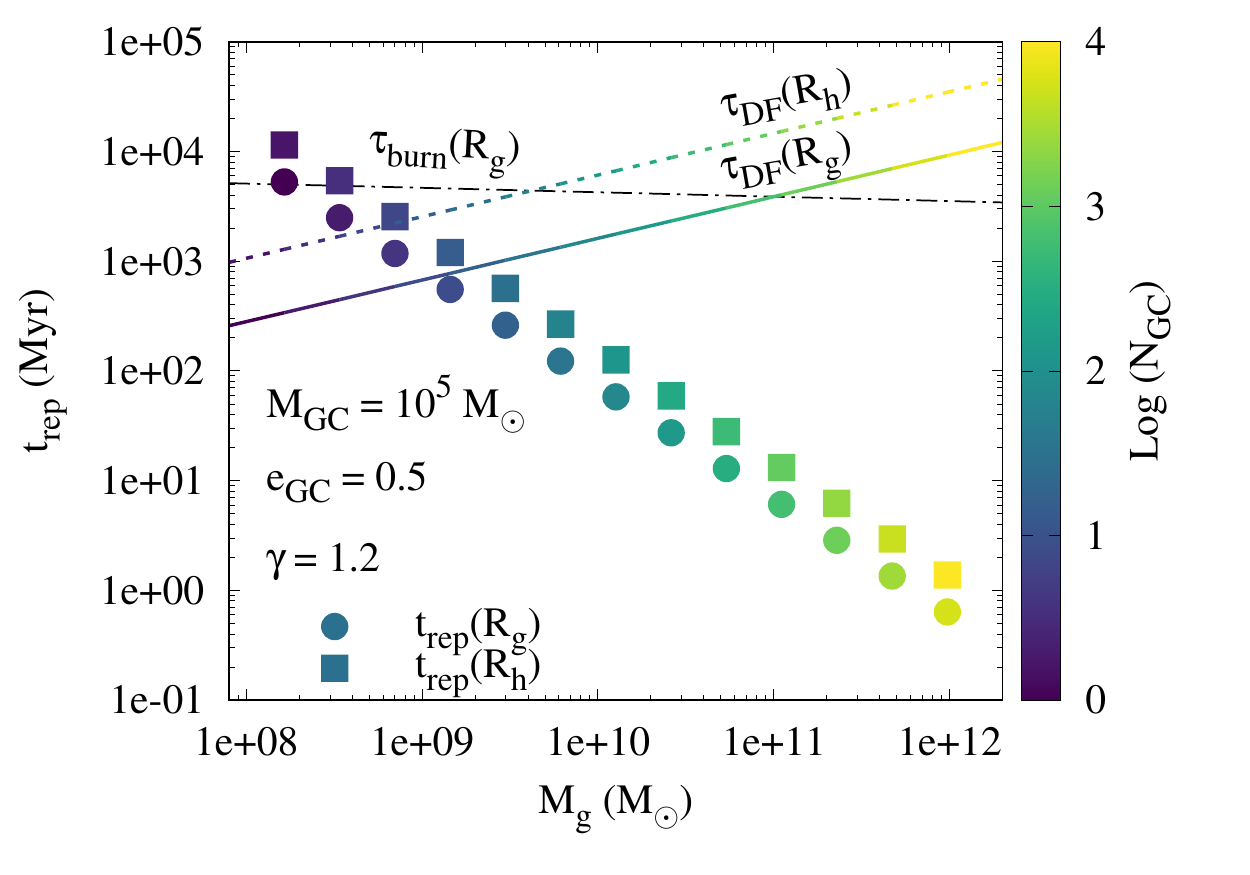}
\caption{Replenishment time as a function of the galaxy mass calculated at $R_g$ (dots) and $R_h$ (squares). The color coded map labels the initial number of clusters. We overplot the average dynamical friction time $\tau_{\rm DF}$ calculated at $R_g$ (straight line) and $R_h$ (dotted line). The black dashed dotted line represent the clusters burning time, $\tau_{\rm burn}$.}
\label{FA1}
\end{figure}

The inverse of the infall rate provides an estimate of the typical time-scale for two subsequent infall episodes to occur, namely a ``cluster replenishment time''
\begin{equation*}
t_{\rm rep}(r_\gc) = \left(\dot{N}_\gc(r_\gc)\right)^{-1}. 
\end{equation*} 
Figure \ref{FA1} shows how this quantity, calculated either at $R_g$ and $R_h$, varies across a range of galaxy masses. In this case, we assume a fixed cluster average mass $M_\gc = 5\times 10^5\Ms$, an average eccentricity $e_\gc = 0.5$, and a fixed slope for the galaxy $g=1.2$.
Note that the dynamical friction time calculated at $R_h$ exceeds a Hubble time for galaxy masses above $\sim 10^{11} \Ms$.

Under the simplest assumption that the infall rate is roughly constant, we can calculate the star clusters {\it burning time}, namely the time after which all the clusters orbiting inside $R_g$ spiralled into the galactic centre
\begin{equation}
\tau_{\rm burn}(R_g) = 1.25\times 10^{-2} \times 2^{\gamma} M_g \left[M_\gc \dot{N}_\gc(R_g)\right]^{-1} ,
\end{equation}
where we used Equation \ref{NGCc} to calculate the number of clusters at $R_g$. As shown in  Figure \ref{FA1}, the cluster burning time ranges between $\sim 4$ and $6$ Gyr, with the lower values attained at larger galaxy masses. Note that this timescale depends on the average cluster mass, its average orbital eccentricity, and the galaxy density slope. For instance, larger $M_\gc$ or lower $e_\gc$ values can increase the burning time up to 10 Gyr, but at the same time can lead to $\tau_{\rm DF}(R_g)>10$ Gyr for galaxies heavier than $10^{11}\Ms$. 

In the delivery scenario for BHB formation, the burning time represents the timescale over which spiralling clusters sustain the BHB reservoir replenishment. Therefore, our analysis suggests that the BHBs deposit via clusters orbital segregation can persist up to $4-6$ Gyr.

\end{document}